\documentstyle[sprocl,here,epsfig,tables]{article}

\def\mycapt#1{\protect\caption{#1}}

\def\journal#1&#2(#3){\unskip, \sl #1\unskip~\bf\ignorespaces #2\rm
(19#3)}

\def\VEV#1{\langle#1\rangle}
\def\mybf#1{ {\bf #1}}

\def\beq{\begin{equation}}
\def\eeq{\end{equation}}
\def\bea{\begin{eqnarray}}
\def\eea{\end{eqnarray}}

\newcommand{\Du}{\hbox{$\Delta u$}}
\newcommand{\Dd}{\hbox{$\Delta d$}}
\newcommand{\Ds}{\hbox{$\Delta s$}}

\def\ZZ{\raise0.12em\hbox{\scriptsize{$
\not
\kern0.15em\not
\kern-0.21em\lower0.2em
\vbox{\hrule width 0.52em height 0.06em depth 0pt}
\kern-0.50em\raise0.65em
\vbox{\hrule width 0.52em height 0.06em depth 0pt}
\,$}}}

\def\A{{\cal A}}

\def\bra#1{\langle #1|}
\def\ket#1{| #1\rangle}

\def\ubu{\bar{u}u}
\def\pbp{\bar{p}p}
\def\dbd{\bar{d}d}
\def\sbs{\bar{s}s}

\def\mc{\,,\ }

\newcommand{\naive}{na\"{\i}ve}
\newcommand{\role}{r\^ole}

\catcode`\@=11 % This allows us to modify PLAIN macros.
\def\lsim{\mathrel{\mathpalette\@versim<}}
\def\gsim{\mathrel{\mathpalette\@versim>}}
\def\@versim#1#2{\vcenter{\offinterlineskip
        \ialign{$\m@th#1\hfil##\hfil$\crcr#2\crcr\sim\crcr } }}

\def\@ccitex[#1]#2{\if@filesw\immediate\write\@auxout
        {\string\citation{#2}}\fi
\def\@ccitea{}\@ccite{\@for\@cciteb:=#2\do
        {\@ccitea\def\@ccitea{,}\@ifundefined
        {b@\@cciteb}{{\bf ?}\@warning
        {Citation `\@cciteb' on page \thepage \space undefined}}
        {\csname b@\@cciteb\endcsname}}}{#1}}

\newif\if@cghi
\def\bcite{\@cghitrue\@ifnextchar [{\@tempswatrue
        \@ccitex}{\@tempswafalse\@ccitex[]}}
\def\ccitelow{\@cghifalse\@ifnextchar [{\@tempswatrue
        \@ccitex}{\@tempswafalse\@ccitex[]}}

\def\@ccite#1#2{{[#1]\if@tempswa\typeout
        {IJCGA warning: optional citation argument
        ignored: `#2'} \fi}}

\catcode`\@=12 % at signs are no longer letters
\def\lappeq{\lsim}
\def\gappeq{\gsim}

\def\slasha#1{\setbox0=\hbox{$#1$}#1\hskip-\wd0\hbox
to\wd0{\hss\sl/\/\hss}}

\def\slashb#1{\setbox0=\hbox{$#1$}#1\hskip-\wd0\dimen0=5pt\advance
       \dimen0 by-\ht0\advance\dimen0 by\dp0\lower0.5\dimen0\hbox
         to\wd0{\hss\sl/\/\hss}}

\def\defeq{\equiv}
\newcommand{\qL}{q_{\scriptscriptstyle L}}
\newcommand{\qR}{q_{\scriptscriptstyle R}}
\newcommand{\qbL}{\bar{q}_{\scriptscriptstyle L}}
\newcommand{\qbR}{\bar{q}_{\scriptscriptstyle R}}

\newcommand{\syst}{~{\rm (syst.)}~}
\newcommand{\stat}{~{\rm (stat.)}~}
\newcommand{\eqref}[1]{(\ref{#1})}   % for referencing eqs by name
\def\Toprel#1\over#2{\mathrel{\mathop{#2}\limits^{#1}}}

\def\etal{{\em et al.}}
\def\PL{{\em Phys. Lett.\ }}
\def\NP{{\em Nucl. Phys.\ }}
\def\PR{{\em Phys. Rev.\ }}
\def\PRL{{\em Phys. Rev. Lett.\ }}

\def\eqr#1{equation~\eqref{#1}}

\begin{document}
\title{
% the title page is controlled by sprocl.sty template, so
% in order to put in report numbers make them part of the title
\begin{flushright}
{\baselineskip12pt \normalsize
CERN-TH/95-334\\
TAUP-2316-96\\
hep-ph/9601280\\
$\phantom{a}$
}
\end{flushright}
THE STRANGE SPIN OF THE 
NUCLEON\protect\footnote{Invited lectures at the
Int. School of Nucleon Spin Structure,
Erice, August 1995.}
}

\author{ JOHN ELLIS}

\address{Theory Division, CERN, CH-1211, Geneva 23, Switzerland,\\
e-mail: johne@cernvm.cern.ch }

\author{MAREK KARLINER}

\address{
School of Physics and Astronomy,\\
Raymond and Beverly Sackler Faculty of Exact Sciences
\\ Tel-Aviv University, 69978 Tel-Aviv, Israel
\\ e-mail: marek@vm.tau.ac.il}


\maketitle\abstracts{
%\noindent{\bf Introduction}
The recent series of  experiments on polarized
lepton-nucleon scattering have provided a strange 
new twist in the story of the nucleon, some of 
whose aspects are reviewed in these lectures. In the 
first lecture,
we review some issues arising in the
analysis of the data on polarized structure functions,
focusing in particular on the importance and treatment
of high-order QCD perturbation theory. 
In the second lecture
some possible
interpretations of the ``EMC spin effect" are reviewed,
principally in the chiral soliton (Skyrmion) approach,
but also interpretations related to the axial $U(1)$ anomaly.
This lecture also discusses other indications from  
recent LEAR data for an $\bar{s} s$ component in the 
nucleon wave function, and discusses test of a model 
for this component. Finally, the third lecture reviews
the implications of polarized structure functions 
measurements for experiments to search for cold dark 
matter particles, such as the lightest supersymmetric
particle and the axion, after reviewing briefly the
astrophysical and cosmological evidence for cold dark matter.
} % end of abstract
\setcounter{figure}{0}
\setcounter{equation}{0}
\section{Polarized Structure Functions
\protect\footnote{This section is updated from 
Ref.~\protect\bcite{Paris}.}
}

%Pol. s.f. stuff to be implanted here
\subsection{Formalism}
 
    The basis for our discussion will be the two spin-dependent structure
functions $G_1$ and $G_2$:
\bea
\frac{d^2\sigma^{\uparrow\downarrow}}{dQ^2d\nu} -
\frac{d^2\sigma^{\uparrow\uparrow}}
{dQ^2d\nu} =
\phantom{aaaaaaaaaaaaaaaaaaaaaaaaaaa}
\nonumber \\
\frac{4\pi\alpha^2}{Q^2E^2}~\bigg[M_N(E+E^{\prime}\cos
\theta )G_1(\nu ,Q^2)
- Q^2G_2(\nu ,Q^2)\bigg]
\label{IE1}
\eea
In the parton model, these structure functions scale as follows in the
Bjorken limit
$x = {Q^2}/{2M_N\nu}$ fixed,  $Q^2\rightarrow\infty$:
\hfill\break
\vbox{
\bea
M_N^2\nu\, G_1(\nu ,Q^2) \defeq g_1(x,Q^2)
\rightarrow g_1(x) \nonumber\\
\label{IE3} \\
M_N\nu^2
G_2(\nu ,Q^2) \defeq g_2(x,Q^2) \rightarrow g_2(x)
\nonumber
\eea
}
\hfill\break
We will discuss the scaling structure function $g_2$ later on, focusing
for now on $g_1$, which is related to the polarized quark distributions
by
\bea
g^p_1(x) &=& {1\over 2} \sum_q~e^2_q[q_{\uparrow}(x) - q_{\downarrow}(x)
+ \bar
q_{\uparrow}(x) - \bar q_{\downarrow}(x)]\\ \nonumber
&=&  {1\over 2} \sum_q~\Delta q(x)
\label{IE4}
\eea
for comparison, the unpolarized structure function $F_2$ is given by
\beq
F_2(x) = \sum_q e^2_qx[q_{\uparrow}(x) + q_{\downarrow}(x) + \bar
q_{\uparrow}(x) -
\bar q_{\downarrow}(x)]
\label{IE5}
\eeq
so that the polarization asymmetry $A_1$ may be written as
\beq
A_1 = \frac{\sigma_{1/2}-\sigma_{3/2}}{\sigma_{1/2} + \sigma_{3/2}}
\label{IE6}
\eeq
in the Bjorken limit, where $\sigma_{1/2}$ and
$\sigma_{3/2}$ are the virtual
photon absorption cross sections.  We will discuss later the
$Q^2$ dependences of the above formulae, as well as the transverse
polarization asymmetry.
 
    Much of the interest in the polarized structure function $g_1$ is due
to its relation to axial current matrix elements:
\bea
\langle p\vert A^q_{\mu}\vert p\rangle = \langle p\vert\bar
q\gamma_{\mu}\gamma_5q\vert
p\rangle =
\langle p\vert\qbR \gamma_{\mu}\qR -
\qbL\gamma_{\mu}\qL\vert
p\rangle =
\nonumber\\
=\Delta q\cdot S_{\mu}(p)
\phantom{aaaaaaaaaaaaaaaaaaaaaaaaa}
\nonumber
\label{IE8}\\
\eea
where $q_{\scriptscriptstyle L,R}
\equiv 1/2 (1 \mp \gamma_5) q$, $S_{\mu}$ is the nucleon
spin
four-vector, and
\beq
\Delta q \equiv \int^1_0 dx[q_{\uparrow}(x) - q_{\downarrow}(x) + \bar
q_{\uparrow}(x) -
\bar q_{\downarrow}(x)]
\label{IE9}
\eeq
Of particular interest is the matrix element of the singlet axial
current
\beq
A^0_{\mu} = \sum_{q=u,d,s} \bar q\gamma_{\mu}\gamma_5q:\quad\quad
\langle p\vert A^0_{\mu}\vert p\rangle =
\sum_{q=u,d,s} \Delta q\cdot S_{\mu}(p)
\label{IE10}
\eeq
which is related in the parton model to the sum of the light quark
contributions to the proton spin.  Prior to the series of measurements
of polarized deep inelastic lepton nuclear scattering, information
was available from charged current weak interactions on some axial
current matrix elements.  For example, neutron beta decay and strong
isospin symmetry tell us that \cite{RPP}
\beq
\Delta u - \Delta d = F+D = 1.2573 \pm 0.0028
\label{IE12}
\eeq
and hyperon beta decays and flavour $SU(3)$ symmetry tell us that
\cite{FoverD}
\beq
\frac{\Delta u + \Delta u - 2\Delta s}{\sqrt{3}}
\equiv {a_8 \over \sqrt{3}}
= {3F -D \over \sqrt{3}} = 0.34 \pm 0.02
\label{IE13}
\eeq
(for a recent discussion of the applicability of $SU(3)$
symmetry see Ref.~\bcite{LL} and references therein).
Equations \eqref{IE12} and \eqref{IE13}
give us two equations for the three unknowns $\Delta u$, $\Delta d$
and $\Delta s$.  In principle, a third piece of information was available
\cite{EK,KaplanManohar}
in 1987 from neutral current weak interactions.  Measurements of
$\nu p$ and $\bar \nu p$ elastic scattering \cite{Ahrens} indicated that
\beq
\Delta s = -0.15 \pm 0.09
\label{IE15}
\eeq
but this information was not generally appreciated before the advent
of the EMC data discussed below. At present there is a new
neutrino experiment under way at Los Alamos \cite{Garvey}
which is expected to significantly improve the precision of
\eqref{IE15} (see Ref.~\bcite{Alberico} for a recent in-depth analysis).

    In the naive parton model, the integrals of the $g_1$ structure
functions for the proton and neutron
\bea
\Gamma_1^p(Q^2) \equiv \int^1_0 dx~g^p_1(x,Q^2)
\nonumber\\
\label{GammaIdefs}\\
\Gamma_1^n(Q^2) \equiv \int^1_0 dx~g^n_1(x,Q^2)
\nonumber
\eea
are related to combinations of the $\Delta q$.
\bea
\Gamma_1^p = {1\over2}\left(
{4\over9}\Delta u
+{1\over9}\Delta d
+{1\over9}\Delta s\right)
\nonumber\\
\label{GammaIDq}\\
\Gamma_1^p = {1\over2}\left(
{1\over9}\Delta u
+{4\over9}\Delta d
+{1\over9}\Delta s\right)
\nonumber
\eea
 
The difference between the proton and neutron integrals yields the
celebrated Bjorken sum rule \cite{BJ}
\beq
\Gamma_1^p(Q^2) - \Gamma_1^n(Q^2) =
{1\over 6}
(\Delta u-\Delta
d)\times (1-\alpha_s(Q^2)/\pi ) + \,\,\ldots
\label{IE16}
\eeq
 
It is not possible to derive individual sum rules for $\Gamma_1^{p,n}$
without supplementary assumptions.  The assumption made
by Ellis and Jaffe in 1973 \cite{EJ}
was that $\Delta s = 0$, on the grounds that very possibly there were
a negligible number of strange quarks in the nucleon wave function,
and if there were, surely they would not be polarized.  With this
assumption, it was estimated that
\bea
\int_0^1 dx g^p_1 (x,Q^2)
{=}{1\over 18} (4\Delta u + \Delta d)~(1-\alpha_s/\pi + \ldots) =
\nonumber\\
\label{IE17}\\
= 0.17 \pm 0.01\,
\phantom{aaaaaaaaaaaaaaaaaaa}
\nonumber
\eea
It should be clear that this was never a rigorous prediction, and was
only intended as a qualitative indication to experimentalists of what
they might find when they started to do polarized electron proton
scattering experiments.
 
    Perturbative QCD corrections to the above relations have been
calculated~\cite{Kodaira}$^-$\cite{Larin}\ :
\bea
\int^1_0 [ \,g_1^p(x,Q^2) - g_1^n(x,Q^2)\,] =
{1 \over 6}\, |g_A| \, f(x)\,\,:
\nonumber\\
\label{bjf}\\
f(x) = 1 - x - 3.58 x^2 - 20.22 x^3 + \,\ldots
\nonumber
\eea
and
\hfill\break
\vbox{
\bea
\int^1_0 \,g_1^{p(n)}(x,Q^2) =
\phantom{aaaaaaaaaaaaaaaaaaaaaaaaa}
\nonumber\\
=\left(\pm{1\over12}\,|g_A| +{1\over36}\,a_8\right)\,f(x)
+{1\over9}\,\Delta\Sigma(Q^2)\,h(x)\,\,:
\label{ejCorr}\\
h(x) = 1 - x - 1.096 x^2 - \,\ldots
\phantom{aaaaaaaaaaaaaaaa}
\nonumber
\eea
}
where $x = \alpha_s(Q^2)/\pi$, and the dots
represent uncalculated higher orders of perturbation theory,
to which must be added higher-twist corrections which we will discuss
later.
The coefficients in \eqref{bjf},\eqref{ejCorr}
are for $N_f{=}3$, as relevant for
the $Q^2$ range of current experiments.
 
%---------------------------------------------------------------------
%  Numerically, these corrections suggest values
% \FH
% at $Q^2 = 3$ GeV$^2$.
%
% don't want to give numerical values here, because they would have
% to be purely perturbative at this point, whereas we used [2/2]
% and people might get confused
%---------------------------------------------------------------------
%
With these corrections, the Bjorken sum rule is a
fundamental prediction of QCD which can be used, for example, to
estimate a value for $\alpha_s(Q^2)$.  On the other hand, the individual
proton and neutron integrals can be used to extract a value of
$\Delta s$.
 
\subsection{The Helen of spin}
 
    Early data on polarized electron-proton scattering from SLAC-Yale
experiments \cite{oldSLACa,oldSLACb,oldSLACc}
were compatible with the prediction of equation
\eqref{IE17} within
large errors.  Over a 1000 theoretical and experimental papers
were launched by the 1987 EMC result \cite{EMC}
\bea
\int^1_0 g^p_1(x,Q^2)=0.126 \pm 0.010 \syst\pm 0.015\stat\ , \quad
\kern-1cm\phantom{a}
\nonumber\\
\hbox{at}~\langle Q^2 \rangle = 10.7~{\rm GeV}^2
\phantom{aaaaaaaaaaaaaaaaaa}
\label{IE19}
\eea
which was in {\em prima facie} disagreement with the dynamical assumption
that \hbox{$\Delta s = 0$}.
 It is worth pointing out that the small-$x$ behaviour
of $g_1^p(x)$ was crucial to this conclusion.  The earlier SLAC-Yale
data had large extrapolation errors, and the EMC data indicated behaviour
different from that in simple dynamical models.  They were, however,
consistent \cite{EK}
with the naive Regge expectation \cite{Heimann}
\beq
g^p_1(x) \simeq \sum_i c_i\,x^{-\alpha_i(0)}
\label{IE20}
\eeq
were the $\alpha_i(0)$ are
the intercepts of axial vector Regge trajectories
which are expected to lie between 0 and -0.5.   A fit to the
EMC data gave \cite{EK}
\beq
g^p_1 \sim x^{-\delta} : \quad \delta = -0.07^{+0.42}_{-0.32}~~{\rm
for}~~x < 0.2
\label{IE21}
\eeq
for $x<0.2$.
 
    Using equations \eqref{IE12},\eqref{IE13},\eqref{IE19}
 and the leading order perturbative QCD corrections
in equation \eqref{ejCorr} it was estimated \cite{BEK} that
\hfill\break
\medskip
\vbox{
\bea
\Delta u &=& \phantom{-}0.78 \pm 0.06 \nonumber \\
\Delta d &=& -0.47 \pm 0.06 \label{IE23} \\
\Delta s &=& -0.19 \pm 0.06 \nonumber
\eea
}
\medskip
Strikingly, these determinations corresponded to a total contribution
of quarks to the proton spin
\beq
\Delta\Sigma =
\Delta u + \Delta d + \Delta s = 0.12 \pm 0.17
\label{IE24}
\eeq
which was compatible with 0.  

 This has sometimes been called the
``proton spin crisis", but we think this is an over-reaction.  The
result equation \eqref{IE24} was certainly a surprise for naive models of non-
perturbative QCD, but it was not in conflict with perturbative QCD.
Moreover, shortly after the first data
became available it was shown \cite{BEK} that $\Delta\Sigma = 0 $
occurs naturally in the Skyrme model, which is
believed to reproduce the essential features of QCD in
the large-$N_c$ limit.
Alternatively, it was suggested
that the $U(1)$ axial anomaly
and polarized glue might provide
an alternative interpretation 
\hbox{\cite{DeltagI}$^-$\cite{DeltagIII}},
or a significant suppression of the QCD topological
susceptibility \cite{SVa}$^-$\cite{NSV}
might play a key \role, which would modify the \naive\ quark
model predictions.
These interpretations are discussed in more detail in Lecture II.
 
In 1972 Richard Feynman wrote
``$\dots$ {\em its [the Bjorken sum rule's] verification,
or failure, would have a most decisive effect on the direction of
future high-energy physics}".  On the other hand, we think that the
verification, or failure, of equation \eqref{IE24}
has only an indecisive effect, though a very interesting one.
 
\subsection{Evaluation of integrals}
 
\begin{figure}[htb]
\begin{center}
\mbox{\epsfig{file=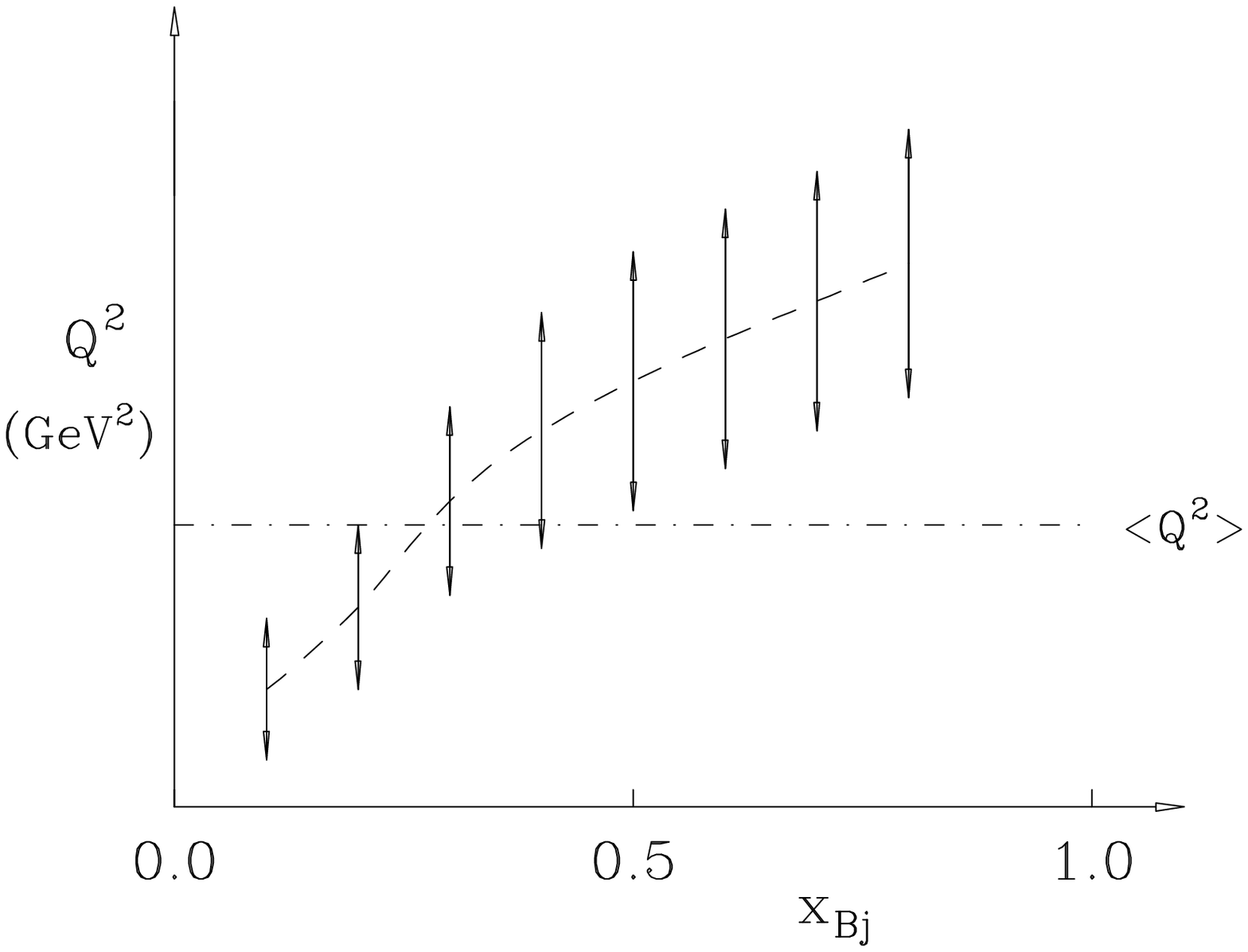,width=9.0truecm,angle=90}}
\mycapt{In any given polarized lepton-nucleon scattering
experiment,
the range of \protect $Q^2$ probed is different in different bins of the Bjorken
variable \protect  $x_{B_j}$.
}
\label{FigI}
\end{center}
\end{figure}

    Before discussing the interpretation of more recent data on polarized
structure functions, we first review a few points that arise in the
evaluation of the integrals $\Gamma_1^{p,n}$.  It should not be forgotten
that the QCD versions of the sum rules are formulated at fixed $Q^2$.
 A generic deep-inelastic sum rule in QCD reads
\beq
\Gamma (Q^2) = \Gamma_{\infty} \bigg[ 1 + \sum_{n \ge 1}
c_n~\bigg({\alpha_s(Q^2)\over\pi}\bigg)^n\bigg] + \sum_{m \geq
1}~{d_m\over (Q^2)^m}
\label{IE33}
\eeq
where $\Gamma_{\infty}$ is the asymptotic value of the sum rule for
$Q^2\rightarrow\infty$, the $c_n$ are the coefficients of the
perturbative corrections,
and the $d_m$ are  coefficients of the so-called mass and higher-twist
corrections.
On the other hand,
the data are normally obtained at values of $Q^2$ that increase
on the average with $x_{Bj}$ as seen in Fig.~\ref{FigI}.
 
 It is therefore necessary
for each individual experiment to interpolate and extrapolate to some
fixed mean value of $Q^2$, as indicated by the dashed horizontal line
in Fig.~\ref{FigI}.
The quantity measured directly is the polarization asymmetry
\eqref{IE6},
which seems experimentally to have only small dependence on $Q^2$.
In particular, a recent analysis by the E143 collaboration
\cite{E143Q2dep} sees no significant $Q^2$ dependence in $A_1$
for $Q^2 \gsim 1$ GeV$^2$, and the $Q^2$ dependence seen at lower
$Q^2$ is compatible with the magnitude of the higher-twist correction
discussed later.
Therefore experiments often assume that $A_1$ is a function of $x$ only,
and then estimate
\bea
g_1(x,Q^2) = {A_1(x,Q^2)F_2(x,Q^2)\over 2x [1 + R(x,Q^2)]}
\simeq
{A_1(x)F_2(x,Q^2)\over 2x [1 + R(x,Q^2)]}
\nonumber\\
\label{IE34}
\eea
where $F_2 (x,Q^2)$ and $R(x,Q^2)$
(the ratio of longitudinal to transverse virtual photon cross-sections)
are taken from parametrizations of
unpolarized scattering data.  Note that these induce a $Q^2$ dependence
in $g_1$ even if $A_1$ is independent of $Q^2$.
 
    The possible reliability of the assumption that $A_1$ is independent
of $Q^2$ can be explored using perturbative QCD models for
$g_1(x,Q^2)$, and  several such studies have been
made \cite{ANR}$^-$\cite{BFR2}.
Leading-order analyses indicated
a small $Q^2$ dependence in $A_1 (x,Q^2)$
(see eg. Ref.~\bcite{ANR}), but higher-order analyses question this
(see eg. Refs.~\bcite{BFR}, \bcite{BFR2},\bcite{GSNLO}). 
The amount of any such $Q^2$ dependence is sensitive to the 
polarization of the gluon density $\Delta G(x,Q^2)$, and
it may soon be possible to use data to constrain this, though 
we do not believe this is yet reliable~\cite{GSNLO}
(see, however, Ref.~\bcite{BFR2}).
It does not appear that such a $Q^2$ dependence in $g_1^{p,n}(x,Q^2)$
would have impact on the integrals $\Gamma_1^{p,n}(Q^2)$
outside the noted statistical and systematic errors. However, it could
become an important effect in the
future, and both theorists and experimentalists should keep their
eyes open.
 
    The old-fashioned assumption of Regge behaviour at low-$x$ also needs
to be checked carefully.  The leading-order perturbative QCD evolution
equations for the non-singlet part of the helicity distributions,
$\Delta q_{NS} (x,Q^2)$, lead us to expect
\cite{DGPTWZ,BEKlowx}
singular behaviour
as $x \rightarrow 0$, so that
\beq
\Delta q_{NS}(x,Q^2) \simeq
C_{NS} \exp(A_{NS}\sigma+ B_{NS} {\sigma\over\rho}
-\ln \rho -{1\over2} \ln \sigma)
\label{DqNS}
\eeq
where $A_{NS}$, $ B_{NS}$ and $C_{NS}$ are some constants
and
\beq
\sigma \equiv \sqrt{\ln{x_0\over x}\ln {t\over t_0}}\,, \quad
\rho \equiv \sqrt{{\ln {x_0\over x}\over {\ln {t\over t_0}}}}\,,
\quad
t \equiv \left(\ln{Q^2\over\Lambda^2}\over\ln{Q^2_0\over\Lambda^2}\right)
\label{DqNSdefs}
\eeq
and we might expect by analogy with the unpolarized structure
functions that $x_0 \sim 0.1$,  $Q^2_0 \sim 1$ GeV and the
leading-order QCD scale parameter $\Lambda \sim 0.25$ GeV, with
\beq
A_{NS} = {4\sqrt 2\over \sqrt{33-2N_f}}\,, \quad B_{NS} = {4\over
33-2N_f} \ .
\label{DqNSdefsII}
\eeq
where $N_f=3$ in the $Q^2$ range of current experimental
interest (see also Ref.~\bcite{BallFortePol}).
%
%\begin{figure}[htb]
%\begin{center}
%\mbox{\epsfig{file=DISfig2.ps,width=9.0truecm,angle=90}}
%$\phantom{a}$
%\vspace{5cm}
%\end{center}
%\mycapt{
%Figure 2 should go here ???
%\label{FigII}
%}
%\end{figure}
%
In principle
Eq.~\eqref{DqNS}
can be applied directly to the low-$x$ behavior of the
integrand of the Bjorken sum rule:
$g^p_1(x,Q^2)-g^n_1(x,Q^2)={1\over6}[\Delta u(x,Q^2)-\Delta d(x,Q^2)]$,
as well as to the other nonsinglet combination $\Delta u(x,Q^2)+
\Delta d(x,Q^2)-2\Delta s(x,Q^2)$ that also contributes to
$g_1^{p,n}(x,Q^2)$.
%As seen in Fig.~\eqref{FigII},
The flavour-singlet
combination of structure functions has a more complicated low-$x$
behaviour, which could be important for the extraction of the
$\Delta q$.  It is not clear whether the behaviour in equation \eqref{DqNS}
is relevant to the data presently available:  one SMC data point may
be in its region of applicability, and could in principle be used
to normalize the perturbative QCD formula, serving as a basis for
extrapolating the integrals to $x=0$.  In practice, it does not seem
at present that this would have a significant effect on the evaluation
of the Bjorken sum rule.
 
    The analysis of the polarized structure function data has often
assumed that the transverse polarization asymmetry
\beq
  A_\perp
%(x,Q^2,\phi)
= \frac{d \sigma^{\downarrow\rightarrow} -
              d \sigma^{\uparrow\rightarrow} }
             {d \sigma^{\downarrow\rightarrow} +
              d \sigma^{\uparrow\rightarrow} }\;.
\label{Atransverse}
\eeq
is negligible.  This is related to the spin-flip photon absorption
asymmetry
\beq
 A_2 = \frac{2\sigma_{TL}}{\sigma_{1/2}+\sigma_{3/2}}
\label{A2def}
\eeq
and the longitudinal $A_1$ asymmetry \eqref{IE6}
through the relation:
\beq
A_\perp = \, d \ \left(A_2-\gamma\,(1-\frac{y}{2})\,A_1\right)\;.
\label{AperpA2}
\eeq
were $\sigma_{1/2}$ and $\sigma_{3/2}$ are the virtual photon--nucleon
 absorption
 cross sections
for total helicity 1/2 and 3/2, respectively, $\sigma_{TL}$
arises from the helicity spin-flip amplitude in forward photon--nucleon
Compton scattering,
$\gamma=2 M x/\sqrt{Q^2}$,
and
$y=\nu/E_{lepton}$,
where $\nu$ is
the energy transfer in the laboratory frame.
The coefficient $d$ is related to the virtual photon depolarization
factor $D$ by
\beq
d = D\frac{\sqrt{1-y}}{1-y/2}
\label{dD}
\eeq
The asymmetries $A_1$ and $A_2$ are subject to the following positivity
conditions~\cite{POS}
\beq
    |A_1|<1\;, \qquad |A_2| \leq \sqrt{R}\; .
\label{PosCond}
\eeq
and are related to the structure functions $g_{1,2}$ by
\begin{eqnarray}
    A_1 = \frac{1}{F_1}(g_1-\gamma^2g_2)\;, \hspace{1cm}
    A_2 = \frac{\gamma}{F_1}(g_1+g_2)\;,
\label{A2}
\end{eqnarray}
where
$F_1=F_2(1+\gamma^2)/2x(1+R)$ is the spin-independent
structure function.
Recently, data on $g_2$ have become available for the first time
\cite{SMCg2,E143g2}.
They indicate that it is considerably smaller than the positivity bound
in equation \eqref{PosCond}, and is very close to the leading twist formula
of Ref.~\bcite{WW}:
\begin{equation}
    g_{2}^{\rm ww}(x,Q^2) = - g_1(x,Q^2) +\int_{x}^{1} g_1(t,Q^2)
\frac{d t}{t}\;,
\label{g2ww}
\end{equation}
The data on $g_2$ are also compatible with the Burkhardt-Cottingham
sum rule \cite{BC}
\begin{equation}
    \int_{0}^{1} g_2(x,Q^2) d x = 0\;,
\label{BCSR}
\end{equation}
which has been verified to leading order in perturbative QCD
\cite{ALNR,KodairaBC}.
The experimental errors are still considerable, in particular because
the low-$x$ behaviour of $g_2$ is less well understood than that of $g_1$.
However, the data already tell us that the uncertainty in $g_2$ is not
significant for the evaluations of the $\Gamma_1^{p,n}$.
 
\subsection{Higher orders in QCD perturbation theory:}
 
    The perturbation series in QCD is expected to be asymptotic with
rapidly growing coefficients:
\beq
S(x) = \sum^\infty_{n=0} c_n x^n~, \quad x \equiv \frac{\alpha_s}{\pi}~,
c_n
\simeq n!K^n {n}^\gamma
\label{GenericSeries}
\eeq
for some coefficients $K, \gamma$ \cite{renmvz,RenormRev}.
This type of behaviour is associated with the presence of the
renormalon singularities, as we shall discuss shortly.  Such series
are often evaluated approximately by calculating up to the ``optimal"
order, implicitly defined by
\beq
| c_{n_{opt}} \, x^{n_{opt}} |
< | c_{n_{opt}+1} \, x^{n_{opt}+1} |
\label{noptDef}
\eeq
and assuming an error of the same order of magnitude as
$c_{n_{opt}}\,x^{n_{opt}}$.  The question arises whether one can approach
or even surpass this accuracy without calculating all the terms up
to order $n_{opt}$.  This possibility has been studied using the
effective charge (ECH) approach
\cite{EffectiveCharge,KataevStarshenko}
and using commensurate scale
relations \cite{BLM,CSR}.
 In this section, we discuss the use of Pad\'e approximants
(PA's) for this purpose \cite{SEK,PBB}.
 
\begin{figure}[htb]
\begin{center}
\mbox{\epsfig{file=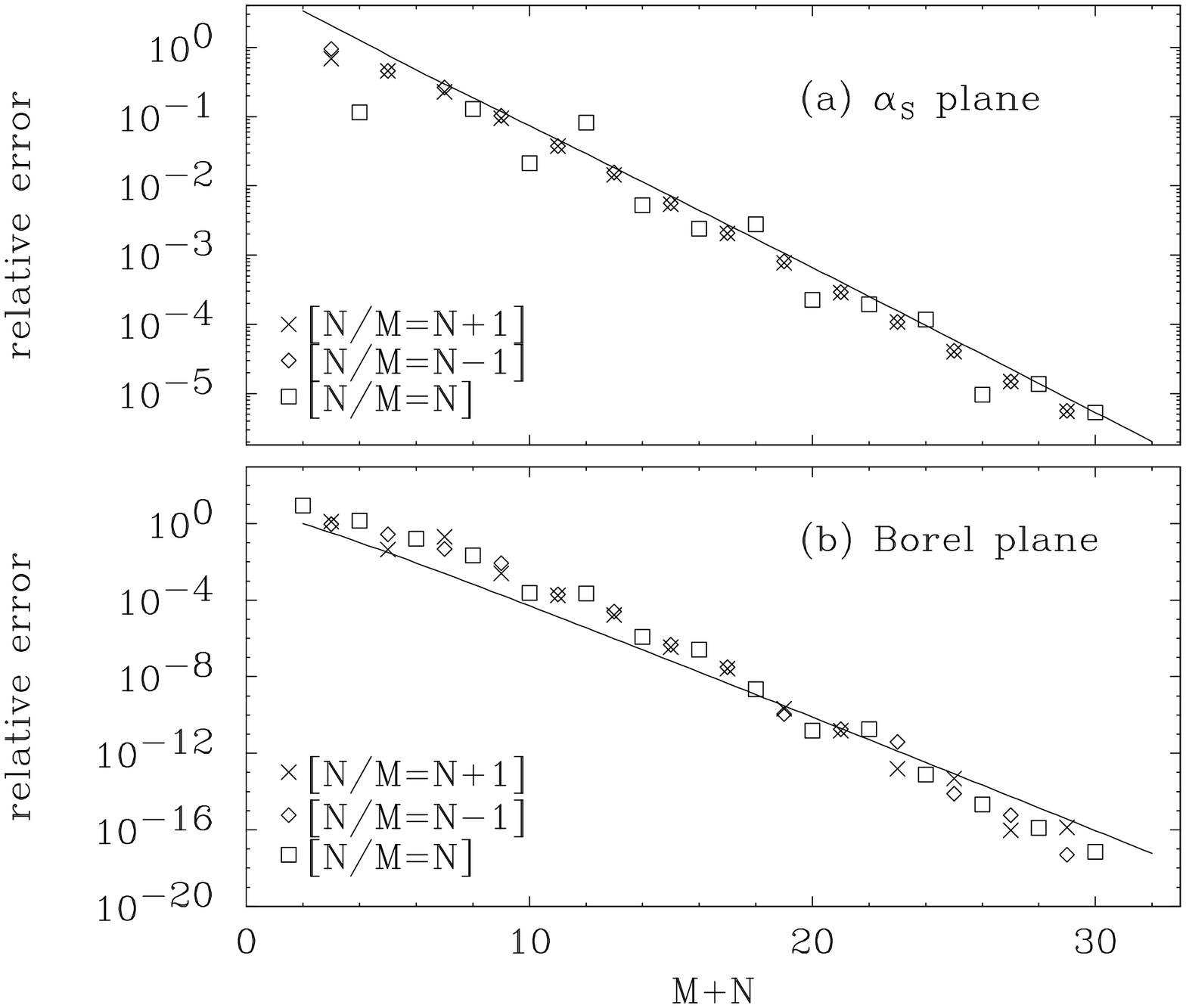,width=10.0truecm,angle=90}}
\end{center}
\mycapt{
Relative errors in the $[N/M]$ Pad\'e approximants
(a) to the
QCD vacuum polarization D-function, evaluated to all
orders
in the large-$N_f$ approximation
\protect\cite{Dfunction}
(the rate of convergence agrees with expectations for
a series with a discrete set of Borel poles),
and
(b) to the Borel transform of the D-function series,
where the convergence is particularly striking.
The straight lines correspond to the error formulae,
eqs.~\protect\eqref{IEfive} and \protect\eqref{IEeight}, respectively.
\label{FigIII}
}
\end{figure}

    Pad\'e approximants \cite{Baker,BenderOrszag}
are rational functions chosen to equal the
perturbative series to the order calculated:
\bea
[N/M] = \frac{a_0 + a_1x + ... +a_Nx^N}{1 + b_1x + ... + b_Mx^M}~:
\nonumber\\
\label{PadeDef}\\
\lbrack N/M] = S + {\cal O}(x^{N+M+1})
\phantom{aaa}
\nonumber
\eea
Under certain circumstances, an expansion of the PA in equation
\eqref{PadeDef}
provides a good estimate,
$c^{est}_{N+M+1}$,
the Pade Approximant Prediction (PAP),
for the next coefficient $c_{N+M+1}$ in the perturbative series
\cite{PAPconvergence}.
  For
example, we have demonstrated that if
\beq
\epsilon_n \equiv \frac{c_{n}\, c_{n+2}}{c^2_{n+1}} - 1
\,\simeq \,{1\over n},
\label{epsDef}
\eeq
as is the case for any series dominated by a finite number of
renormalon singularities, then $\delta_{[N/M]}$ defined by
\beq
\delta_{[N/M]} \equiv \frac{c^{est.}_{N+M+1} -
c_{N+M+1}}{c_{N+M+1}}
\label{IEfour}
\eeq
has the following asymptotic behaviour
\beq
\delta_{[N/M]} \simeq
-\,\frac{M!}{L^M  }\,, \quad~{\rm where}~ L = N+M+a\,M
\label{IEfive}
\eeq
and where $a$ is a number of order 1 that depends on the series under
consideration.  This prediction agrees very well with the known
errors in the PAP's for the QCD vacuum polarization D function
calculated in the large $N_f$ approximation \cite{Dfunction},
as seen in Fig.~\ref{FigIII}a.
 
Large-$N_f$ calculations of the perturbative corrections to the
Bjorken sum rule \cite{largeNfBjSR}
indicate the presence of only a finite number of
renormalon singularities, so that the PAP's should be accurate.
Using the known terms in equation \eqref{bjf}, the [1/2] and [2/1] PAP's
yield the following estimates for
the fourth-order coefficient \cite{PBB}:
\bea
c^{Bj}_{4[PA]}\approx {-}111\qquad (\,\hbox{[1/2] \ PA)}
\nonumber\\
\label{bjpa}\\
c^{Bj}_{4[PA]}\approx {-}114\qquad (\,\hbox{[2/1] \ PA)}
\nonumber
\eea
and the error estimator in equation \eqref{IEfive} with $a=1$ yields
\beq
\delta_{[1/2]} \simeq {-}1/8; \qquad
\delta_{[2/1]} \simeq {-}1/4
\label{bje}
\eeq
These results can be combined to obtain
\beq
c^{Bj}_{4[PA]}=
{1\over2}\left(\,
{-111\over 1 + \delta_{[1/2]}}
\,+\,
{-114\over 1 + \delta_{[2/1]}}
\,\right)
\approx  {-}139
\eeq
which is very close to the ECH estimate \cite{KataevStarshenko}
\beq
c^{Bj}_{4[ECH]} \simeq {-}130
\label{bjech}
\eeq
 
\begin{figure}[htb]
\begin{center}
\mbox{\epsfig{file=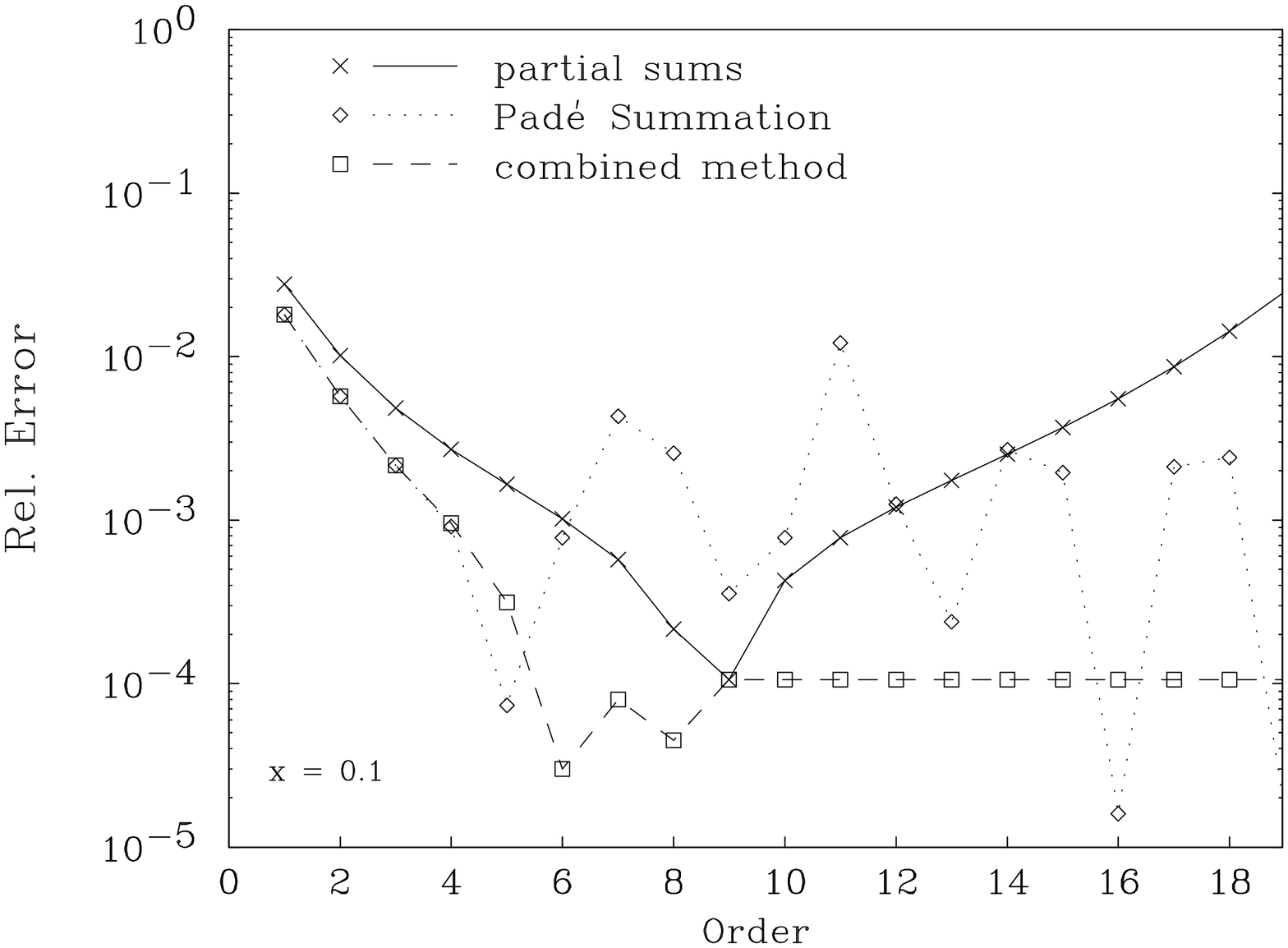,width=10.0truecm,angle=90}}
\end{center}
\mycapt{
The relative errors between partial sums of the series
$S(x) = \Sigma n! x^n$ and the Cauchy principal value of the series
(solid line)
is compared with the relative errors of Pad\'e \ Sums \ (dotted line).
\ We \ see \ that \ the \ relative
errors \ of \ the \ Pad\'e \ Sums are smaller
than those of the partial sums in
low orders, \ fluctuate in an intermediate r\'egime, \ and \ are again
\ more \ accurate than the partial sums \ in \ higher orders. \ The
fluctuations \ are associated with nearby poles in the Pad\'e Sums,
that may be treated by the ``combined method"
mentioned in the text, shown as the dashed line.
\label{FigIV}
}
\end{figure}

    A second application of PA's is to ``sum" the full perturbative
series.  The latter is ambiguous if the perturbative series possesses
an infrared renormalon singularity, i.e. a divergence of the form
in equation \eqref{GenericSeries} with $K > 0$.
Consider the following toy example:
\beq
\sum\limits_0^\infty n! \, x^n =
\int^\infty_0~{e^{-t}\over 1-xt}~~dt = {1\over x}
\int^\infty_0~{e^{-y/x}\over
1-y}~dy
\label{IEseven}
\eeq
which exhibits an infrared renormalon pole at $y = 1$.  One possible
way to define the ambiguous integral on the right hand side of
equation \eqref{IEseven} is via the Cauchy principle
value prescription \cite{PvaluePrescription}.
We see
in Fig.~\ref{FigIV}
that the errors in the Pad\'e "Sums" (PS's) $[N/M] (x)$ are
smaller than the truncated perturbative series
\beq
\sum\limits_{n=0}^{N+M} n! \, x^n
\eeq
when $n < n_{opt}$,
which is 5 in this example.  You will notice in
Fig.~\ref{FigIV} that the errors in the PS's become
unstable for large $n$:  this is
because of nearby poles in the denominator of equation
$\eqref{PadeDef}$ which are
not important for small $n$.  Also shown in Fig.~\ref{FigIV}
 as ``combined method"
is a systematic approach to treating these poles and optimizing the PS's
for large $n$, which is described elsewhere \cite{next}.
 
\begin{figure}[htb]
\begin{center}
\mbox{\epsfig{file=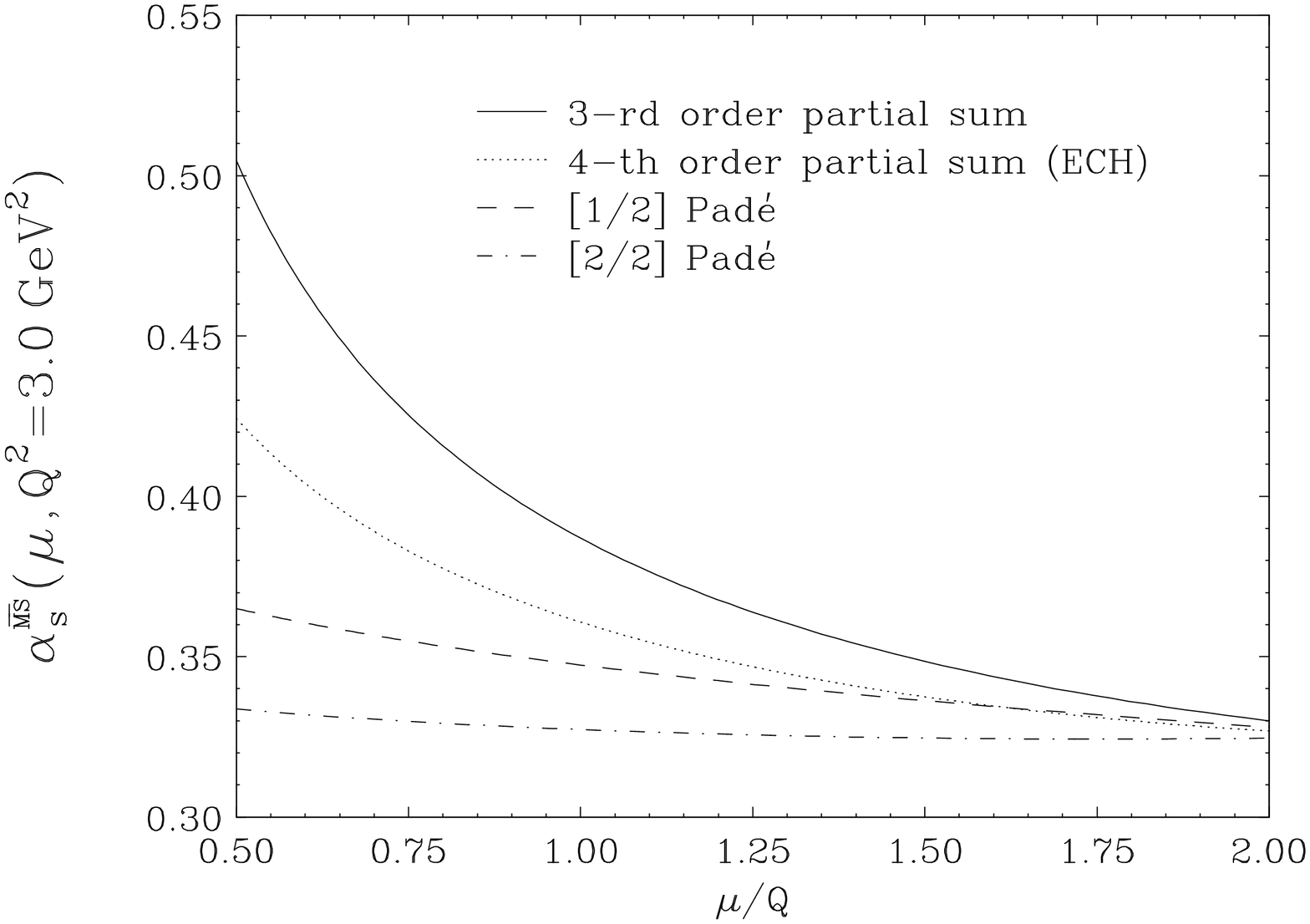,width=11.9truecm,angle=90}}
\end{center}
\mycapt{
%\hyphenpenalty=-1000
The scale dependence of $\alpha_s(3 {\rm GeV}^2)$ obtained from
a \ fixed \ value
\ \ $f(x) = (6/ g_A) \times 0.164 =0.783$,
\ ({\em cf.} eq.~\protect\eqref{N}$\,$)\,,
\ for $Q/2 < \mu < 2Q$, \ \ using \ the \ naive \ third- \ and
\ fourth-order
per\-turbative series and the [1/2] and [2/2] PS's.
\label{FigV}
}
\end{figure}
 
    Evidence that the PS's for the Bjorken sum rule provide a good
estimate of the perturbative correction factor in equation
\eqref{bjf} is
provided by the study of the renormalization scale dependence.  We see
in Fig.~\ref{FigV}
that the renormalization scale dependence of the [2/2]
PS is much smaller than that of the [2/1] and [1/2] PS's, which is
in turn much smaller than that of the naive perturbation series
evaluated to third order.  We recall \cite{CollinsRG}
that the full correction factor
should be scale-independent, and interpret Fig.~\ref{FigV}
as indicating that the PS's may be very close to the true result.
 
\begin{figure}[htb]
\begin{center}
\mbox{\epsfig{file=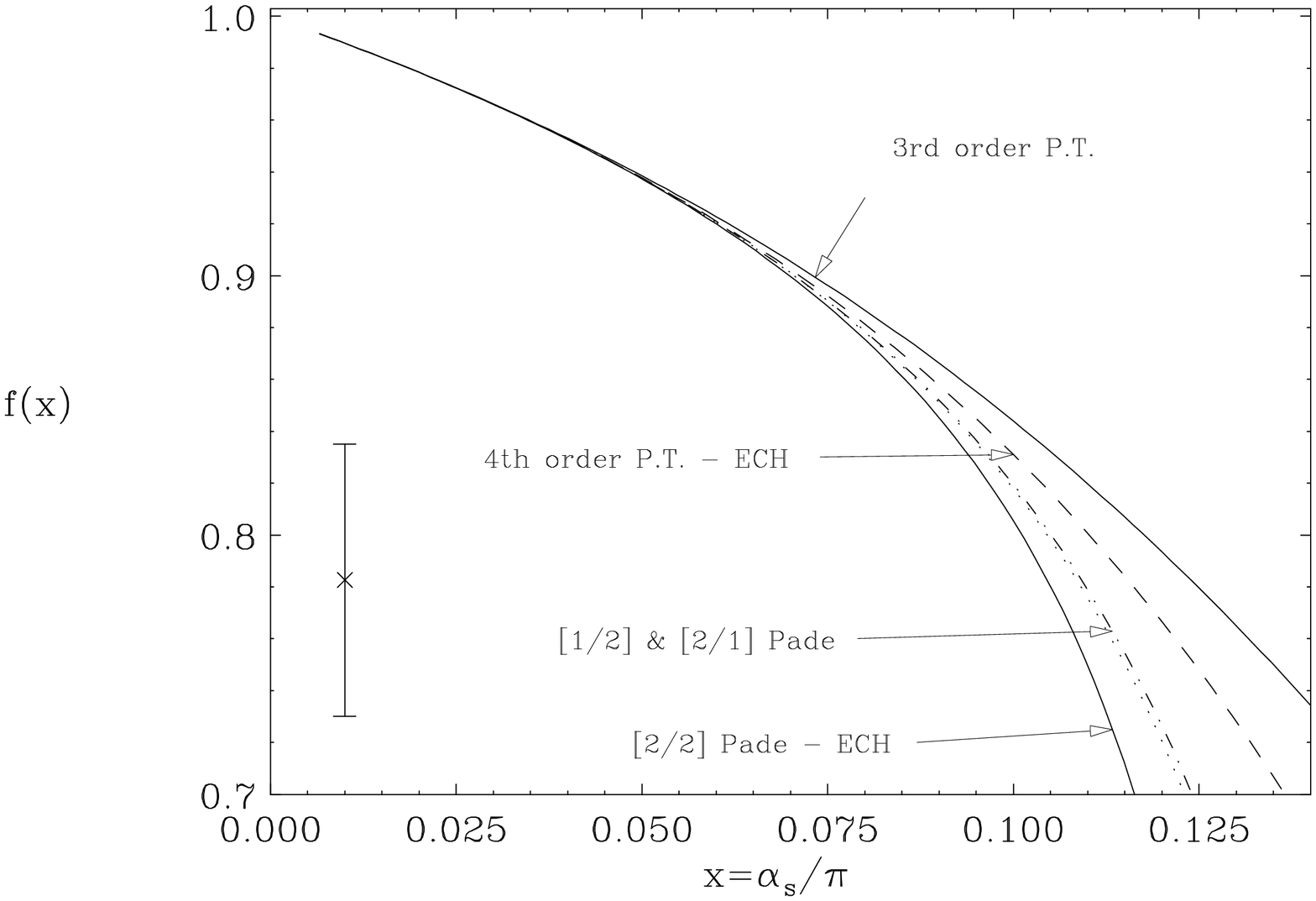,width=4.7truein,angle=90}}
\end{center}
\mycapt{
%\hyphenpenalty=-1000
Different \ approximations \ to \ the \ Bjorken \ sum \ rule correction
factor $f(x)$, third-order and fourth-order perturbation theory,
\ [1/2], \ [2/1] and \ [2/2] \ Pad\'e \ Sums
\ are compared. Also shown
as a vertical error bar is the value of $f(x)$
we extract from the available polarized structure data
\protect\eqref{N}.
\label{FigVI}
}
\end{figure}
 
    Fig.~\ref{FigVI} shows the estimates of the perturbative QCD correction
to the Bjorken sum rule obtained in various different approximations,
including third-order perturbation theory, fourth-order perturbation
theory estimated using the ECH technique and the [2/1], [1/2] and
[2/2] PS's.  We interpret the latter as the best estimator, and
take the difference between it and the [2/1] and [1/2] PS's as a
theoretical uncertainty.  Also shown in Fig.~\ref{FigVI} is the experimental
error on this quantity, as extracted from the combined analysis
of the available experimental data discussed in the next section.
 
    More information can be extracted by considering PA's in the
Borel plane.  The Borel transform of a perturbative series is
defined by
\bea
S(x) = \sum_{n=0}^\infty c_n x^n
\mathop{\quad\longrightarrow\quad}\limits_{\hbox{\scriptsize Borel}}
\phantom{aaaaaaaaaaaaa}
\nonumber\\
\tilde S(y) \equiv \sum^\infty_{n=0} \tilde c_n y^n~:~\tilde c_n =
\frac{c_{n+1}}{n!}\,\left({4\over \beta_0}\right)^{n+1}
\label{IEsix}
\eea
where $\beta_0 = (33- 2N_f)/3$.
A discrete set of renormalon singularities
would show up as a set of finite-order poles in this plane
\beq
{r_k\over (y-y_k)^P}
\label{rPoles}
\eeq
The PA's in equation \eqref{PadeDef}
are clearly well suited to find the locations
$y_k$ and the residues $r_k$ of such poles.
In the case of a perturbative series dominated by a finite set of
$L$
renormalon singularities, a sufficiently high-order PA will be {\em exact}
\beq
[M/N](y) = \tilde S(y):\qquad {\rm for}~M+N > L_0
\label{PadeLimit}
\eeq
for some $L_0\propto L$.
Generically, in any case where the quantity analogous
to eq.~\eqref{epsDef}, $\tilde\epsilon_n\simeq 1/n^2$,
the error analogous to \eqref{IEfour}  is given by
\beq
\tilde \delta_{[M/M]} \simeq - \,\frac{(M!)^2}{L^{2M}}
\label{IEeight}
\eeq
This prediction of very rapid convergence is confirmed in
Fig.~\ref{FigIII}b
\cite{SEK}
in the case of the QCD vacuum polarization D function evaluated in
the large $N_f$ limit \cite{Dfunction},
which has an infinite number of renormalon poles.
 
\begin{figure}[htb]
\begin{center}
\mbox{\epsfig{file=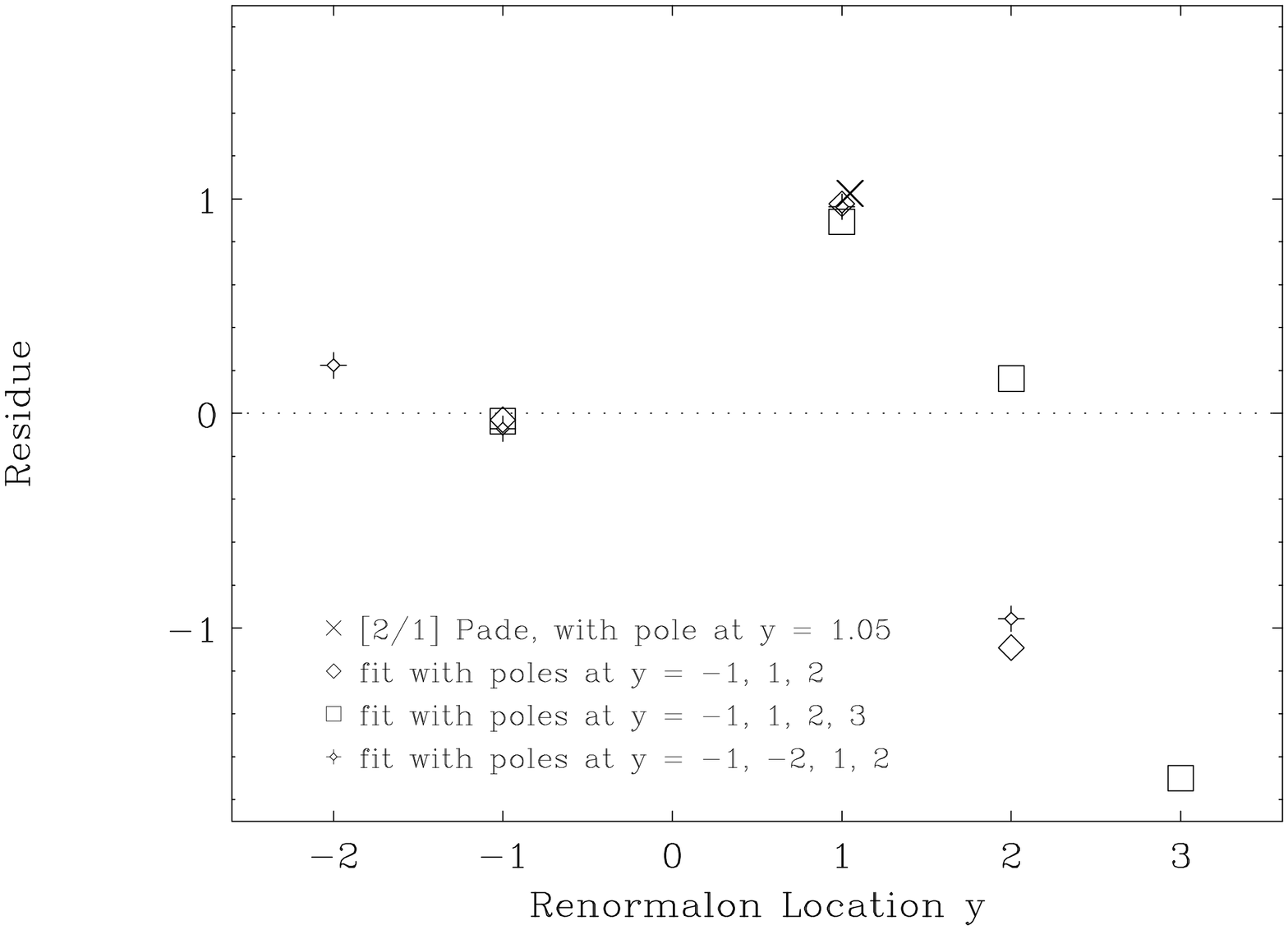,width=11.9truecm,angle=90}}
\end{center}
\mycapt{
The locations and residues of
poles \ in \ the \ [2/1] PA and \ in \ rational-function
fits to the Borel transform \ of
\ the first four terms \ in \ the perturbation series \ for
the Bjorken sum rule.
We note that the location of the lowest-lying infrared renormalon
pole is estimated accurately \ by Pad\'e Approximants in the Borel
plane, and that its residue is stable in the different fits.
\label{FigVII}
}
\end{figure}

    The power of the application of PA's in the Borel plane is
shown in Fig.~\ref{FigVII},
where we see that the [2/1] PA of the Borel-transformed
Bjorken series has a pole at
$y=1.05$\ .
The agreement with the expectation of a first infrared renormalon pole
at $y = 1$ is striking.  Fig.~\ref{FigVII} also shows other rational-function
approximations to the Borel-transformed Bjorken series.  We see evidence
that this is dominated by a strong infrared renormalon pole at $y = 1$,
that there may be a weak ultraviolet renormalon pole at $y = {-}1$ and
possibly another pole at $y = 2$.
 
    The ambiguity in the definition of the perturbative Bjorken series
associated with the $y = 1$ renormalon singularity corresponds to a
possible $1/Q^2$ correction of magnitude \cite{PBB}
\beq
\Delta\left(\Gamma^p_1 - \Gamma^n_1\right) = \pm \frac{|g_A|}{6}\,
0.98\,\pi\,\frac{\Lambda^2}{Q^2}
\label{bjrenamb}
\eeq
It is expected that the QCD correction to the Bjorken sum rule should
include a higher-twist correction of similar form with magnitude
\cite{HTrefs,BjSRalphas,BraunMoriond}
\beq
\Delta_{HT}\,\left(\Gamma_1^p - \Gamma_1^n\right)
= - \,\frac{0.02\pm0.01}{Q^2}
\label{bjht}
\eeq
The perturbative ambiguity in equation
\eqref{bjrenamb} is cancelled by a corresponding
ambiguity in the definition of the higher-twist contribution.  In the
next section we will treat equation \eqref{bjht} as a correction (with error)
to be applied to the perturbative QCD factor shown in Fig.~\ref{FigVI}.
 
    We have also compared PA's to the predictions of commensurate
scale relations within the framework of Ref.~\bcite{CSR}.  The predictions
of the two approaches are numerically very similar, and we give
formal reasons in Ref.~\bcite{next}  why we believe that this should be so.
However, we shall not use commensurate scale relations in the data
analysis of the next section.
 
\subsection{Numerical analysis of the Bjorken sum rule}
 
     Table I shows the data on the integrals
$\Gamma_1^{p,n}$ currently available from experiments at CERN
and SLAC \cite{EMC}$^,$\hbox{\cite{PSFdata}$^-$\cite{Roblin}}\ .
 We do not attempt to correct these numbers for any of the
effects discussed in section 1.4, such as the $Q^2$-dependence of
the asymmetry $A_1$, the extrapolation to low-$x$, or the
transverse polarization asymmetry. We do not believe that any
of these effects will change any of the data outside their
quoted errors. We choose to evaluate the Bjorken sum rule at
$Q^2 = 3$ GeV$^2$, which requires rescaling all the data as
described in Ref.~\bcite{PBB}.
 
\medskip
\def\pls{\phantom{{+}}}
\thicksize=1pt
\vbox{
\begin{center}
Table I \\
{\footnotesize
$\Gamma_1^{p,n,d}$ currently available from experiments at CERN
and SLAC.}
\end{center}
\medskip
\begintable
experiment | target | $\Gamma_1$                \crthick
E142       | $n$    | ${-}  0.045 \pm 0.009$    \cr
E143       | $p$    | $\pls 0.124 \pm 0.011$    \cr
E143       | $d$    | $\pls 0.041 \pm 0.005$    \cr
SMC        | $d$    | $\pls 0.023 \pm 0.025$    \cr
\ SMC ('94) \ | $d$    | $\pls 0.030 \pm 0.011$    \cr
SMC        | $p$    | $\pls 0.122 \pm 0.016$    \cr
EMC        | $p$    | $\pls 0.112 \pm 0.018$
\endtable
\begin{center}
{\footnotesize
All experimental data have been evolved to
$Q^2=3$ GeV$^2$.
}
\end{center}
} % end of \vbox
 
The following is the combined result
that we find for the Bjorken sum rule:
\beq
\Gamma_1^p(3 {\rm GeV}^2) - \Gamma_1^n(3 {\rm GeV}^2) = 0.164 \pm 0.011
\label{N}
\eeq
which is indicated by a vertical error bar
in the lower left corner of Fig.~\ref{FigVI}. Comparing this value
with the [2/2] PS estimate also shown there, we find
\beq
\alpha_s (3~{\rm GeV}^2) = 0.328^{{+}0.026}_{{-}0.037}
\label{asatQBj}
\eeq
which becomes
\beq
 \alpha_s(M_Z^2) = 0.119_{{-}0.005}^{{+}0.003}\,\, \pm \,\,\dots~\qquad,
\label{N+2}
\eeq
when we run $\alpha_s$ up to $M_Z^2$ using the three-loop
renormalization group equation. The errors quoted in equations
\eqref{asatQBj} and \eqref{N+2}
are purely experimental, and the second $\pm$ sign in
equation \eqref{N+2} indicates that further theoretical systematic errors
must be estimated. Those we have evaluated include that
associated with the renormalization scale dependence shown in
Fig.~\ref{FigV} ($\pm 0.002$), the difference between the [2/2] and
[2/1], [1/2] PS's ($\pm 0.002$), and the correction due to
the higher-twist estimate in equation
\eqref{bjht} ($ -0.003 \pm 0.002$),
whereas the error in the running of $\alpha_s$ is found to be
negligible. Combining these estimates with equation \eqref{N+2}, we find
\cite{PBB}
\beq
\alpha_s (M_Z^2) = 0.116_{-0.005}^{+0.003} \pm 0.003 ~,
\label{N+6}
\eeq
 
\begin{figure}[H]
\begin{center}
\mbox{\epsfig{file=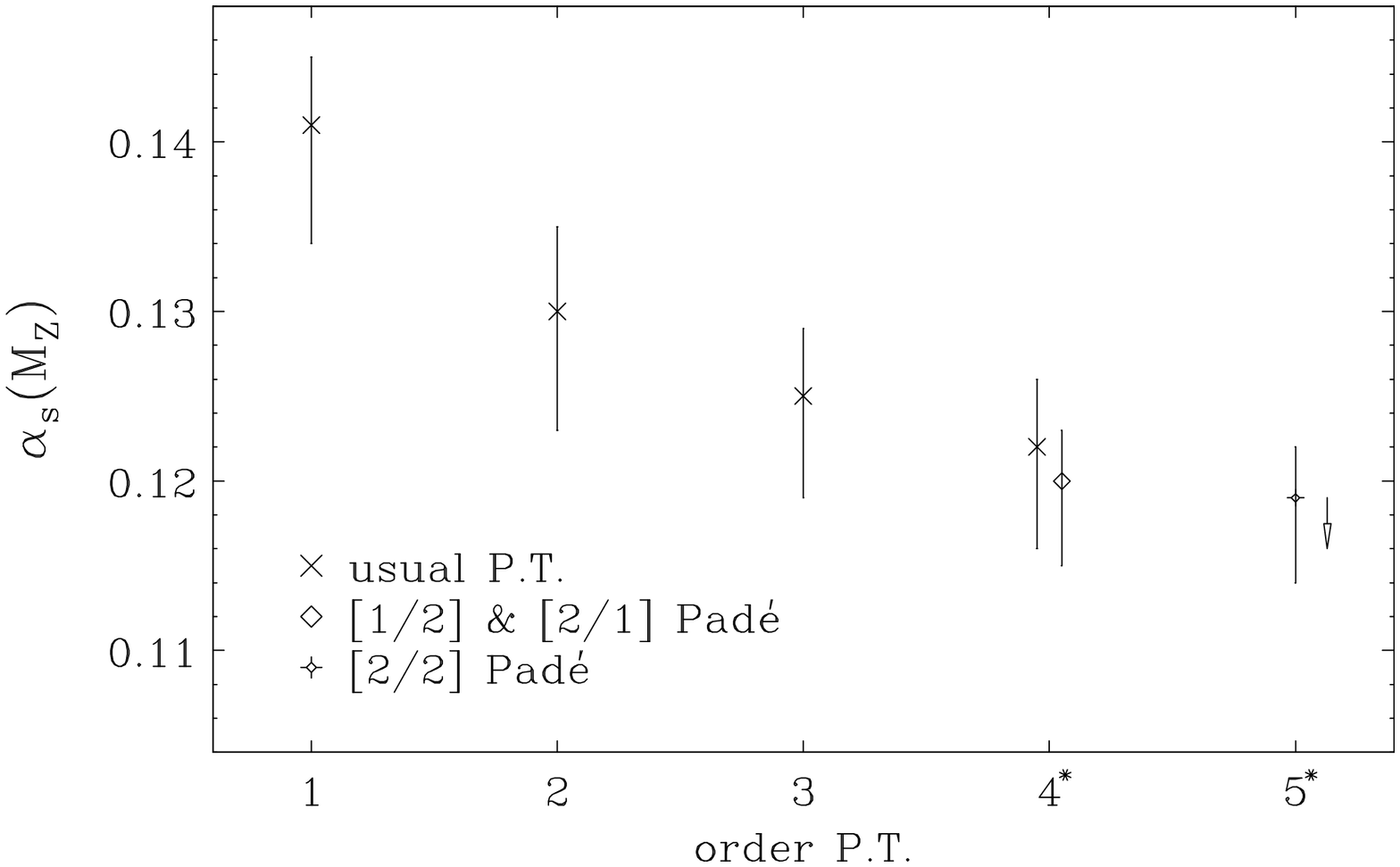,width=10.0truecm,angle=90}}
\end{center}
\mycapt{
Values of $\alpha_s(M_Z^2)$ obtained
using different orders \ of perturbation theory, \ compared \ with
\ our result \protect\eqref{N+2}, \ obtained using the [2/2] PS.
\ The size of the shift induced by higher-twist correction is
\protect\eqref{bjht}
indicated by a downward arrow to the right of the [2/2] point.
\label{FigVIII}
}
\end{figure}

The stability of this result is indicated in Fig.~\ref{FigVIII},
where we exhibit the values of $\alpha_s(M_Z^2)$ obtained
using different orders of perturbation theory, compared with
our result \eqref{N+6} obtained using the [2/2] PS. Also indicated is the
shift induced by the higher-twist correction, which lies within
our error bars.
 
\begin{figure}[H]
\begin{center}
\mbox{\epsfig{file=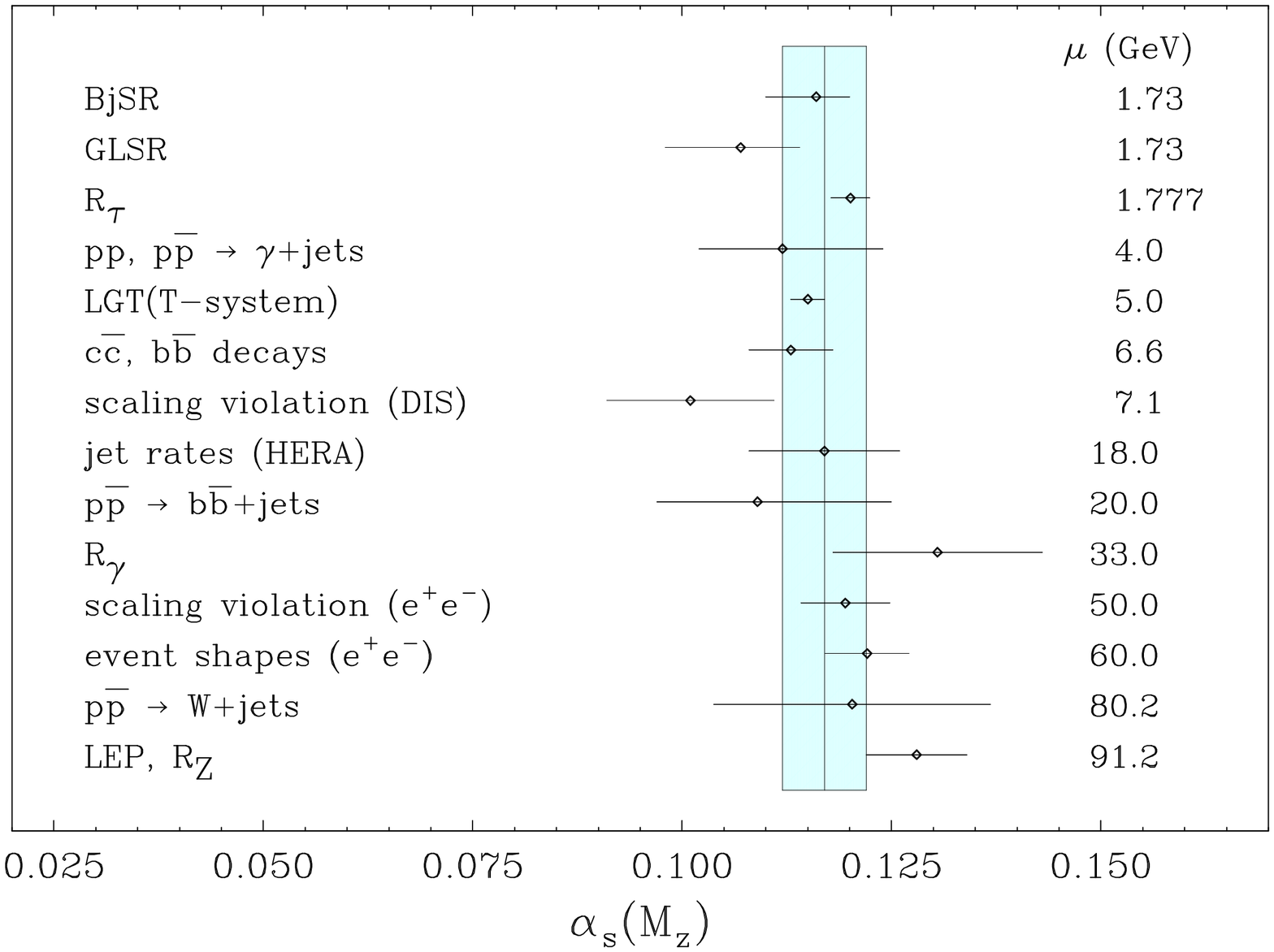,width=9.0truecm,angle=90}}
\end{center}
\mycapt{
\ \ Compilation \ of \ world data \ on
\ $\alpha_s$ \ from \ different sources
(adapted from  Ref.~\protect\bcite{Schmelling}).
\label{FigIX}
}
\end{figure}

    As can by seen from the compilation in Fig.~\ref{FigIX}, our central
value for $\alpha_s(M_Z^2)$ is quite compatible with other
determinations and the world average, which is quoted to be
\cite{RPP,Bethke,Schmelling}
\beq
 \alpha_s(M_Z^2) = 0.117\pm 0.005
\label{alphasWorld}
\eeq
Indeed, the error quoted in equation \eqref{N+6}
is quite competitive
with the most precise determinations of $\alpha_s(M_Z^2)$ that
are available. Moreover, we note that plenty of precise new
data will soon be available from the SMC experiment at CERN,
the E154 and E155 experiments at SLAC, and the HERMES
experiment \cite{HERMES}
at DESY. In the longer run, experiments with
a polarized proton beam at HERA will provide valuable
information on the behaviour of $g_1$ at low-$x$, as well as
on its $Q^2$-dependence at fixed $x$.
 
\vfill\eject
\subsection{Decomposition of the Nucleon Spin}
 
\begin{figure}[H]
\begin{center}
\mbox{\epsfig{file=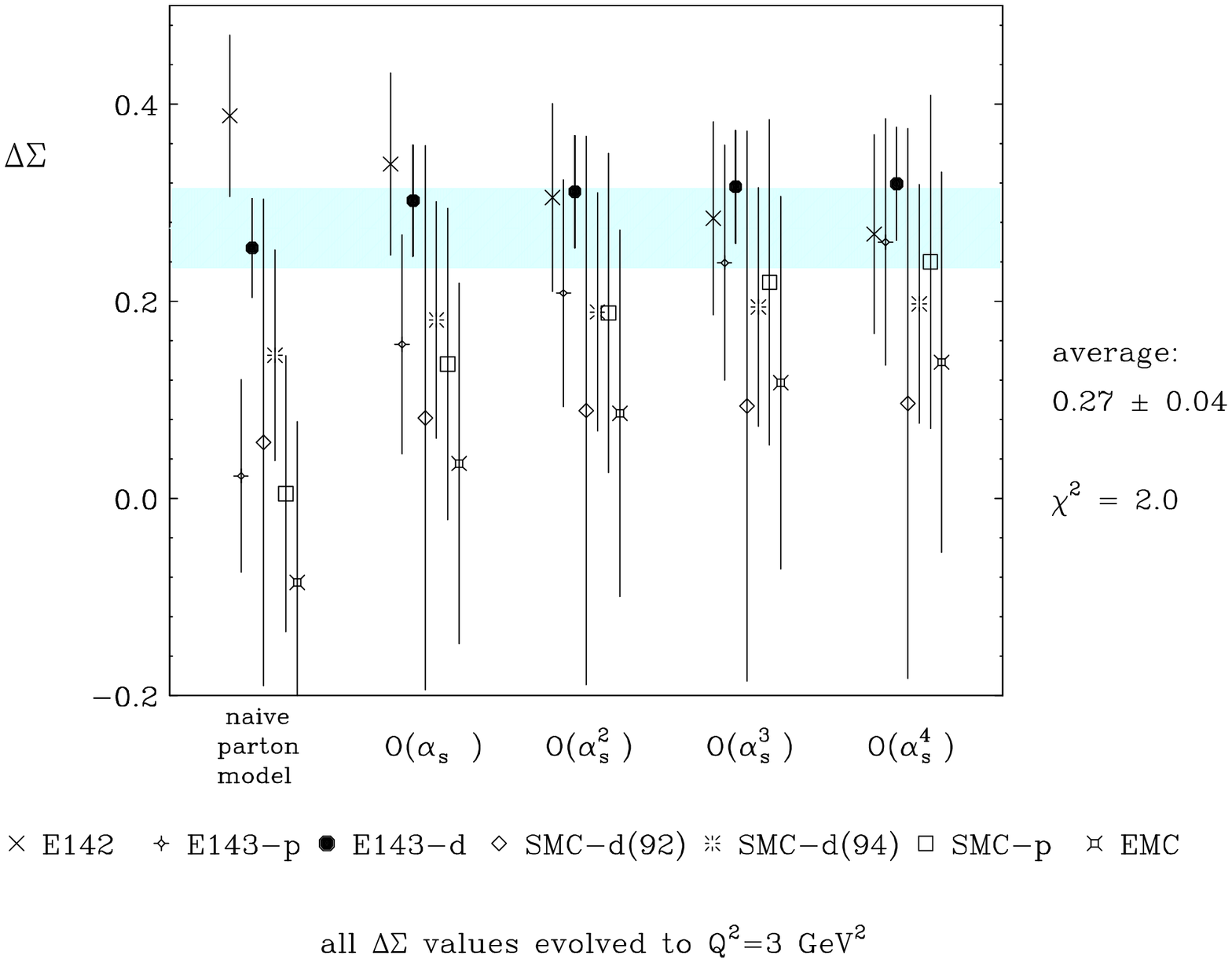,width=11.9truecm,angle=90}}
\end{center}
\mycapt{
The values of $\Delta\Sigma(Q^2{=}3\ \hbox{GeV}^2)$
extracted from each experiment, plotted as functions of the
increasing order of QCD perturbation theory
used in obtaining  $\Delta\Sigma$ from the data \
(from Ref.~\protect\bcite{BjSRalphas}, updated with most recent data).
\label{FigX}
}
\end{figure}

So far, we have only discussed the combination $\Gamma_1^p -
\Gamma_1^n$ which enters in the Bjorken sum rule. The individual
$\Gamma_1^{p,n}$ can be used as in equation \eqref{GammaIDq}, though not
forgetting the perturbative QCD corrections in equation \eqref{ejCorr}, to
extract the individual $\Delta q$. As is seen in Fig.~\ref{FigX}, the
different experiments on both proton and neutron targets are
all highly consistent, {\it once the perturbative QCD corrections
are taken into account}. 
Some time ago, it appeared as if the
neutron data from E142 might be at variance with the other
data points. However, this is no longer the case if all the
higher-order corrections in equation \eqref{ejCorr} are taken into account,
and the latest evaluations \cite{Roblin}
of the E142 data indicate a different
preliminary value of $\Gamma_1^n$, as seen in Table I. Making a
global fit, we find
\hfill\break
\vbox{
\bea
\Delta u &=& \phantom{-}0.82 \pm 0.03 \pm\ldots\nonumber \\
\Delta d &=& -0.44 \pm 0.03\pm\ldots\label{globalDqs} \\
\Delta s &=& -0.11 \pm 0.03\pm\ldots\nonumber
\eea
}
\hfill\break
and
\beq
\Delta\Sigma =
\Delta u + \Delta d + \Delta s = 0.27\pm 0.04\pm\ldots
\label{globalDsigma}
\eeq
where the second $\pm$ sign indicates that further theoretical
and systematic errors remain to be assigned.
These include
higher-twist effects, errors in the extrapolation to low-$x$
which is more complicated than for the nonsinglet combination
of structure functions appearing in the Bjorken integrand, the
possible $Q^2$-dependence of $A_1$, etc.. We believe that these
errors may combine to be comparable with the errors quoted in
equations \eqref{globalDqs},\eqref{globalDsigma},
but prefer not to quote definitive ranges for the
$\Delta q$ until all these errors are controlled as well as
those appearing in the Bjorken sum rule.
 
The SMC Collaboration has presented at this school a global
analysis of the $\Delta q$ \cite{Hughes}, in which they combine 
SLAC and CERN measurements of the structure functions to arrive
at new estimates of the $\Gamma_1^{p,n}$, which differ from
the values we have used (see Table I) principally because of their
treatment of the low-$x$ data. An analysis of a subset of the the 
available data has also been presented in Ref.~\bcite{BFR2}. Both 
these analyses differ from ours in the treatment of higher-order
perturbative QCD effects, but agree within the stated errors.

\begin{figure}[htb]
\begin{center}
\mbox{\epsfig{file=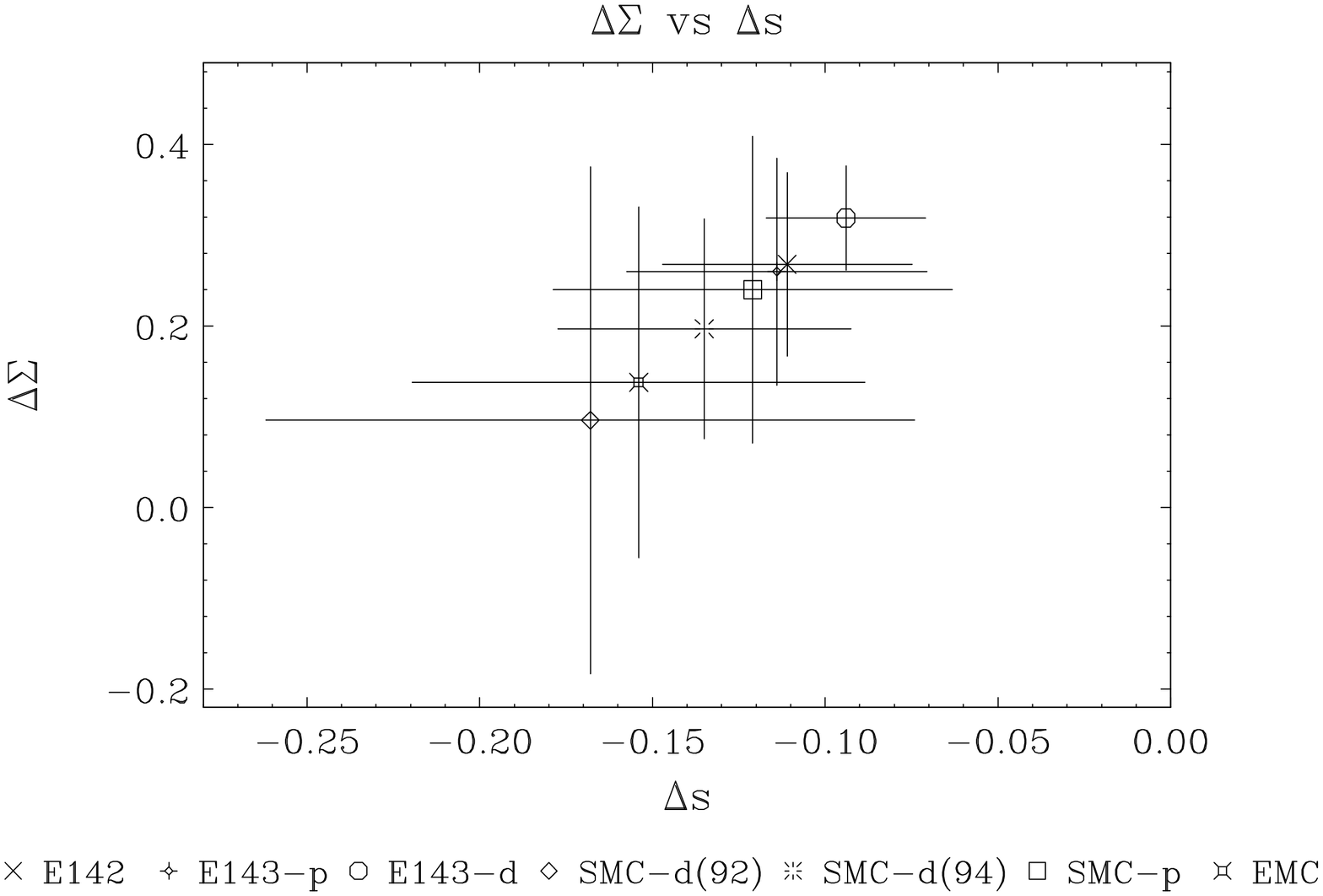,width=11.9truecm,angle=90}}
\end{center}
\mycapt{
\ \ The values \ of \ $\Delta\Sigma$ \ and \ $\Delta s$
extracted from each experiment, \ plotted \ against \ each other.
\ All \ data \ have been evolved \ to common $Q^2=3$ GeV$^2$.
The clear linear \ correlation
%\hbox{
between \ \ $\Delta\Sigma$
\ \ and \ \ $\Delta s$ \ results \ \ from \ \ the \ \ linear \ \ relations
\protect\eqref{IE12},\protect\eqref{IE13},\protect\eqref{ejCorr}.
%}
\label{FigXI}
}
\end{figure}
 
One may also get a feeling for the expected range of
$\Delta\Sigma$ and $\Delta s$ by plotting the
results for these two observables
extracted from each of the existing experiments,
as shown in Fig.~\ref{FigXI}.
 
\subsection{Outlook}
 
    In this lecture we have concentrated on the phenomenological
analysis of the data on polarized structure functions presently
available. As we have seen, these tell a remarkably consistent
story, once higher-order QCD corrections are included. We have
not addressed in great detail here the
theoretical interpretation of the data,
nor their spin-offs in hadron physics and elsewhere,
nor possible future developments in this field. In
fact, these measurements provide valuable insights into important
issues in non-perturbative QCD, such as the role of chiral
symmetry in nucleon structure \cite{BEK}, the axial anomaly and the $U(1)$
problem \cite{DeltagI}$^-$\cite{NSV},
and the relationship between current
and constituent quarks \cite{KaplanQ}$^-$\cite{FHK}
which are provoking lively
theoretical debates (see Ref.~\bcite{ConstPion} for a recent
application to the pion structure).
Some of these issues are reviewed in Lecture II.
 
The polarized structure function data support
previous indications from the $\pi$-nucleon $\sigma$-term and
elsewhere that strange quarks in the nucleon wave function cannot
be neglected, with interesting implications for the analysis of
recent data from LEAR on $\phi$ production in proton-antiproton
annihilation \cite{EKKS}, as also discussed in Lecture II.
Among other spin-offs, we recall
that the axial-current matrix elements extracted from polarization
data determine scattering matrix elements for candidate dark
matter particles such as the lightest supersymmetric particle
\cite{SmallSpinI} and the axion \cite{BjSRalphas}, as discussed
in Lecture III.
 
In the future, we look
forward to the completion of the SMC programme and its possible
HMC successor at CERN, the E154 and E155 experiments at SLAC, data
from the HERMES experiment at HERA, the polarized proton programme
at RHIC, and possible polarized electrons and protons in the HERA
ring. The tasks of these experiments will include the determination
of the gluon contribution to the nucleon spin, the elucidation of
the $Q,^2$ dependence of $A_1$, and the low-$x$ behaviour of $g_1$.
These will continue to fuel activity in this interesting field for
the foreseeable future, which will lead us to a deeper understanding
of the nucleon, an object we thought we knew so well, but which
reveals a new face when it spins.
 
\vfill\eject

\setcounter{figure}{0}
\setcounter{equation}{0}
\section{Chiral Solitons and Strangeness
in the Nucleon}
In this lecture, we review possible interpretations of the
polarized structure function data, with particular emphasis
on the chiral soliton (skyrmion) proposal \cite{BEK}, but with comments
on other possibilities  \cite{DeltagI}$^-$\cite{NSV}. We also
discuss indications from LEAR for a pattern of violations of the
Okubo-Zweig-Iizuka (OZI) rule in $\phi$ production in
$\bar{p} p$ annihilation, motivating a phenomenological model for
$\bar{s} s$ component in the nucleon wave function, whose
possible tests we also review.

\subsection{Chiral soliton interpretation of EMC spin effect}

    In this section, we will describe a possible interpretation of
EMC spin effect in terms of a chiral soliton model of hadrons.  This
has the double interest of being, as far as we know, the only model
which {\em explains}~\cite{BEK}
 the small experimental value of
$\Delta \Sigma$ and can also be derived from the underlying QCD theory,
as we now describe (see Ref.~\bcite{ChengLi} for a general
introduction to QCD and its symmetries).

    The classical QCD Lagrangian
\beq
{\cal L}_{QCD}=
{\cal L}_{YM}
%-{\textstyle{1\over2}} \hbox{tr} F_{\mu\nu}F^{\mu\nu}
- \sum_{q=1}^{N_f} (\bar q_L\slasha{D} q_L + \bar q_R\slasha{D} q_R)
-\sum_{q=1}^{N_f} m_q (\bar q_L q_R + \bar q_R q_L)
\label{E1}
\eeq
is invariant under the following global chiral transformations
\beq
q_L \rightarrow L q_L
\quad\hbox{and}\quad
q_R \rightarrow R q_R
;\quad L,R \in U(N_f)
\label{E2}
\eeq
which constitute the symmetry group
$SU(N_f)_L \times SU(N_f)_R \times U(1)_V \times U(1)_A$,
where $V = L + R$ and $A = R - L$.  As will be discussed later, the
$U(1)_A$ part of the chiral symmetry is broken by quantum effects,
and we concentrate for now on the rest of the chiral symmetry
group.  It is believed that non-perturbative QCD dynamics break
this chiral symmetry spontaneously to the subgroup  $SU(N_f)_V$.
In agreement with the Goldstone theorem, this spontaneous chiral
symmetry breaking is accompanied by the appearance of $N_f^2 - 1$
Goldstone bosons, which are the $\pi^{\pm, 0}$, $K^{\pm, 0}$,
$\bar K^0$,
$\eta_8$ in the case of $SU(3)$ with the $u$, $d$, and $s$ quarks.
In the limit of small $u, d, s$ quark masses, where $SU(3)_V$
symmetry becomes exact, the
interactions of these light pseudo-scalar bosons are described
by the following effective Lagrangian:
\bea
{\cal L}_{PS} &=&
{f_\pi^2\over 16} \hbox{Tr}
\,(\partial_\mu U \partial^\mu U^\dagger)
+{1\over 32 e^2}
\hbox{Tr}\left[
 (\partial_\mu U)U^\dagger,\partial^\mu U U^\dagger\right]^2 +
\nonumber\\
\label{E3}\\
&+&{f_\pi^2 m_{PS}^2 \over 8}\,
(\hbox{Tr}\, U -2 ) + {\cal L}_{WZ} + \ldots
\nonumber
\eea
where $U$ is a matrix which represents the pseudoscalar Goldstone boson
fields $\phi_{PS}^a$, $a=1\ldots 8$:
\beq
U = \exp\left[ {2 i \over f_\pi} \phi_{PS}^a \lambda_a \right]
\label{E4}
\eeq
The first term in \eqref{E3}
provides the $P$-wave $\pi\pi$ scattering lengths, as well
as the canonical kinetic term for the boson fields.
The second term contains higher order in the field derivatives,
such as
may be generated by vector-meson exchange, or by quark loops.  The
third term is related to the light quark masses, and yields
\beq
m_{PS}^2 \propto \Lambda_{QCD} \,m_q\,.
\label{E5}
\eeq
When $N_f \ge 3$ there is an additional Wess-Zumino term
${\cal L}_{WZ}$ which does not concern us here.

    The above effective Lagrangian \eqref{E3}  is believed to describe
accurately QCD dynamics at energies $E \ll \Lambda_{QCD}$ for the light
quarks \cite{GL}:
\beq
m_{u,d} \lsim 10\ \hbox{MeV},\qquad m_s \lsim 150\ \hbox{MeV}
\label{E6}
\eeq
It may be regarded as the first step in a systematic bosonization of
QCD, i.e., the expression of QCD dynamics in terms of meson fields.
This bosonization can be derived formally in the $1/N_c$ expansion,
which is often treated in the lowest-order approximation, though this
is not necessary.  In the $1/N_c$ expansion, one considers
a replacement $SU(3)_c \rightarrow SU(N_c)$, in which
\beq
\alpha_s(Q^2) \simeq {12 \pi \over (11 N_c - 2 N_f)\,\ln Q^2/\Lambda^2}
\label{E7}
\eeq
\begin{figure}[htb]
\begin{center}
\mbox{\epsfig{file=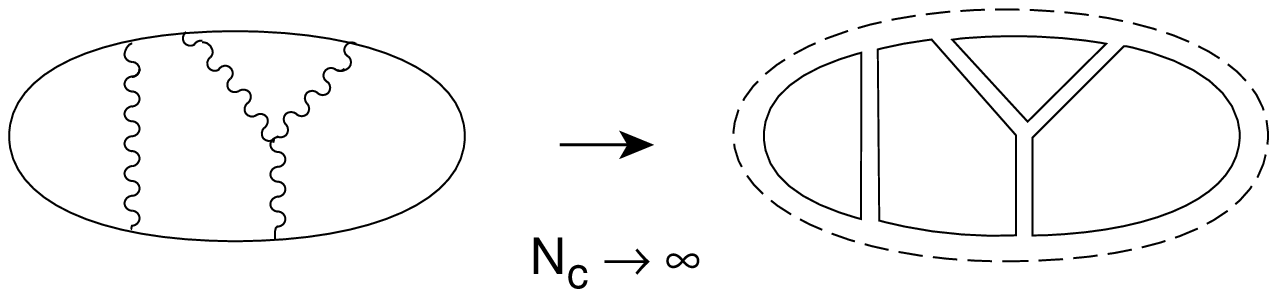,width=11.0truecm}}
\end{center}
\mycapt{Planar diagrams for mesons in the large-$N_c$ limit. The
solid lines in the right-hand part represent colour lines, and
the dashed line flavour.
\label{Fig2.1}
}
\end{figure}
We see explicitly that
for large $N_c$
 the gauge coupling strength $g^2 \sim 1/N_c$.
In this expansion, QCD is regarded as a theory of an infinite number
of mesons, whose dynamics is described in leading order by
planar diagrams such as those in Fig.~\ref{Fig2.1}.  It is easy to verify that
these are all of the same order in the limit $N_c\rightarrow\infty$,
$g^2 N_c$ fixed.  Higher orders in the $1/N_c$ expansion may be treated
systematically using diagrams of higher topological order.  In two
dimensions this programme has been realized explicitly; indeed $QCD_2$
has been ``solved" for any value of $N_c$ \cite{tH}.

    All well and good, but where are the baryons?  The answer is that
they are solitons in this apparently bosonic
theory \cite{Skyrme}$^-$\cite{WittenLewes}.
The energy of
any field configuration described by the effective Lagrangian
\eqr{E3} can be written as
\beq
E = \int d^3  \vec x \left\{
{f_\pi^2\over 16} \hbox{Tr}\, (\vec\partial U{\cdot}\vec\partial U^\dagger)
+{1\over 32 e^2} \hbox{Tr} \left[(\partial_i U) U^\dagger,
(\partial_j U) U^\dagger \right]^2
+\ldots
\right\}
\label{E8}
\eeq
This expression is finite if the following condition is satisfied:
\beq
U(\vec x ) \rightarrow 1\qquad\hbox{as}\qquad |\vec x |\rightarrow
\infty
\label{E9}
\eeq
Elementary topology tells us that such field configurations are
classified by the group
\hbox{$\Pi_3(SU(N_f)) = \ZZ$},
which characterizes all
the possible ways of mapping the 3-sphere into the chiral symmetry
group.  These different field configurations may be labeled by the
quantity
\beq
B = {1\over 24 \pi^2} \epsilon_{i j k}
\int d^3 \vec x \,\hbox{Tr}\, (\partial_i {U} U^\dagger
\partial_j U^\dagger \partial_k {U} U^\dagger )
\label{E10}
\eeq
which a topologist would identify as the winding number of the
configuration.  The quantity $B$ is related in the normal way to a
conserved current:
\beq
B = \int d^3\vec x\, J_0\,\,:\quad \partial_\mu J^\mu = 0
\label{E11}
\eeq
As discussed extensively elsewhere \cite{WittenWZ,ANW},
$B$ can be identified with baryon
number.  In the case of a spherically-symmetric soliton configuration:
\beq
U(\vec x) = \cos \theta(r) + i{\vec \tau \cdot \vec x \over r}
\sin \theta(r)\,\,:\quad r \equiv |\vec x |
\label{E12}
\eeq
one has:
\beq
B={1\over\pi} \left[\theta(0) - \theta(\infty)\right]
\label{E13}
\eeq
which is unity if $\theta(0) = \pi$ and $\theta (\infty) = 0$, which
is the case for the lowest-energy soliton with $B = 1$.  Also as discussed
elsewhere \cite{WittenWZ,ANW}
one can demonstrate that this $B = 1$ soliton is a fermion.

    The above discussion may sound rather high-falutin', but the
construction of such chiral solitons has been carried out explicitly
in QCD in two dimensions \cite{DFS}
as we will discuss shortly.

    An essential complication in the treatment of these chiral solitons
is the fact that there are many equivalent soliton configurations:
if $U_0$ is a solution, then so is
\beq
V_A \equiv V U_0 V^{-1}
\label{E14}
\eeq
where $V$ is an arbitrary spatially-constant $SU(N_f)$ matrix.  It is
essential to take account of this when one quantizes the theory.  This
is done by parametrizing $V$ in terms of its internal variables, the
 collective coordinates, and considering soliton
wave functions $\chi (V)$ in the space of such collective coordinates:
\beq
\ket{N} = \int d V \chi(V) \ket{V}
\label{E15}
\eeq
\begin{figure}[htb]
\begin{center}
\mbox{\epsfig{file=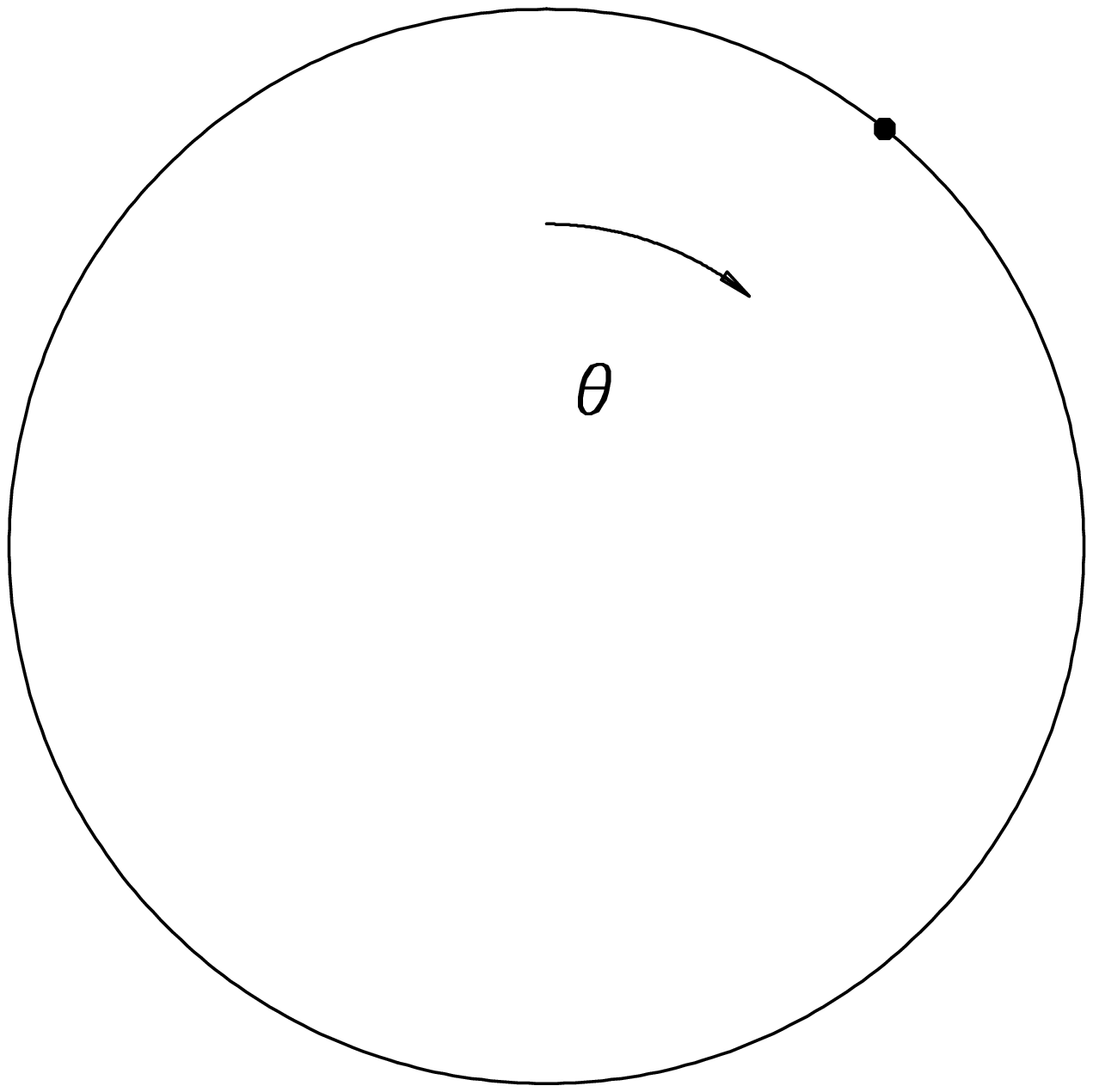,width=5.0truecm,angle=90}}
\end{center}
\mycapt{A one-dimensional analogue of the collective coordinate
$V$: particle constrained to move on a circular ring. Classical
ground state corresponds to a particle at rest at some fixed angle
$\theta$. In quantum mechanics this is no longer true and we must
have an eigenstate of the angular momentum operator
$L_\theta = {-}i{\partial\over\partial\theta}\,$.
\label{Fig2.1extra}
}
\end{figure}
At the quantum level, any state $|N\rangle$ must have
definite isospin $I$ and
spin $J$, and this requires a superposition in the internal $V$ space.
The simplest way to see this is to consider the analogous case of a
point particle located on a ring: $U_0 = e^{i \theta}$ in this case.
  Classically, the particle may sit
at any fixed value of the angle $\theta$.  However, at the quantum level
the only consistent state is an eigenstate of angular momentum,
i.e., the point particle must rotate around the ring with an angular
velocity $\omega$, which is described by $V = e^{i \omega t}\,$!
Exactly the same phenomenon occurs with the
$SU(N_f)$ chiral soliton: the quantum wave function must correspond
to a rotation in the internal $V$ space.  This rotation $\omega$ is slow
($\sim 1/N_c$)
in the large-$N_c$ limit, in which a semi-classical treatment of the
quantum state is sufficient, but one need not in principle restrict
oneself to this assumption.  In the limit of large $N_c$, the physical
parameters of the soliton behave as follows \cite{WittenB,ANW}\ :
\beq
m_N \sim N_c\,,\quad R_N \sim N_c^0\,,\quad I\sim N_c\,,\quad
J={1\over2} \sim N_c^0
\label{E16}
\eeq
where $R_N$ is the nucleon radius, and $I$ its moment of inertia.
The corresponding picture of the baryon that emerges is one of a rotating
collective state of pseudoscalar mesons. The rotation velocity
$\omega = J / I$ is slow in the limit of large $N_c$, but the underlying
physical picture is valid in principle for any value on $N_c$.

As already mentioned, the above picture has been realized
explicitly in QCD in two space-time dimensions for arbitrary
values of $N_c$ and $N_f$ \cite{DFS}. The solitons have precisely the form
of \eqr{E14}, with the matrix $U_0$ taking the form
\bigskip
\beq
U_0=\pmatrix{
1 &  &  &  &  &\cr
& &1 &  &  &  &\cr
& &  &. &  &  &\cr
& &  &  &. &  &\cr
& &  &  &  &. &\cr
& &  &  &  &  & \exp{\,-i\sqrt{4\pi\over N_c}\phi_0(x)\,\,\,}
}
\label{E17}
\eeq
\bigskip\noindent
where
\bigskip\noindent
\beq
\phi_0 (x) = \sqrt{{4 N_c \over \pi}} \tan^{-1}
\left[ \exp \sqrt{{8 \pi \over N_c}} m x \right ]
\label{E18}
\eeq
The $SU(N_f)$ rotation matrix $V$ can be written in the form
\beq
V = \pmatrix{
\tilde V & \matrix{z_1\cr z_2 \cr . \cr . \cr . \cr z_{N_f} }
}
\label{E19}
\eeq
where the form of $\tilde V$ does not concern us here, while
the parameters
\break
\hbox{$z_i : i = 1, \ldots , N_f$ }
play the role of the
collective coordinates introduced earlier. They characterize
the soliton solutions as follows:
\beq
\psi(z_i, z_j^*) =
z_1^{n_1} \times
z_2^{n_2} \times
\ldots \times
z_{N_f}^{n_{N_f}} \times
{z_1^*}^{m_1} \times
\ldots
{z_{N_f}^*}^{m_{N_f}} \,\,:
\sum_{i=1}^{N_f}\,(n_i-m_i) = N_c
\label{E20}
\eeq
The lowest-lying states lie in a {\bf 10} representation
of $SU(3)$ analogous to that containing the $\Delta$ and $\Omega$ baryons
in four dimensions.  Thus we indeed have an explicit realization of
the soliton ideas discussed previously.  In particular, it is possible
in this model to calculate the ratios of the matrix elements
$\langle B|\bar{q}q|B\rangle$ in these baryon states \cite{FK}.
In the general case of $N_f$ flavors and $N_c$ colors, one obtains
\beq
\VEV{(\bar{q}q)_{sea}}_B={1\over N_f+N_c},
\label{GenCaseSea}
\eeq
where
$(\bar{q}q)_{sea}$ refers to the non-valence quarks in the
baryon $B$.
One can also compute flavor content of valence
quarks. Consider a baryon $B$ containing $k$ quarks
of flavor $v$. The $v$-flavor content of such a baryon is
\beq
\VEV{\bar{v} v}_B={k+1\over N_f + N_c}
\label{GenCaseV}
\eeq

For $N_f=3$, $N_c=3$ one finds
\bea
\bra{\Delta^{++}}\ubu\,,\,\dbd\,,\,\sbs\ket{\Delta^{++}}
\ \propto \ 4\,,\,1\,,\,1
\nonumber\\
\label{E21}\\
\bra{\Omega^{-}}\ubu\,,\,\dbd\,,\,\sbs\ket{\Omega^{-}}
\ \propto\ 1\,,\,1\,,\,4
\nonumber
\eea
We see explicitly that the $\sbs$ matrix element in the
non-strange baryon $\Delta^{++}$ is not
negligible, and neither are the $\ubu$, $\dbd$ matrix
elements in the triply-strange baryon $\Omega^-$.

Unfortunately, it is not possible to explore axial-current
matrix elements directly in this two-dimensional model, so we
return to four dimensions.

In the picture of spontaneously-broken chiral symmetry, the matrix
elements of axial currents are given in general by the PCAC hypothesis,
namely that they are dominated by couplings of the pseudoscalar Goldstone
boson with the same quantum numbers.  We know from $\pi^{\pm}$ decay
that
\beq
\bra{0} A_\mu \ket{\pi} = i f_\pi p_\mu
\label{E22}
\eeq
and similarly for the $K^{\pm}$, and the same is believed to be true for
other members of the pseudoscalar-meson octet.  Matrix elements of
the $SU(3)$ octet axial currents are related to the couplings of the
octet pseudoscalars:
\beq
\bra{X} A^i \ket{Y} \propto f_{\pi} g_{\phi^i_{PS} X Y}\,.
\label{E22.5}
\eeq
The ninth axial current is special, because it
is not conserved at the quantum level \cite{Anomaly}:
\beq
\partial^\mu A_\mu^0 \ \propto\ g^2 F_{\mu\nu}\tilde F^{\mu\nu}
\label{E23}
\eeq
For this reason, the ninth pseudoscalar
$\eta^0$ is only a pseudo-Goldstone
boson, whose mass would be non-zero even if $m_{u,d,s} = 0$
\cite{largeNeta}\ :
\beq
m_{\eta_0} \ \propto\ 1/N_c
\label{E24}
\eeq
It should also be noted that the ninth pseudoscalar decouples from the
other eight in the limit of large $N_c\,$ 
since its couplings to them proceed via intermediate gluons
\cite{GL}\ :
for example, there is a coupling
\beq
{\cal L}_{\pi,\eta_0} =
{f_\pi^2\over 16} \,\hbox{Tr}\, (\partial_\mu U \partial^\mu U^\dagger)
{\eta_0^2\over N_c^2}
\label{E25}
\eeq
As already commented, the chiral soliton exists because of the
topology of $SU(N_f)$, not that of $U(1)$.  This means that the baryon
is to be regarded as a ``lump" of the $\pi/K/\eta_8$ mesons, {\em not}
the singlet $\eta_0$.  Moreover, \eqr{E25} indicates that the
$\eta_0$ decouples from the baryon at leading order in $1/N_c$.

    Chiral soliton calculations are generally made in the relatively
easy limits $N_c\rightarrow\infty$, $m_{u,d,s}\rightarrow 0$.  One
should beware of $N_c$-dependent predictions, which depend on the
truncation of the {\em a priori} infinite-dimensional meson theory.
Examples of such bad $N_c$-dependent quantities include
\beq
\mu_p\mc \mu_n\mc g_A\mc f_\pi\mc g_{\pi{N}N}\mc g_{\pi N \Delta}\,,
\ldots
\label{E26}
\eeq
On the other hand, calculations of $N_c$-independent quantities such as
\beq
\matrix{\,\,\,\langle r^2 \rangle_N\mc\cr (19 \%)}
\matrix{\,\,\mu_p/\mu_n \mc\cr (2\%) }
\matrix{\,\,f_\pi^2/g_A \mc\cr (3 \%)}
\matrix{\,\,g_{\pi N \Delta}/g_{\pi{N}N }\cr ( 1\% )}
%\label{E27}
\eeq
may not be so unreliable,
as indicated by the accuracy of the predictions
listed above \cite{KM.86}.
 Another example of a quantity which may be reliably
estimated in the $SU(3)_f$ limit is
\cite{DN}
\beq
{\bra{p} \sbs \ket{p} \over
\bra{p} \ubu + \dbd + \sbs \ket{p}}
={7\over30}
\label{E28}
\eeq
and this number is indeed consistent with determinations of the
$\pi$-nucleon $\sigma$ term.
The chiral soliton model has also demonstrated
successes in calculations of meson-baryon scattering phase
shifts \cite{ourPiN,SiegenPiN}.

    The chiral soliton model offers three ways of seeing that
$\langle p|A_\mu^0|p\rangle \simeq 0$, i.e. that
$\Delta u + \Delta d + \Delta s \simeq 0$ \cite{BEK}.
The brute-force method is by direct calculation using
the following standard representation of $A_a^i (x)$ in the soliton
state
\beq
\int d^3 \vec x \, A_a^i(\vec x)
\ \propto\ \hbox{Tr} \left[ \lambda_a V^{-1} \lambda_i V \right]
\label{E29}
\eeq
it is easy to see that the matrix element of $A_a^0$ vanishes because
$\hbox{Tr}(\lambda_a) = 0$.

The second way to see the vanishing of $\langle p|A_\mu^0|p\rangle$
is to consider the Goldberger-Treiman relation for the
baryonic matrix element of the ninth axial current. As discussed
above, the PCAC hypothesis relates this to the baryonic coupling
of the ninth pseudoscalar meson $\eta_0$. To see that this
coupling vanishes, it suffices to make another trivial trace
over $SU(N_f)$ indices:
\beq
g_{\eta_0{N}N} \ \propto\ \hbox{Tr}
\left( \lambda_i \hat U (\vec x)\right) = \hbox{Tr}\left(
\lambda_i V \hat U_0 V^{-1}\right)
\label{E30}
\eeq
This vanishing result can be understood at a more basic level
by remembering that the baryonic soliton consists of
$\pi$, $K$ and $\eta_8\ $
mesons alone in the large-$N_c$ limit, and that the $\eta_0$ meson decouples
from the others in this same limit.

    The third way of seeing that
$\langle p|A_\mu^0|p\rangle \simeq 0$ is to consider
the underlying physics of the soliton spin, by analogy with the particle
on a ring discussed earlier.  At the classical level, the baryon mass
is given simply by
\beq
M = \int d^3 \vec x \,\Theta_{0 0} (\vec x)
\label{E31}
\eeq
where the energy-momentum tensor $\Theta_{\mu \mu}$ is derived from a
static solution $U_0(x)$.
However, we recall that this is not an eigenstate
of spin or isospin:  these rotate with an angular velocity $\omega \sim
1/N_c$.  The resulting expression for the soliton angular momentum is
\beq
J_i = \int d^3 \vec x \,\epsilon_{i j k} x_j \,\Theta_{0 k}
(\vec x) = \omega_i\, I
\label{E32}
\eeq
where
\beq
I = {2\over3} \int d^3 \vec x \,\Theta_{0 0} \,r^2
\label{E33}
\eeq
is its moment of inertia.  Including the corresponding rotational energy,
equation \eqref{E31} is modified to become
\beq
M = \int d^3 \vec x \,\Theta_{0 0} (\vec x)
+ {J (J+1) \over 2 I}
\label{E34}
\eeq
where the quantization of angular momentum imposes
\beq
J (J+1) = {3\over4}\ \hbox{for}\ N,\quad
{15\over4}\ \hbox{for}\ \Delta
\label{E35}
\eeq
We see explicitly from this construction that {\em all} the nucleon
angular momentum is orbital in origin.  If we consider the angular
momentum sum rule
\beq
{1\over2} = {1\over2}\sum_q \Delta q + \Delta G + L_z
\label{E36}
\eeq
this argument tells us that
\beq
\Delta\Sigma = \sum_q \Delta q = 0,\qquad
L_z = {1\over2}
\label{E37}
\eeq
Simple models which extend the chiral soliton model  to include gluons
suggest also that \cite{EK}
\beq
\Delta G = 0
\label{E38}
\eeq
These results have been
 derived in the limit of large $N_c$ and small $m_{u,d,s}$.
Attempts have been made to calculate corrections to this double limit,
but these are incomplete so far.  Nevertheless, \naive\ guess and
incomplete calculations suggest that the result \eqr{E37} might be
accurate to within 30\% or so.  Since this range includes the present
experimental value discussed in Lecture I, we consider this an
encouraging success for the chiral soliton model.

    An alternative approach \cite{DeltagI}$^-$\cite{DeltagIII}
to interpreting the polarized structure
function data is based on the axial $U(1)$ anomaly, which should be
taken into account in computing matrix elements of the singlet axial
current.  The definition of the polarized quark distribution
$\Delta q(x,Q^2)$ is ambiguous at the 1-loop level.  One possible
interpretation replaces
\beq
\Delta q \rightarrow \widetilde{\Delta q} \equiv
\Delta q - {\alpha_s \over 2 \pi} \Delta G
\label{E39}
\eeq
where the second term on the right-hand side is derived from the gluon
contribution $\Delta G$ to the proton spin.  In this interpretation
all the previous determinations of $\Delta q$ should be rephrased as
determinations of the $\widetilde {\Delta q}$,
and it is possible, in principle,
to resurrect the original quark model assumption by postulating that
\beq
\widetilde{\Delta s} = \Delta s - {\alpha_s \over 2 \pi} \Delta G
\neq 0 \,:
\qquad
\Delta s = 0,\ \Delta G \neq 0
\label{E40}
\eeq
In order for this mechanism to work, the value of $\Delta G$ must be
quite large
\beq
\Delta G \simeq 2
\label{E41}
\eeq
at $Q^2 \simeq 3$ GeV$^2$.
This requires what might at first sight appear to be a rather bizarre
decomposition of the proton angular momentum:
\beq
{1\over 2} \simeq {1\over 2} + ({\simeq} 2) + ({\simeq}{-} 2)
\label{E42}
\eeq
However, it should be noted that a leading-order compensation between
large values of $\Delta G$ and $L_z$ is natural in perturbative QCD.
Nevertheless, this anomalous gluon mechanism does not really
explain the EMC result, though it may accommodate it.  It is not yet
possible to test definitively this model.  For one thing, the
$x$-dependence of the possible $\Delta G$ contribution requires
further specification.  So far we are aware of just one experimental
result which bears on $\Delta G$: a Fermilab fixed-target experiment
has searched for an asymmetry in multiple $\gamma$ production in
polarized proton-proton collisions \cite{FNALgamma}.
\begin{figure}[H]
\begin{center}
\mbox{\epsfig{file=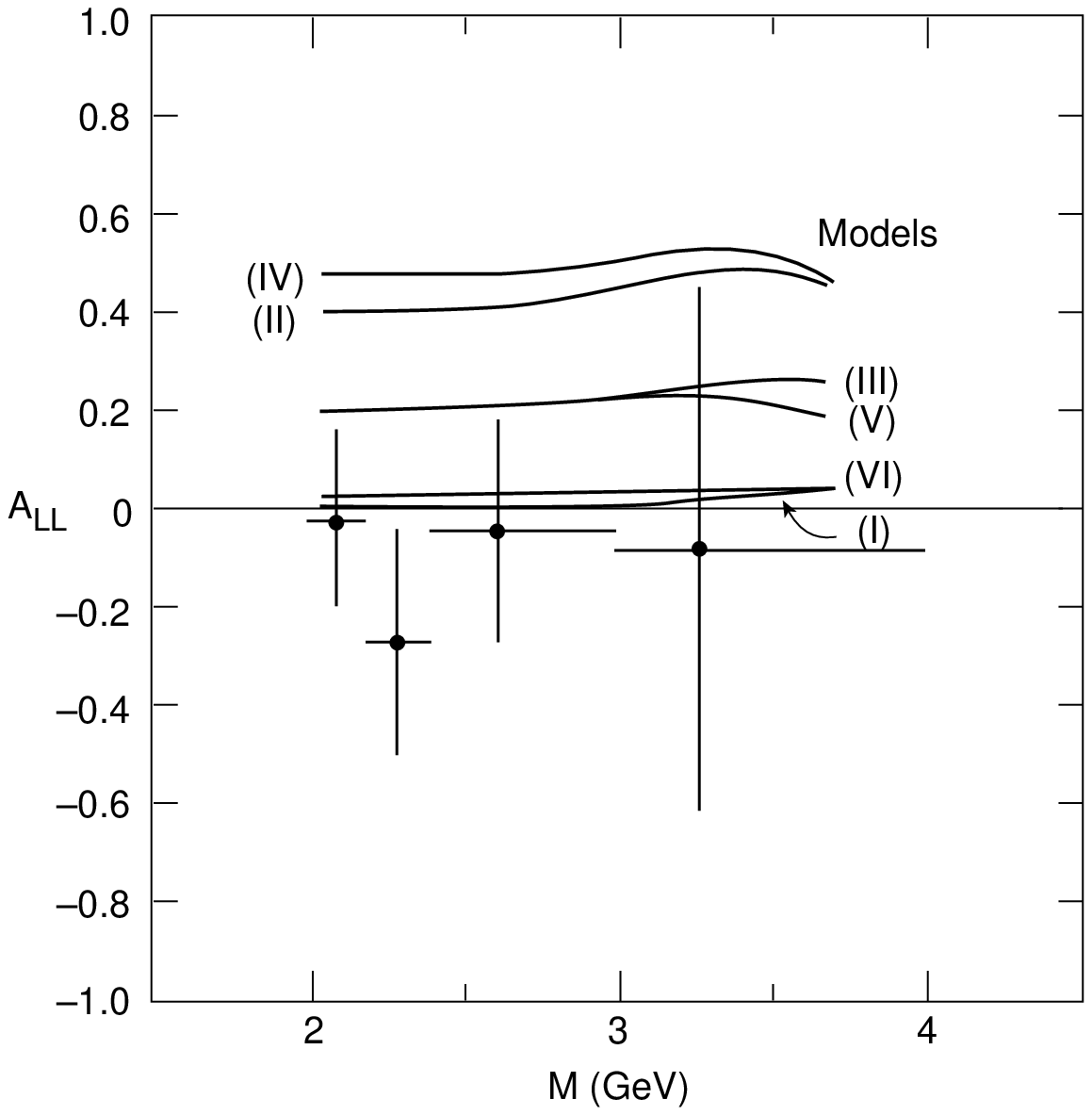,width=8.0truecm}}
\end{center}
\mycapt{The particle production asymmetry $A_{LL}$ measured in 
Ref.~\protect\bcite{FNALgamma},
compared with various phenomenological models labeled by the
curves (I) to (VI), as functions of the effective mass $M$ of
the produced system.
\label{Fig2.2}
}
\end{figure}
  As seen in Fig.~\ref{Fig2.2}, they see
no significant asymmetry, which is in conflict with some
polarized-gluon models, but not all \cite{JaffeGlue}.
Unravelling the polarization of the gluons
in the proton is one of the primary objectives of the next round of
polarization experiments such as the HMC and RHIC.

In addition to the soliton and anomaly interpretation,
it was also suggested \cite{SVa}$^-$\cite{NSV} that
a significant suppression of the QCD topological
susceptibility might play a key \role, which would modify the \naive\ quark
model predictions (see also the discussion in Lecture I).

\subsection{The Okubo-Zweig-Iizuka Rule}

    The polarized structure function experiments are not alone in
indicating that there may be strange quarks inside the proton wave
function.  Other indications come from experimental violations of
the Okubo-Zweig-Iizuka (OZI) rule \cite{OZI},
as discussed in the rest of this
lecture.  

\begin{figure}[H]
\begin{center}
\mbox{\epsfig{file=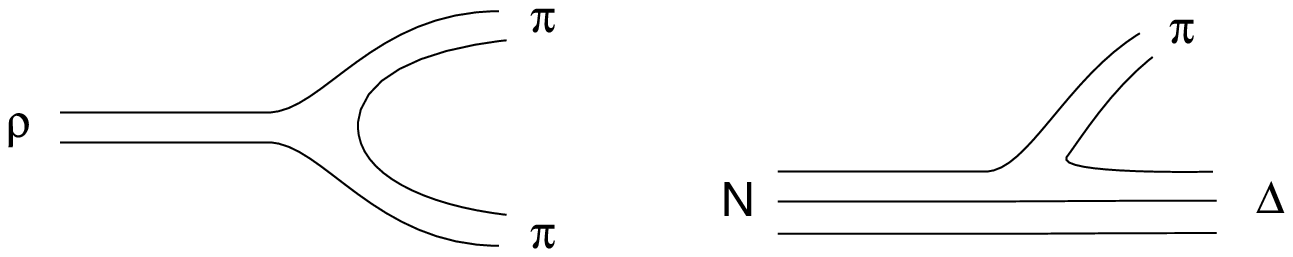,width=11.0truecm}}
\end{center}
\mycapt{Connected  quark line diagrams for $\rho$ decay and the
$\pi n \Delta$ coupling, using the naive quark model content
of the particles, and hence corresponding to processes
conventionally allowed by the OZI rule.
\label{Fig2.3}
}
\end{figure}

According to the OZI rule, the only strong-interaction
processes allowed are those which can proceed via connected quark
line diagrammes such as those shown in Fig.~\ref{Fig2.3}.
This assumption must
be supplemented by Ans\"atze for hadron wave functions, such as
the \naive\ quark model contents
\beq
\ket{\phi} = \ket{\sbs}\,,\quad \ket{p} = \ket{ u{u}d\,}
\label{E43}
\eeq
Indications from meson mass formulae and the observation of
$\phi \rightarrow 3\pi$ decay are that the $\phi$ wave function is
actually a mixture:
\beq
\ket{\phi} = \cos\,\delta\,\ket{\sbs}\,+\,\sin\,\delta\,
\ket{\ubu\,+\,\dbd\,}/\sqrt{2}
\label{E43a}
\eeq
where $\delta \ll 1$.
\begin{figure}[htb]
\begin{center}
\mbox{\epsfig{file=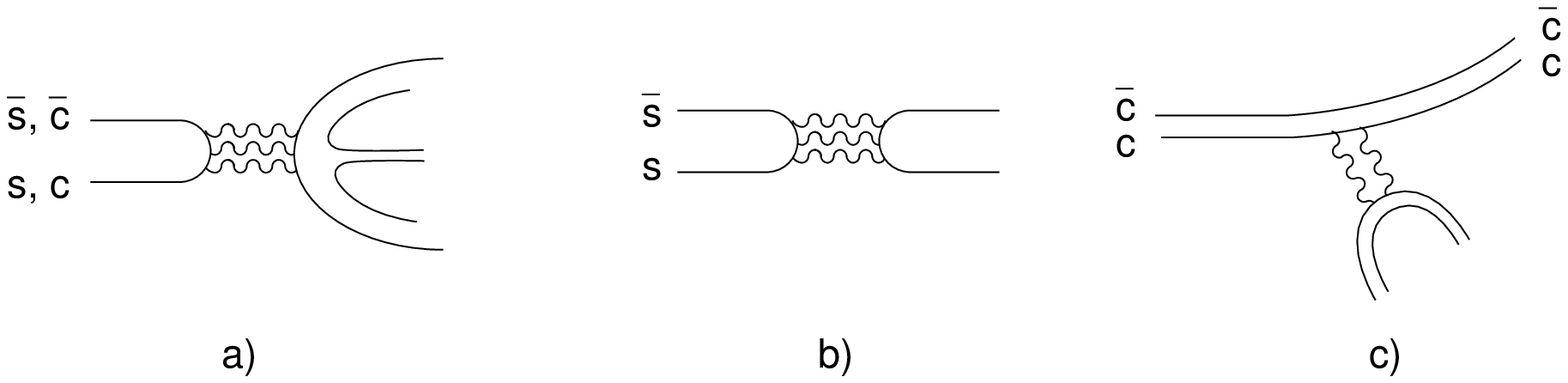,width=11.9truecm}}
\end{center}
\mycapt{Disconnected diagrams mediated by gluon exchange, relevant
to (a) $\phi$ and $J/\psi$ decays, (b) a departure from ideal
mixing in the $\phi$ wave function, and (c) transitions between
different charmonium states.
\label{Fig2.4}
}
\end{figure}
Disconnected diagrams may be mediated by gluon exchange, as shown in
Fig.~\ref{Fig2.4},
which are subject to dynamical suppression which depends on the
process considered:
\beq
{g^2_{\phi 3 \pi}\over g^2_{\omega 3 \pi}} \sim 10^{-2}\,,\ \
{g^2_{f^{\prime} \pi\pi}\over g^2_{f \pi\pi}} \sim 10^{-3}\,,\ \
{\Gamma_{J/\psi}\over\Gamma_{had}}\sim 10^{-4}\,,\ \
{\Gamma_{J/\psi\pi\pi}\over\Gamma_{had}}\sim 10^{-2}
\label{E44}
\eeq
The OZI rule finds some justification in the large-$N_c$ limit of QCD
considered earlier.  Simple $N_c$ power-counting for the meson diagrams
in Fig.~\ref{Fig2.5}
 shows that they are suppressed by $1/N_c$ and $1/N_c^2$,
respectively.
\begin{figure}[htb]
\begin{center}
\mbox{\epsfig{file=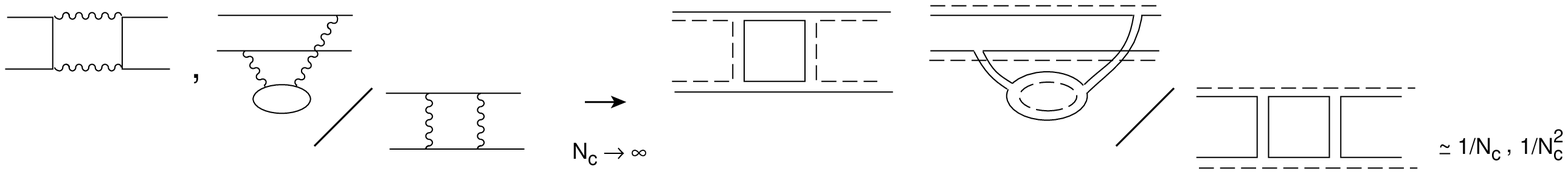,width=11.9truecm}}
\end{center}
\mycapt{Meson diagrams that are suppressed in the large-$N_c$ limit.
As in Fig.~\protect\ref{Fig2.1}, 
the solid lines represent colour, and the dashed
lines flavour.
\label{Fig2.5}
}
\end{figure}

     However, the applicability of the OZI rule to baryons
is more questionable~\cite{EGK}.
 It is more difficult to justify in the large
$N_c$ limit, because the baryon wave function contains ${\cal O}(N_c)$
quarks.  Moreover, it does not seem to work very well experimentally.
Historically, one of the first indications of a discrepancy with
the OZI rule was the $\pi$-nucleon 
$\sigma$-term \cite{Cheng,Gasser}\ :
\beq
\Sigma^{\pi N} = {1\over2}(m_u + m_d)
\bra{p} \ubu + \dbd \ket{p}
\label{E45}
\eeq
Using the Gell-Mann-Okubo mass formula and the OZI assumption that
$\langle p | \bar{s} s | p \rangle = 0$ one estimates
\beq
\Sigma^{\pi N} \simeq 25\ \hbox{MeV}
\label{E46}
\eeq
to be compared with the experimental value of about 45 MeV,
corresponding to
\beq
{\bra{p} \sbs \ket{p} \over
\bra{p} \ubu +\dbd + \sbs \ket{p}}
\simeq 0.2
\label{E47}
\eeq
The data on polarized structure functions discussed in Lecture I
provide another example of an $\bar{s}s$ matrix element in the proton
which is non-negligible.

    In addition to these static matrix elements, there have long been
indications of OZI ``violation" in the production of the $\phi$ and
other supposedly pure $\bar{s}s$ mesons.  For example, old bubble chamber
data indicated that
\beq
{\sigma(\pbp \rightarrow \phi \pi^+ \pi^-)\over
\sigma(\pbp \rightarrow \phi \pi^+ \pi^-)}
=(19 \pm 5) \times 10^{-3}
\label{E48}
\eeq
which is much larger than what could be expected from an admixture of
$\bar{u}u$, $\bar{d}d$ components in the $\phi$ wave function, which is
estimated to be around $(1\div4)\times 10^{-3}$ on
the basis of mass formulae.  The
bubble chamber result eq.~\eqref{E48} corresponds to
\beq
0.05 <
\left|
{\sqrt{2} \,\A(\pbp \rightarrow \sbs+ X) \over
\A(\pbp\rightarrow \ubu+X)+ \A(\pbp\rightarrow \dbd+X)}
\right| < 0.22
\label{E49}
\eeq
In addition to this observation, early data indicated possible OZI
violations in $\bar p n \rightarrow \phi \pi^-$, \
$\pbp \rightarrow f_2^\prime(1520) \pi^+ \pi^-$, \
$p{p}\rightarrow p {p} \phi \ldots$ \
and $\Omega^* \rightarrow \Omega \pi \pi$ decay.

\begin{figure}[t]
\begin{center}
\mbox{\epsfig{file=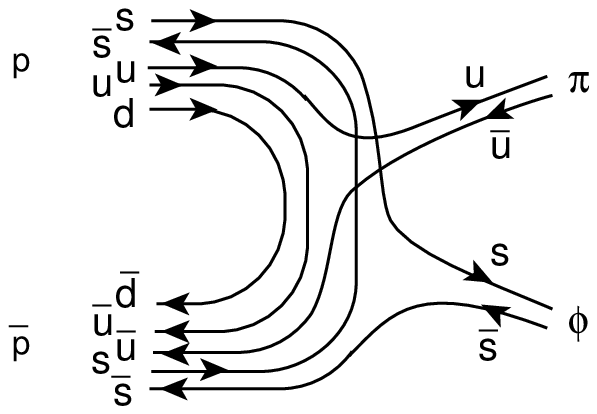,width=6.0truecm}}
\end{center}
\mycapt{New class of connected quark line diagrams responsible
for OZI ``evasion" in $\bar p p \rightarrow \phi \pi$.
\label{Fig2.6}
}
\end{figure}

\begin{figure}[h]
\begin{center}
\mbox{\epsfig{file=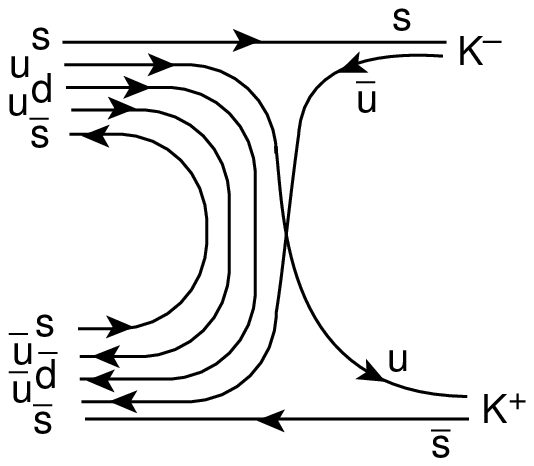,width=5.5truecm,angle=0}}
\end{center}
\mycapt{New class of connected quark line diagram which could
account for the appearance of a backward peak in the reaction
$\bar p p \rightarrow K^- K^+$.
\label{Fig2.7}
}
\end{figure}

    Among the proposed interpretations was OZI ``evasion"
\cite{EGK} due to an
$\bar{s}s$ component in the proton wave function, which provides
the possibility of a new class of connected quark line diagrams as
seen in Fig.~\ref{Fig2.6}. 
Additional evidence for this hypothesis may come from
the presence of a backward peak in the reaction
$\bar{p}p \rightarrow K^- K^+$ which could be due to the connected
quark line diagram shown in Fig.~\ref{Fig2.7}, which involves the intrinsic
strange component of the proton wave function.

  An alternative explanation for ``excess" $\phi$ production
that has been
proposed is rescattering through intermediate $K,K^*$ states, but
it has also been argued that any such
effect would be too small, and subject
to systematic cancellations.  Yet another interpretation was proposed
in terms of an exotic $1^{--}$ resonance $C(1480)$ previously reported
in the $\phi \pi^0$ mass spectrum in the reaction $\pi^- p \rightarrow
C n$.  However, it seems difficult to reconcile this interpretation with
the final-state channel dependence seen more recently, and LEAR data
do not confirm the existence of such a resonance.

\vfill\eject
It is also worth noticing that dispersion relation analyses of
the proton vector isoscalar form factor also found suggestions of
a surprisingly large $\phi\pbp$ coupling 
\cite{Hoh76}$^-$\cite{Jaf89}\ :
\beq
{g^2_{\phi p{p}}\over g^2_{\omega p{p}}}
\simeq 0.211\ \hbox{to}\ 0.276
\label{E50}
\eeq

\subsection{New Data from LEAR}

     The LEAR ring at CERN has provided large numbers of
nucleon-antinucleon annihilations at or close to rest,
which have enabled
the OZI rule to be tested in $\phi$ production in association with
many different final states $X$.  The ASTERIX collaboration has
measured the ratios of
$\phi X$ to $\omega X$ production for the
states $X = \pi, \eta, \omega, \rho$ and $\pi \pi$, as reported in
Table II. 
%
%------------- Table II ----
\bigskip
\begin{table}[H]
\vbox{
\begin{center}
TABLE II
\end{center}
\centerline{\parbox{11.9truecm}
{\footnotesize
The ratios $R=\phi X/\omega X$ for production of
the $\phi$ and $\omega$ - mesons in
antinucleon annihilation at rest.
The parameter $Z$ of the OZI-rule violation is
calculated for $\delta=\Theta-\Theta_i=3.7^0$, assuming
identical phases of the $\phi$ and $\omega$ production amplitudes.
The data are given for annihilation in
liquid hydrogen target (percentage of annihilation from P-wave
is $\sim 10-20 \%$),
gas target ($\sim$61\% P-wave) and
 LX-trigger \cite{Rei.91} ($\sim$86-91\% P-wave).
} % end of footnote
} % end of centerline

\def\bakA{\kern-3em}
\def\bakB{\kern-0.7em}
\def\bakC{\kern-0.9em}
\def\bakD{\kern-0.5em}
\bigskip
% \\~\\
\centerline{\footnotesize
\begin{tabular}{llllll}
\hline
\bakB
Final state
& \bakC Initial states & \bakC B.R.$\cdot10^{4}$& $R\cdot 10^{3}$
&$\left| Z\right|~(\%)$ &\bakD Comments\\
\hline
\bakB
$\phi\gamma$ & \bakC $^1S_0,^3P_J$ & \bakC $0.17\pm0.04$  & $250\pm89$
&  $42\pm8$ & \bakD liquid, \cite{Fae.93}\\
\hline
\bakB
$\phi\pi^0$ & \bakC $^3S_1,^1P_1$ &\bakC $5.5\pm0.7$  & $96\pm15$
&  $24\pm2$ & \bakD liquid, \cite{Fae.93}\\
$\phi\pi^0$ & &\bakC $1.9\pm0.5$ &
&   & \bakD gas, \cite{Rei.91}\\
\bakB
$\phi\pi^0$ & &\bakC $0.3\pm0.3$ &
&   & \bakD LX-trigger, \cite{Rei.91}\bakA\\
\hline
\bakB
$\phi\pi^-$ &\bakC $^3S_1,^1P_1$ &\bakC $9.0\pm1.1$
& $83\pm25$
&  $22\pm4$ & \bakD liquid, \cite{Biz.74}$^-$\cite{Bet.67}\\
\bakB
$\phi\pi^-$ & &\bakC $14.8\pm1.1$  & $133\pm26$
&  $29\pm3$ & \bakD $\bar p d$, \cite{Abl.93}\ $^\dagger$\\
\bakB
$\phi\pi^-$ & & & $ 113\pm30$
&  $27\pm4$ & \bakD $\bar p d$, \cite{Abl.93}\ $^\ddagger$ \\
\bakB
$\phi\pi^+$ &  & & $110\pm15$
&  $26\pm2$ & \bakD $\bar n p$, \cite{Abl.93} \\
\hline
\bakB
$\phi\eta$ &\bakC $^3S_1,^1P_1$ & \bakC $0.9\pm0.3$ & $6.0\pm2.0$
&  $1.3\pm1.2$ & \bakD liquid, \cite{Fae.93}\\
\bakB
$\phi\eta$ & &\bakC $0.37\pm0.09$ &
&  & \bakD gas, \cite{Rei.91}\\
$\phi\eta$ & &\bakC $0.41\pm0.16$ &
&   & \bakD LX-trigger, \cite{Rei.91}\bakA\\
\hline
\bakB
$\phi\rho$ & \bakC $^1S_0,^3P_J$ & \bakC $3.4\pm0.8$ & $6.3\pm1.6$
&  $1.4\pm1.0$ & \bakD gas, \cite{Rei.91}$^,$\cite{Wei.93}\\
\bakB
$\phi\rho$ & &\bakC $4.4\pm1.2$ & $7.5\pm2.4$
&  $2.1\pm1.2$ & \bakD LX-trigger, 
\cite{Rei.91}$^,$\cite{Wei.93}\bakA\\
\hline
\bakB
$\phi\omega$ & \bakC $^1S_0,^3P_{0,2}$ &\bakC $6.3\pm2.3$  & $19\pm7$
& $7\pm4$ & \bakD liquid, \cite{Biz.71}$^,$\cite{Ams.93}\\
\bakB
$\phi\omega$ & &\bakC $3.0\pm1.1$  &
&          & \bakD gas, \cite{Rei.91}\\
\bakB
$\phi\omega$ & &\bakC $4.2\pm1.4$  &
&          & \bakD LX-trigger, \cite{Rei.91}\bakA\\
\hline
\bakB
$\phi\pi^0\pi^0$ &\bakC $^{1,3}S_{0,1},^{1,3}P_J$ 
&\bakC $1.2\pm0.6$  & $6.0\pm3.0$
& $1.3\pm2.0$ & \bakD liquid, \cite{Fae.93}\\
\bakB
$\phi\pi^-\pi^+$ & &\bakC $4.6\pm0.9$  & $7.0\pm1.4$
& $1.9\pm0.8$ & \bakD liquid, \cite{Biz.69}\\
\bakB
$\phi X,
X=\pi^+\pi^-, \rho$ & &\bakC $5.4\pm1.0$  &$7.9\pm1.7$
&$2.4\pm1.0$  & \bakD gas, \cite{Rei.91}$^,$\cite{Wei.93}\\
\bakB
$\phi X,
X=\pi^+\pi^-, \rho$ & &\bakC $7.7\pm1.7$  &$11.0\pm3.0$
&$4.0\pm1.4$  & \bakD LX-trigger, \cite{Rei.91}$^,$\cite{Wei.93}\bakA\\
\hline
\multicolumn{6}{l}{\bakB $^\dagger$ $p < 200$ MeV/c}\\
\multicolumn{6}{l}{\bakB $^\ddagger$ $p > 400$ MeV/c}\\
\hline\\~\\
\end{tabular}
} % end reduced font size
%----------------- end of Table II
}% end of \vbox
\end{table}
\noindent
We see that there are large enhancements in $S$-wave
annihilations as compared to \naive\ OZI rule, 
especially for the case
$X =\pi$, whereas there are little or no enhancements in $P$-wave
annihilations for
$X=\pi,\eta,\rho,\pi\pi$.  The Crystal Barrel
collaboration has measured the corresponding ratios for $X = \pi^0,
\eta, \pi^0 \pi^0, \gamma$.  As also seen in Table II, they confirm
an enhancement over the OZI rule in the case $X = \pi^0$, see no big
effect in the cases $X = \eta,\pi^0\pi^0$, and find a very large
enhancement in the case $X = \gamma$, which is about a hundred times
the OZI prediction!  The OBELIX collaboration also finds large
enhancements in the cases $X = \pi^{\pm}$, as also seen in Table II.
%\clearpage
%\noindent

\begin{figure}[H]
\begin{center}
\mbox{\epsfig{file=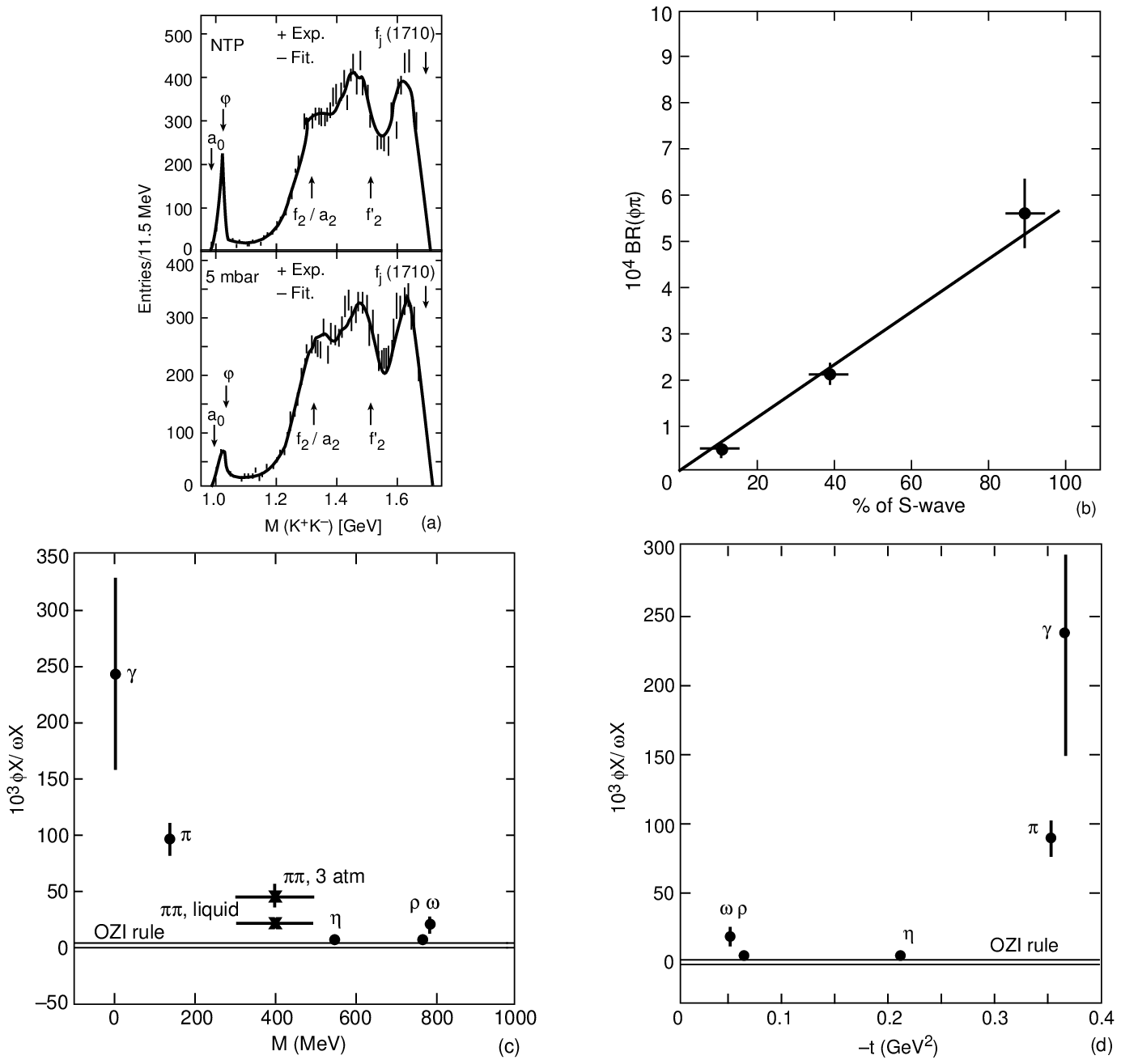,width=11.9truecm,angle=0}}
%\mbox{\epsfig{file=dummyfig.ps,width=7.0truecm,angle=0}}
\end{center}
\mycapt{
Recent experimental evidence from LEAR
for breakdowns of the \naive\ OZI rule. (a) A $\phi$ peak is seen
clearly in annihilations at NTP, but is less prominent at a
pressure of 5 mbar, indicating (b) that $\phi\pi$ production
is suppressed in P-wave annihilations. The amount of OZI
``violation" depends on (c) the invariant mass of the system
produced in association with the $\phi$ meson, and (d) the
invariant momentum transfer to the $\phi$.
\label{Fig2.8}
}
\end{figure}

Some interesting features of the available data are shown in
Fig.~\ref{Fig2.8}.
We see in Fig.~\ref{Fig2.8}a
that $\phi\pi$ production is much smaller in $P$-wave
annihilations than in $S$-wave annihilations, and Fig.~\ref{Fig2.8}b shows
that
the $\pi\pi$ spectrum also depends on the partial wave.  Fig.~\ref{Fig2.8}c
shows the dependence of $\phi X /\omega X$ production ratios on the
invariant mass of the system $X$, where we see progressively larger
enhancements at smaller masses.  Fig.~\ref{Fig2.8}d
shows the corresponding  ratios
plotted versus the invariant momentum transfer $t$, where we see a
large effect at large $t$.

    These very interesting data exhibit the following features which
need to be explained or accommodated in any model of OZI violation:

--   Non-universal enhancement factors, which are strong for $X = \pi,
\gamma$, smaller for $X = \rho,\omega,\pi\pi$, and not apparent for
$X = \eta$.

--   Larger enhancements in proton-antiproton annihilation at or near
rest than in higher-energy annihilations, or in $\pi\pi$ and ${p}p$
scattering.

--   When the initial $\pbp$ state is known, large enhancements are
seen in $S$-wave annihilations into $X = \pi, \pi\pi$, but not in
$P$-wave annihilations.

The next section introduces a model \cite{EKKS}
which accommodates and
explains these key features, and also makes some predictions for
possible future experiments.
\subsection{Model for a Polarized Strange Component in the Proton Wave
Function}

    We consider \cite{EKKS}
the likelihood that the Fock-space decomposition of
the proton wave function contains an $\sbs$ component  which we
parametrize as follows:
\begin{equation}
|p> = x \sum_X\ket{u {u}d X}  +
z \sum_X\ket{u {u}d\sbs X}
\label{E51}
\end{equation}
\begin{figure}[htb]
\begin{center}
\mbox{\epsfig{file=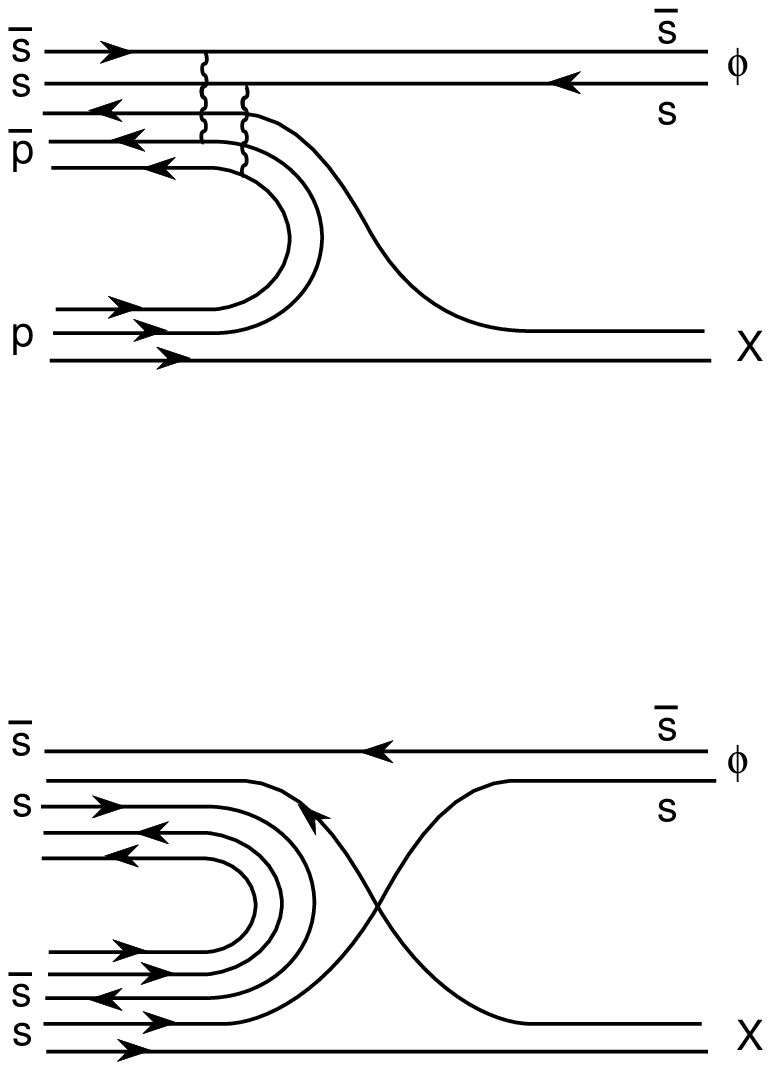,width=6.0truecm,angle=0}}
\end{center}
\mycapt{
``Shakeout" and ``rearrangement" diagrams responsible for
$(\bar{p} p \rightarrow \sbs + X)$ in presence of 
an $\sbs$ component in the proton wave function.
\label{Fig2.9}
}
\end{figure}
where $X$ denotes any combination of gluons and light $\bar{q}q$ pairs.
As seen in Fig.~\ref{Fig2.9},
  two important new classes of connected quark line
diagrams become possible, the shake-out diagrams illustrated
in Fig.~\ref{Fig2.9}a
and the rearrangement diagrams illustrated in Fig.~\ref{Fig2.9}b.
A typical shake-out amplitude may be estimated as
\begin{equation}
\A(\bar{p} p \rightarrow \sbs + X) \simeq 2 Re(x z^*)\ P(\sbs),
\label{E52}
\end{equation}
where $P(\sbs)$ is a projection factor which may depend on the
final state, and perhaps also the initial state.  A typical rearrangement
amplitude may be estimated as
\begin{equation}
\A(\bar{p} p \rightarrow \sbs + X) \simeq |z|^2\ T(\sbs)
\label{E53}
\end{equation}
where $T(\sbs)$ is another projection factor which is very likely
to depend on the initial state: for example, \naive\ quark models might
suggest that $\phi$ production would be enhanced in rearrangement of
an initial $S$-wave $\pbp$ state.

\noindent
    If we define the generic amplitude ratio
\begin{equation}
Z = \frac{ \sqrt{2} \,\A(A+B\rightarrow \sbs + X)}
{\A(A+B\rightarrow \ubu +X) + \A(A+B\rightarrow \dbd +X)}
\label{E54}
\end{equation}
the measured production ratios are given by
\beq
R_X\equiv{\sigma(\phi X) \over \sigma(\omega X)} \approx
\left( \frac{ Z+ \tan \delta}{1-Z\tan \delta} \right)^2 \times
(\hbox{phase-space ratio})
\label{E55}
\eeq
where $\delta$ is the angle representing the departure from ideal
mixing in the vector meson nonet, which is a measure of the ``expected"
deviation from the OZI rule.  We see from equation \eqref{E51}
that one might expect generic shake-out diagrams to yield
\begin{equation}
|Z| = 2 \left|{z \over x}\right|
= 2 {|z| \over \sqrt{1-|z|^2}}
\label{E56}
\end{equation}
assuming similar projection factors $P$, and generic rearrangement
diagrams to yield
\begin{equation}
|Z| =  \left|{z\over x}\right|^2 = {|z|^2 \over {1 - |z|^2}}
\label{E57}
\end{equation}
assuming similar projection factors $T$.  Data on $S$-wave $\pi$
production correspond to
\beq
|Z(\phi\pi/\omega\pi)| = 0.24\pm0.02
\label{E58}
\eeq
leading to the prediction
\beq
0.01 \leq |z|^2 \leq 0.19
\label{E59}
\eeq
to the extent that the reactions are dominated by the
shake-out and rearrangement mechanisms, respectively. These
estimates are both compatible with data on $K$ production in
$\bar{p}p$ annihilation, which can be expected to contain a contribution
from the higher Fock-space states in \eqref{E51} as well as $\sbs$
production in the final state:
\begin{equation}
Y_K=(4.74\pm 0.22)\% \simeq 4[Re (x z^*)]^2
\label{E60}
\end{equation}

\begin{figure}[htb]
\begin{center}
\mbox{\epsfig{file=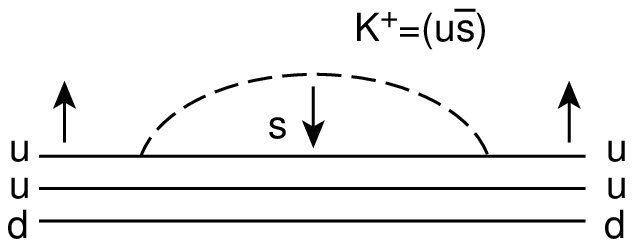,width=7.0truecm,angle=0}}
\end{center}
\mycapt{Emission of a $K^+$ meson in a chiral model, which may
explain the negative polarization of $s$ quarks in the proton
wave function.
\label{Fig2.10}
}
\end{figure}
    As we have seen in Lecture I, polarized structure function data
indicate that the strange component of the proton wave function is
polarized negatively: $\Delta s < 0$, and this feature should be
incorporated in \eqref{E51}. Negative polarization of $s$ quarks
is understandable in a chiral model, as seen in Fig.~\ref{Fig2.10}.
The emission
of a $K^+$ flips the positive helicity of a $u$ quark into negative
polarization for an $s$ quark. This argument does not provide
immediately an indication on the possible polarization of an
$\sbs$ antiquark, but the simplest allowed possibility is
\beq
\ket{u{u}d}_{{1\over2},{1\over2}} \,\,\oplus_{1,1}\,\,
\ket{\sbs}_{1,-1}
\label{E61}
\eeq
where there is a relative orbital angular momentum $|1,1 \rangle$ between
the two indicated components.  The internal state of the $\bar{s}s$
pair is motivated by the fact that condensation in the vacuum occurs
in a $^3P_0$ state.

    Within this picture, one could expect that the rearrangement
diagrams from a spin-triplet initial $\bar{p}p$ state would yield
preferentially spin-triplet $\sbs$ states such as the $\phi$.
One could also expect that $S$-wave $\pbp$ annihilations would
yield preferentially final states containing $S$-wave $\sbs$
pairs.  These arguments favour maximum $\phi$ production in $^3S_1$
annihilations, and less enhancement in $^1S_0$
and $P$-wave annihilations.
A corollary would be the dilution of the $\phi$ enhancement at higher
energies, where more partial waves contribute. Predictions for final
states containing $\eta$ mesons are more complicated, because more
diagrams contribute since the $\eta$ wave function contains both
$\bar{q}q$ and $\sbs$ components.

\subsection{Future Tests of the Model}

    The following are some possible tests of these ideas for
OZI evasion that could be tested in future experiments.

\begin{enumerate}
\item
Production of the P-wave quark-model state $f^\prime(1525)$ could be
enhanced in $P$-wave $p \bar p$ annihilations, at least to the
extent that they receive contributions from rearrangement
diagrams. This suggestion already seems to be compatible with
preliminary LEAR data.

\item
A reduction in the enhancement of $\phi\pi/\omega\pi$ as the
$\bar p$ momentum is increased, associated with the decreasing
fraction of $S$-wave annihilations -- for example, the $S$-wave
fraction at $P_{\bar p} = 600$ MeV is between 14 and 20 \%.
This suggestion also seems to be compatible with
data from the Crystal Barrel collaboration.

\item
There should be considerable spin-dependence in the cross
section for $p \bar p \rightarrow \phi \phi$, which should, for
example, be higher from spin-triplet initial states.

\item
Both $S$- and $P$-waves may contribute to the $\phi \pi \pi$
final state, and we would expect dominance by the $^3S_1$ initial
state, which seems once more to be supported by recent data.

\item
We would expect strong spin correlations in
$p \bar p \rightarrow K^* \bar K^*$,
which should be dominated by the $L=0, S=2$ state.

\item
The large enhancement of $\phi \gamma /\omega \gamma$ observed
should be dominated by the $^3P_{0,1,2}$ initial states, rather
than the $^1S_0$ state.

\item
The angular distribution of the $e^+e^-$ pairs produced by
$\phi$ decay in reactions of the type $p \bar p \rightarrow \phi + X$
should be
\beq
\sim (1 + \cos^2\,\theta)
\label{E62}
\eeq

\item
There should be significant dependence on the beam and target
polarizations in the reaction $p + p \rightarrow p + p + \phi$,
specifically, the cross section should be maximal when their
polarizations are parallel. This effect should also be seen in
the reaction $p + d \rightarrow {}^3\hbox{He} + \phi$, where there are
recent indications of a substantial enhancement of $\phi$
production relative to $\omega$ production and the \naive\ OZI rule.

\item
The constituent counting rules
\beq
\left.
{d \sigma(A+B\rightarrow C +D) \over d t}\,
\right\vert_{\Toprel{\theta_{CM}}\over{\hbox{\scriptsize fixed}}}
\,\,\simeq\,\,{f(\theta_{CM})\over
{s^{\phantom{!}\kern-0.2em}}^{n_A + n_B + n_C +
n_D -2^{\phantom{!}\kern-0.2em}}}
\label{E63}
\eeq
suggest different behaviours for the reactions $\pi + p \rightarrow
\phi, \omega + n$ at large momentum transfers. Specifically, the
cross section for $\pi + p \rightarrow \phi + n$ should behave
as $s^{-12}$ if it is dominated by production from the
$\ket{u{u}d\sbs}$
Fock state, or as $s^{-9}$ if it is dominated by gluonic production.
In either case, the $\phi/\omega$ ratio should decrease at higher
energies.

\item
The different production mechanisms for $\phi \pi$ and $\omega
\pi$, and for $\phi\, \pi\pi$ and $\omega \,\pi\pi$ final states, could
lead to different angular distributions, as may be seen experimentally.

\item
There should be a large enhancement in the Pontecorvo reaction
\break
\hbox{$\bar p + d \rightarrow \phi + n$}.
\end{enumerate}

\subsection{Extension of the Model to $\Lambda$ Production}

We finish this lecture by mentioning a couple of tests of these
ideas \cite{AEK,EKhKo}
using data on $\Lambda$ production in $p \bar p$ annihilation
and elsewhere.

The total cross section for $p \bar p \rightarrow
\Lambda \bar{\Lambda}$ and its angular distribution have been
measured at LEAR by the PS 185 collaboration \cite{PS185}.
These have been
described in terms of phenomenological models based on quarks
and gluons, and on meson exchanges. The PS 185 data \cite{PS185}
indicate that
the $\Lambda \bar{\Lambda}$ pairs are produced mainly in a spin-triplet
state, as would be expected in our model \eqref{E61}, or in
a gluon-exchange model \cite{gex}, which produces
a spin-triplet $s \bar s$ pair. Spin-triplet dominance can also be
accommodated in a meson-exchange model \cite{mex}, with the appropriate
combination of $K$ and $K^*$ exchanges. Measurements of the $\Lambda$
depolarization $D_{nn}$, i.e., the transfer to the final-state $\Lambda$
of polarization from a polarized $p$ target, may help distinguish
between different models \cite{AEK}.

Models which explain the EMC spin effect in terms of polarized
gluons \cite{DeltagI}$^-$\cite{DeltagIII}
would naturally expect that the gluon polarization is the
same as the target $p$, and hence also that of the $s \bar s$
pair, and thus that of the $\Lambda$, so that $D_{nn} > 0$. On the
other hand, the meson-exchange model has been used to predict
$D_{nn} < 0$ \cite{mex}. On the other hand,
in our polarized-strangeness model we would
expect that the $s$ in the proton wave function would have
negative polarization, and hence that $D_{nn} < 0$
\cite{AEK}. Thus the
polarized-gluon and polarized-strangeness explanations of the
EMC spin effect seem to predict opposite signs.

\begin{figure}[H]
\begin{center}
\mbox{\epsfig{file=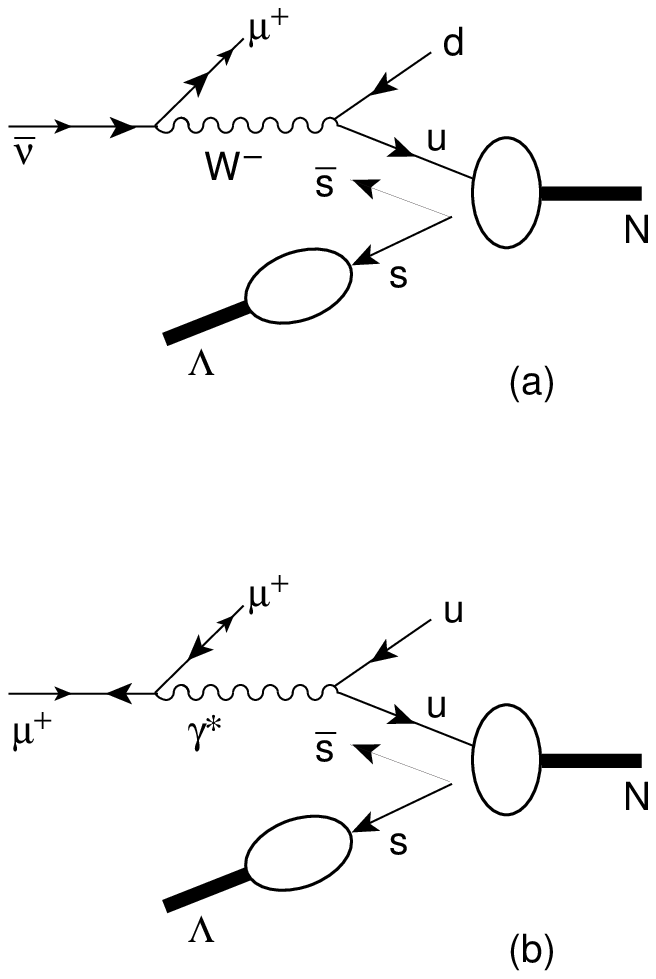,width=3.5truecm,angle=0}}
\end{center}
\mycapt{Diagrams which suggest that $\Lambda$ baryons observed in
the target fragmentation regions of deep-inelastic
(a) $\bar \nu$ and (b) polarized $\mu^+$ collisions should
be polarized. The solid triangles represent the spin states of
the various particles.
\label{Fig2.11}
}
\end{figure}

This idea can be extended to $\Lambda$ production in the target
fragmentation region of deep-inelastic lepton-nucleon scattering.
Briefly stated, the proposal made in Ref.~\bcite{EKhKo}\ is that a
polarized lepton beam (a $\nu$, $\bar{\nu}$, or a polarized $\mu$
or $e$) couples to the nucleon target preferentially through a
particular boson ($W$ or $\gamma$) polarization state, which then
picks out preferentially a particular quark (or antiquark)
polarization state. The target nucleon remnant ``remembers"
the spin that was removed, without the need for a polarized
target. For example, if a positively-polarized $u$ quark is
removed from the proton wave function, it will tend to leave behind
a negatively-polarized $s \bar s$ pair, which may lead to negative
polarization for $\Lambda$'s produced in the target fragmentation
region \cite{EKhKo}, as also seen in Fig.~\ref{Fig2.11}.

Just such an effect has been seen in the reaction $\bar{\nu} + N
\rightarrow \mu^+ + \Lambda + X$ by the WA59 collaboration \cite{wa59}.
In the
kinematic region $0.2 < x_{Bj} < 1$, $z < 0$, they measure
\beq
P_{\Lambda} = - 0.85 \pm 0.19
\label{E64}
\eeq
to be compared with the postdiction
\beq
P_{\Lambda} = - 0.94 D
\label{E65}
\eeq
in our model, where $D$ is an uncalculable dilution factor. The
comparison between equations \eqref{E63} and \eqref{E64} above indicates that
$D = 0.9 \pm 0.2$. Similar predictions can be made, for example,
for $\Lambda$'s produced in the NOMAD \ $\nu$ experiment at CERN.

The fact that, according to the WA59 experiment,
there does not seem to be dilution
by a large factor encourages the extrapolation of these ideas to
polarized $\mu(e)^+$ scattering, where we predict \cite{EKhKo}
that in the target fragmentation region
\beq
P_{\Lambda} = 0.7 P_{\mu(e)} D
\label{E67}
\eeq
This should be observable in the proposed HMC experiment at CERN, in
the E665 experiment at Fermilab \cite{Ashery}, and
perhaps also in the HERMES \cite{HERMES} experiment at DESY.

%-------- lecture III -------------------------------
\vfill\eject
\setcounter{figure}{0}
\setcounter{equation}{0}
\section {Cosmological Spin-offs}

In this lecture we discuss, as an example of the relevance of
polarized structure function measurements and the spin
decomposition of the nucleon to other areas of physics, their
applications to experimental searches for dark matter particles.
This discussion is prefaced by a brief review of the
motivations for such particles.

\subsection {How Much Dark Matter?}
\begin{figure}[htb]
\begin{center}
\mbox{\epsfig{file=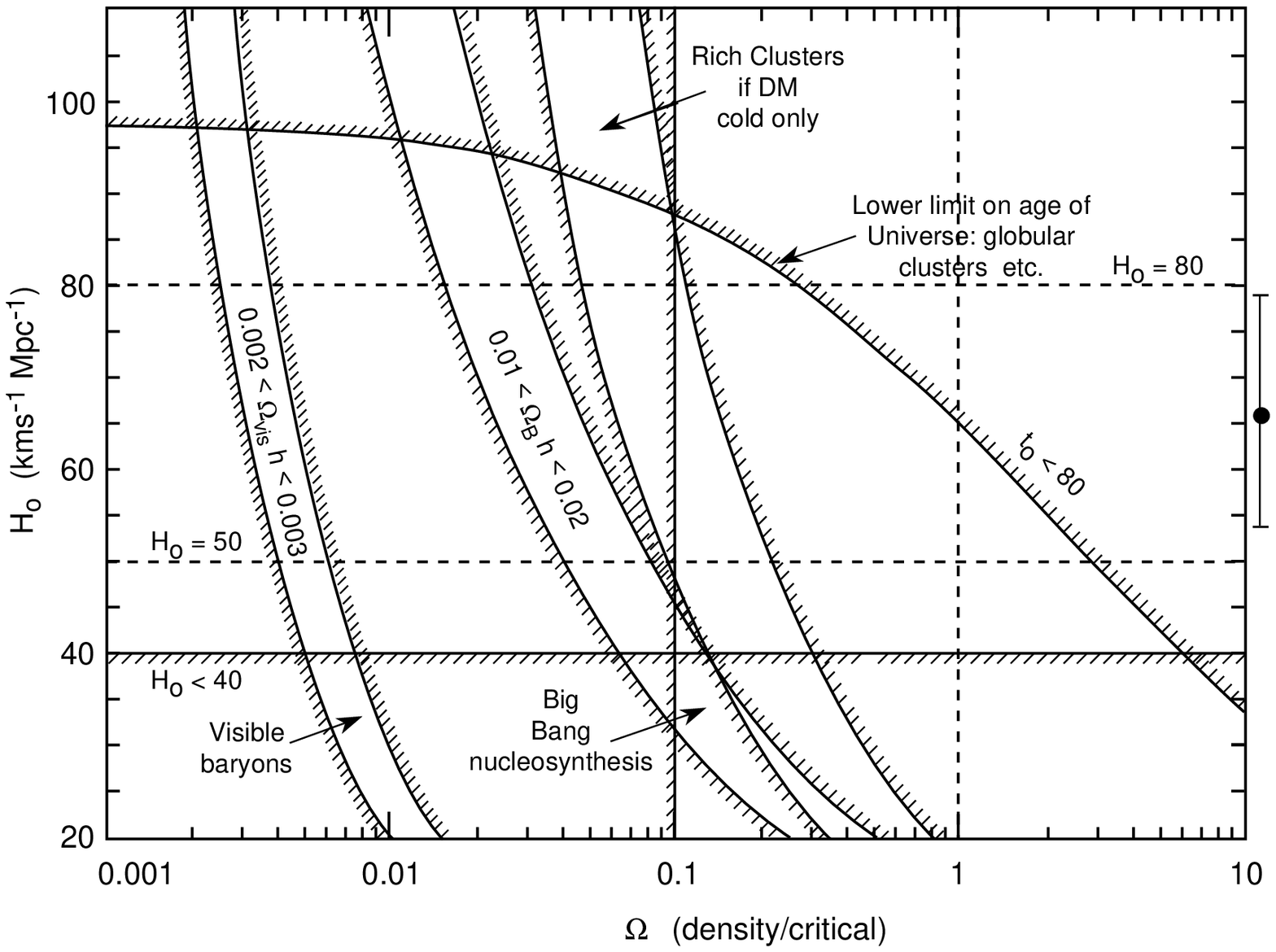,width=11.9truecm,angle=0}}
\end{center}
\mycapt{The ($\Omega$, $H_0$) plane, adapted from
\protect\cite{copi},
showing that there is no serious discrepancy between the average
 measured value of $H_0$, $\Omega = 1$, and an age for the Universe
 of $10^{10}$ years. This plot also
shows the estimates of the present baryon density
$\Omega_{baryons}$
obtained from visible features in the Universe, from Big Bang
Nucleosynthesis \protect\cite{bbn}
and from rich clusters. All the indications are that
$\Omega_{baryons} \protect\lsim 0.1$, 
so that at least $90 \%$ of the matter
in the Universe is non-baryonic dark matter.
\label{Fig3.1}
}
\end{figure}

    Naturalness and inflation \cite{infl}
suggest that the density averaged over the
universe as a whole should be very close to the critical density, which
marks the boundary between a universe that expands forever and one which
eventually collapses, i.e. $\Omega\defeq\rho/\rho_c\simeq$1.  On the
other hand, the matter we can see shining in stars, in dust, etc.
amounts only to $\Omega \simeq 0.01$ \cite{copi},
as seen in Fig.~\ref{Fig3.1}.
The agreement
between big bang nucleosynthesis calculations \cite{bbn}
and the observed
abundances of light elements suggests that $\Omega_{baryons}
\lappeq 0.1$, as also seen in Fig.~\ref{Fig3.1}.  
This is to be compared with
observations of rotation curves, which suggest according to the virial
theorem that the amount of matter in galactic haloes $\Omega_{halo}
\simeq 0.1$.  A similar abundance of baryonic matters is suggested
\cite{copi} by observations of rich clusters of galaxies.

    Mathematically, the galactic haloes could in principle be purely
baryonic, although they seem unlikely to be made out of gas, dust or
``snow balls" \cite{hegyi}.
However, there has recently been considerable interest
in the possibility that haloes might be largely composed of ``brown
dwarfs".  There have been several searches for such ``failed stars"
via microlensing of stars in the Large Magellanic Cloud \cite{lmc}
which indicate that
\beq
f = 0.20^{+0.33}_{-0.14}
\label{E71}
\eeq
be composed of brown dwarfs.  This suggests that most of the dark matter
present locally in our galactic halo
\beq
\rho_{halo} = 0.3 \ \hbox{GeV/cm}^3 \times 1.5^{0\pm1}
\label{E72}
\eeq
is not baryonic in nature.  Moreover, models of galaxy formation suggest
that it is unlikely to be composed of massive neutrinos \cite{EllSik}.
This leaves us only with
the alternative of cold dark matter, whose detection involves in an
essential way the spin decomposition of the nucleon, as we shall see
later in this lecture.

    Before addressing in more detail the nature of the dark matter,
we first comment on the age and Hubble expansion rate of the Universe,
which have recently generated some controversy.  Globular clusters
seem to be at least $14 \pm 3$ Gyr old, and nucleocosmochronology
suggests an age of $13 \pm 3$ Gyr \cite{copi}.
The question is whether these
ages are compatible with current estimates of the Hubble constant
$H_0$ km/sec/Mpc. Recent determinations of $H_0$ may be combined
\cite{mrr} to
yield an estimate of $66 \pm 13$, which is shown on the vertical axis
of Fig.~\ref{Fig3.1}.  
We see from this that there is no incompatibility between
the age of the Universe being above $10^{10}$ yr old and
$\Omega = 1$ as wanted by inflation \cite{infl}.
However, if this is indeed the
case, at least $90 \%$ of the matter in the universe must be unseen
dark matter.

\subsection{Hot or Cold Dark Matter?}

    These terms refer to whether the dark matter was relativistic or
non-relativistic at the cosmological epoch when structures such as
galaxies and clusters began to form.  Whether you favour hot or cold
dark matter depends on your favourite theory of structure formation.
If you believe that its origins lie in an approximately scale-invariant
Gaussian random field of
density perturbations, as suggested by inflationary models
\cite{perts}, then
you should favour cold dark matter.  This is because it enables
perturbations to grow on all distance scales, whereas relativistic
hot dark matter escapes from small-scale perturbations, whose growth
via gravitational instabilities is thereby stunted
\cite{sformation}.  Thus galaxies
form later in a scenario based on Gaussian fluctuations and hot dark
matter then they would in a scenario with cold dark matter.  For this
reason, the latter has commonly been regarded as the ``standard model"
of structure formation.  However, if you believe that structures
originated from seeds such as cosmic strings
\cite{strings}, then you should prefer
hot dark matter, because cold dark matter would then give too much
power in perturbations on small distance scales.
\begin{figure}[H]
\begin{center}
\mbox{\epsfig{file=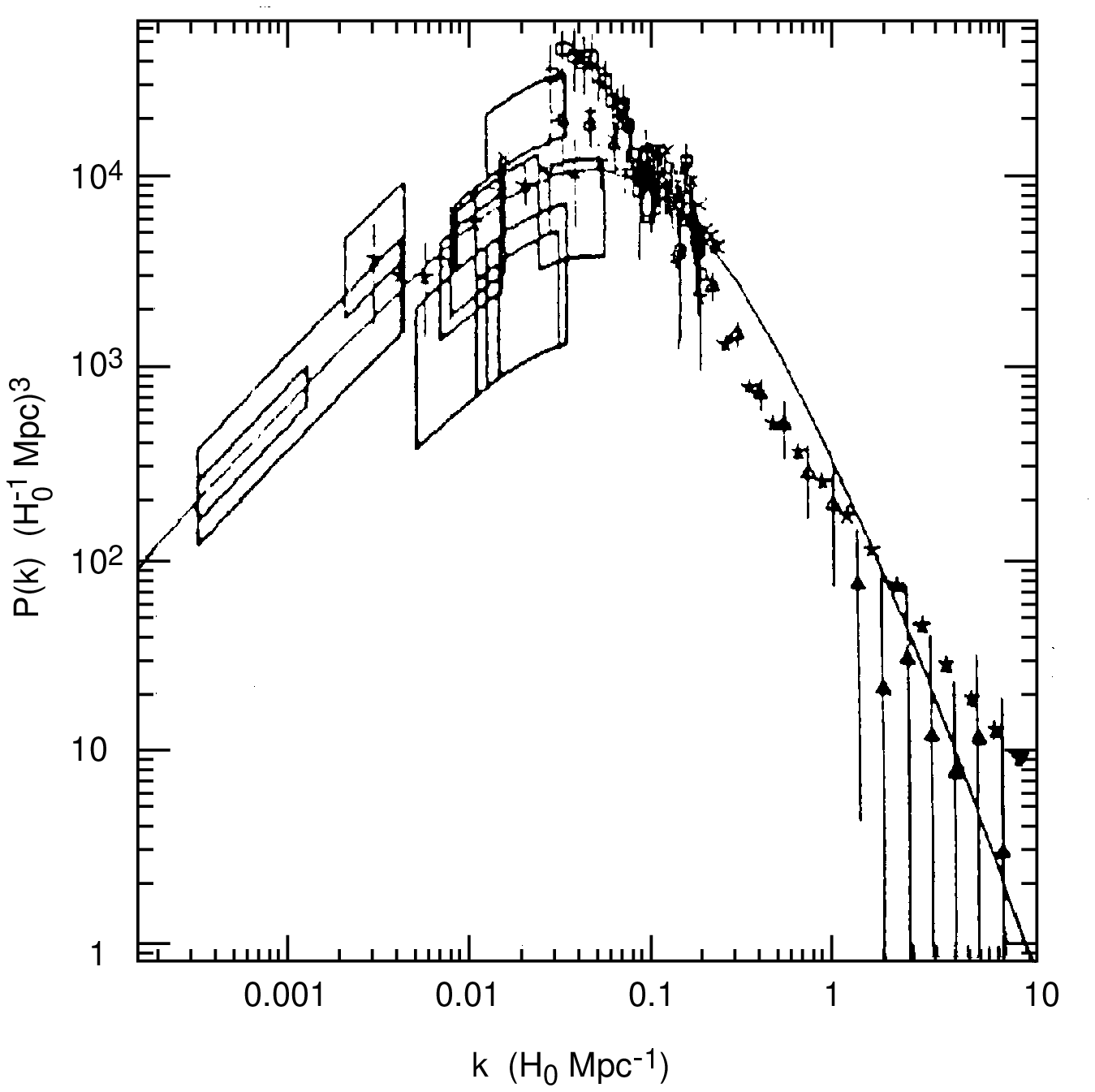,width=8.0truecm,angle=0}}
\end{center}
\mycapt{A compilation of data on the primordial perturbation spectrum
\protect\cite{comp},
 compared with a cold dark matter simulation assuming an
initially scale-invariant spectrum of Gaussian fluctuations.
\label{Fig3.2}
}
\end{figure}
\noindent
    Fig.~\ref{Fig3.2} shows a compilation \cite{comp}
of data on the power spectrum of
astrophysical perturbations, as obtained from COBE
\cite{COBE} and other observations
of the cosmic microwave background radiation, and direct astronomical
observations of galaxies and clusters.  The solid line which does not
quite pass through all the points is one calculated in the
above-mentioned standard model of Gaussian fluctuations and
cold dark matter.
The discrepancies from this curve indicate that there is less
perturbation power at small distances than would be expected in this
theory, given the COBE normalisation at large distance scales.

    This and other observations have suggested that it may be necessary
to modify the pure cold dark matter model.  Several suggestions have
been offered, including a non-zero cosmological constant and a
deviation of the spectrum of Gaussian perturbations from scale
invariance.  However, the preferred scenario seems to be an admixture
of hot dark matter together with the cold, resulting in the following
cocktail recipe for the Universe \cite{mdm}:
\beq
\Omega_{cold} \simeq 0.7\,,\qquad \Omega_{hot} \simeq 0.2,\,
\qquad \Omega_{baryons} \lsim 0.1
\label{E73}
\eeq
\begin{figure}[htb]
\begin{center}
\mbox{\epsfig{file=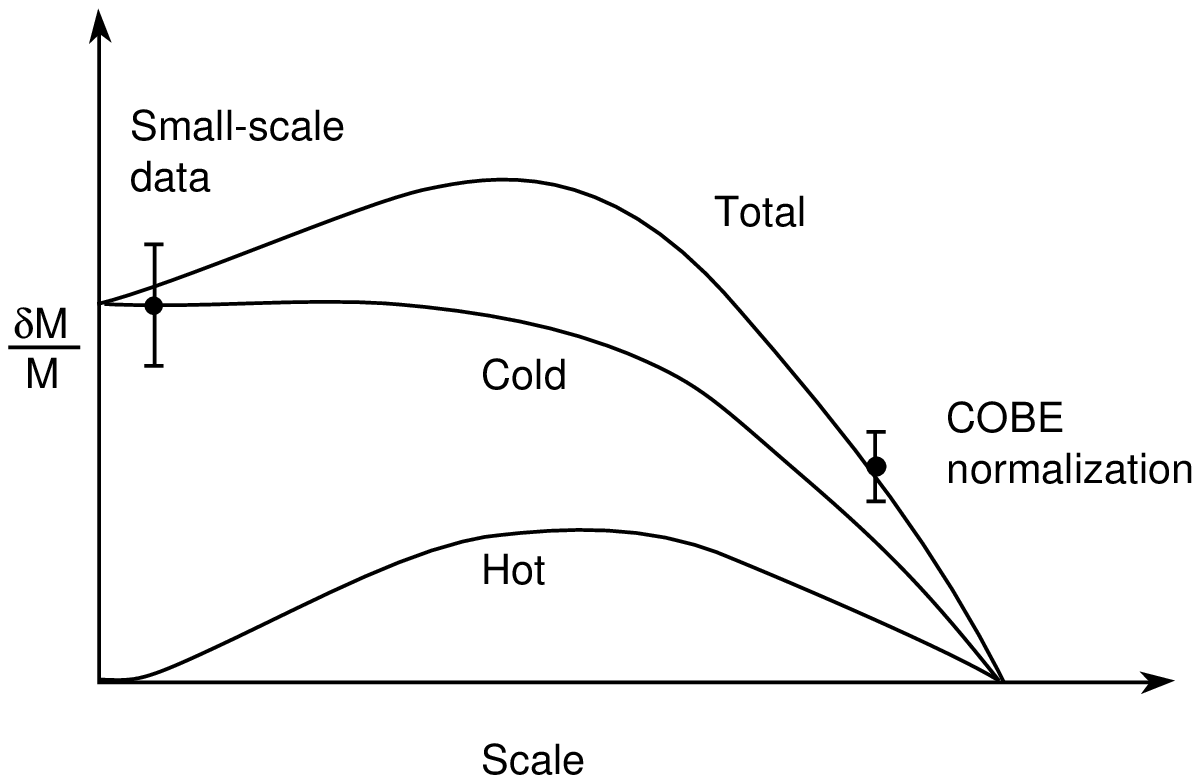,width=11.0truecm,angle=0}}
\end{center}
\mycapt{Illustration how a mixed dark matter scenario 
\protect\cite{mdm} may
reconcile the large-scale perturbations seen by COBE \protect\cite{COBE}
with the relatively small magnitude of the perturbations seen at
small scales.
\label{Fig3.3}
}
\end{figure}
The way in which this scenario works
is illustrated in Fig.~\ref{Fig3.3}.  Hot dark
matter alone would give a spectrum of perturbations that dies out
at small scales, whereas hot dark matter does not.  Combining the two,
one can reconcile the relatively high COBE normalisation at large scales
with the relatively small perturbations seen at small scales.

    The only realistic candidate that particle physicists are able to
offer for the hot dark matter is a neutrino weighing about 10 eV
\cite{hdm}, but
there are many candidates for the cold dark matter, of which we discuss
two in the rest of this lecture.

\subsection{Lightest Supersymmetric Particle}

    This is our favourite candidate for cold dark matter
\cite{ehnos}, and one for
which the polarized structure function measurements are particularly
relevant, as we shall see in the next section. The lightest
supersymmetric particle (LSP) is expected to be stable in many
models, and hence present in the Universe as a cosmological relic
from the Big Bang. This is because supersymmetric particles possess
a multiplicatively-conserved quantum number called $R$ parity
\cite{Fayet}, which takes
the values $+1$ for all conventional particles and $-1$ for all
their supersymmetric partners.  Its conservation is a consequence
of baryon and lepton number cancelation, since
\beq
R = (-1)^{3B + L + 2 S}
\label{E74}
\eeq
There are three important consequences of $R$ conservation:
\begin{enumerate}
\item  Sparticles should always be produced in pairs.
\item  Heavier sparticles should decay into lighter ones.
\item  The LSP should be stable, since it has no legal decay mode.
\end{enumerate}

    If the LSP had electric charge or strong interactions, it would
have condensed into galaxies, stars and planets such as ours, where
it could in principle be detected in searches for anomalous heavy
isotopes.  Electromagnetic or strong interactions would bind such
an LSP deeply inside some conventional nuclear species.  Various
searches have excluded such bound LSPs above the abundances
\cite{isotopes}
\beq
{n(\hbox{relics})\over n(\hbox{protons})}\,\sim\,
10^{-15}\,\ \hbox{to}\, \ 10^{-30}
\label{E75}
\eeq
for  1 GeV $\lappeq m_{LSP} \lappeq 10$ TeV, which are far below
the abundances calculated for such relic particles, which usually
lie in the range
\beq
{n(\hbox{relics})\over n(\hbox{protons})}\,\gsim\,
10^{-6}\,(\hbox{em})\,\,\,\ \hbox{to} \ \,\,10^{-10}\,(\hbox{strong})
\label{E76}
\eeq
We conclude \cite{ehnos} that any supersymmetric relic LSP should be
electromagnetically neutral and possess only weak interactions.
Scandidates in the
future sparticle data book include the sneutrino $\tilde\nu$ of spin
$0$, some form of ``neutralino" of spin $1/2$, or the gravitino
$\tilde G$ of spin $3/2$.  The sneutrino is essentially excluded by
the LEP experiments which measured the decay of the $Z^0$ into
invisible particles, which have
counted the number of light neutrino species: $2.991 \pm  0.0016$
\cite{Renton}, which does not leave space for any
sneutrino species weighing less than
${1\over2}M_{Z}$, and by underground experiments to be discussed in the
next section, which exclude a large range of heavier sneutrino masses.
Since the gravitino is probably impossible to discover, and is anyway
theoretically disfavoured as the LSP, we concentrate on the
neutralino \cite{ehnos}.

    The neutralino $\chi$ is a mixture of the photino $\tilde \gamma$,
the two neutral higgsinos $\tilde H_{1,2}^0$
expected in the minimal supersymmetric
extension of the Standard Model, and the zino
$\tilde Z$.  This is characterized essentially by three parameters,
the unmixed gaugino $m_{1/2}$, the Higgs mixing parameter $\mu$, and
the ratio of Higgs vacuum expectation values $\tan\,\beta$.  The
phenomenology of the lightest neutralino is quite complicated in general,
but simplifies in the limit $m_{1/2} \rightarrow 0$, where $\chi$ is
approximately a photino state \cite{Goldberg},
and in the limit $\mu\rightarrow 0$,
where it is approximately a higgsino state.  
\begin{figure}[htb]
\begin{center}
\mbox{\epsfig{file=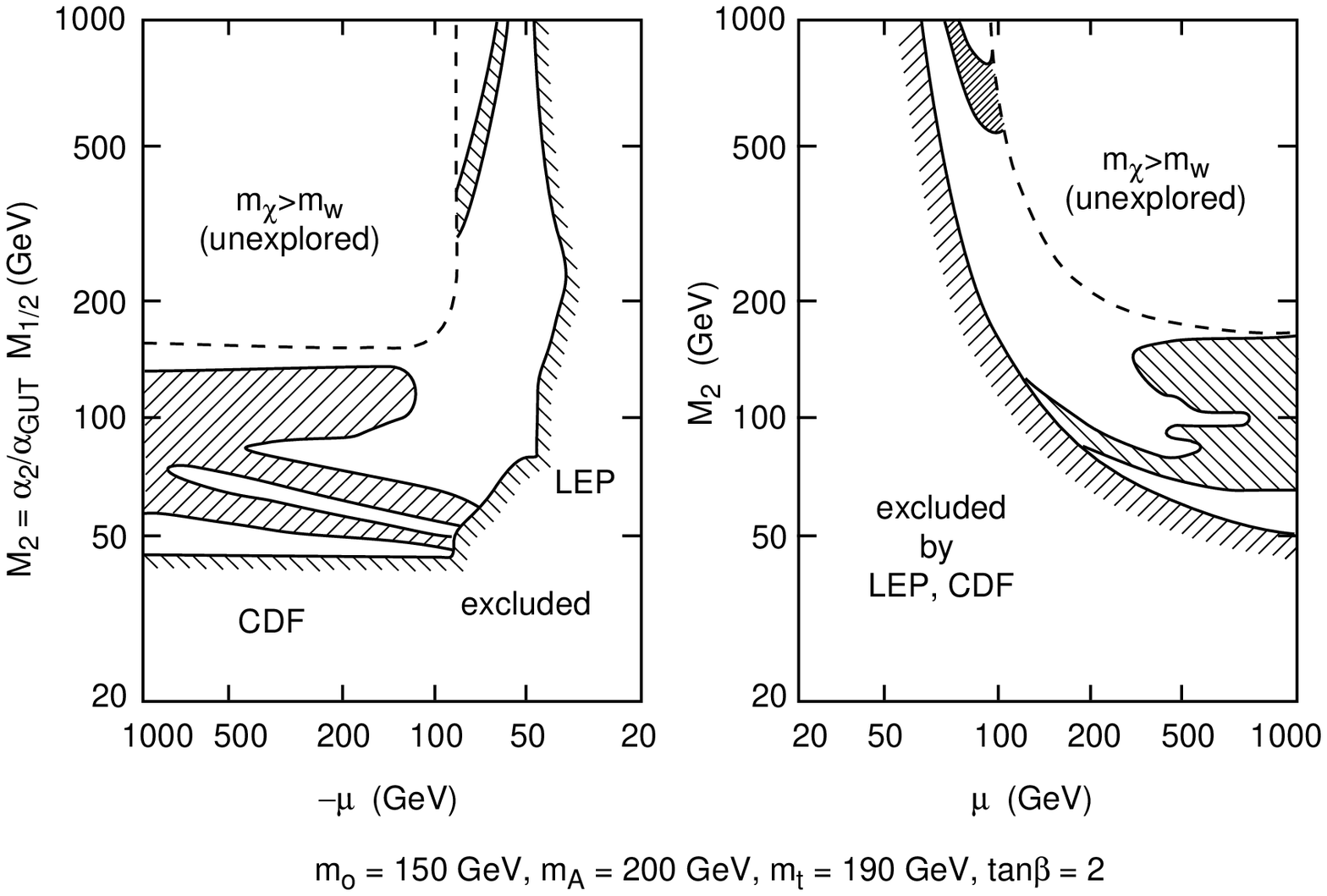,width=11.0truecm,angle=0}}
\end{center}
\mycapt{The cosmological relic density of neutralinos $\chi$ 
\protect\cite{ehnos}
may well (shaded regions)
lie in the range of interest to astrophysicists and
cosmologists \protect\cite{density},
 namely $0.1 \protect\lsim \Omega_{\chi} \protect\lsim 1$.
\label{Fig3.4}
}
\end{figure}
As seen in Fig.~\ref{Fig3.4},
experimental constraints from LEP and the Fermilab collider in fact
exclude these two extreme limits \cite{erz}, so that
\beq
m_\chi\gsim (10\ \,\hbox{to}\ \,20) \,\hbox{GeV}
\label{E77}
\eeq
Fig.~\ref{Fig3.4} also indicates that there are generic domains of parameter
space where the LSP may have an ``interesting" cosmological relic
density \cite{density}, namely
\beq
0.1 \lsim \Omega_\chi H_0^2 \lsim 1
\label{E78}
\eeq
for some suitable choice of supersymmetric model parameters.  
\begin{figure}[htb]
\begin{center}
\mbox{\epsfig{file=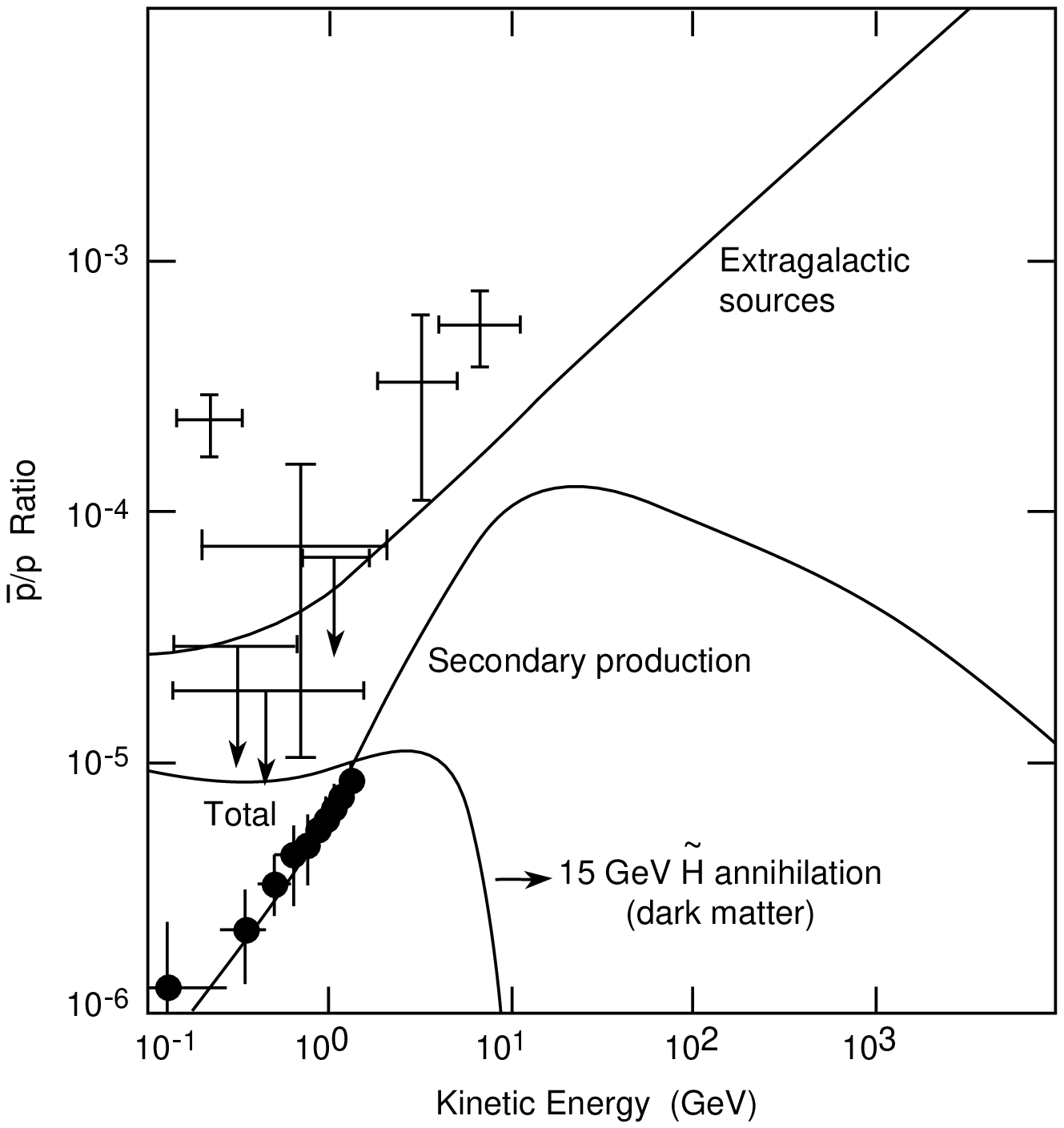,width=9.0truecm,angle=0}}
\end{center}
\mycapt{The results of experimental searches for cosmic-ray
antiprotons
\protect\cite{previous}, compared with the fluxes expected from
primary matter cosmic rays 
\protect\cite{Gaisser} and a supersymmetric
model\protect\cite{Freese}.
The lower points with error bars are what should be 
obtainable with the AMS\break
 experiment\protect\cite{AMS}.
\label{Fig3.5}
}
\end{figure}
Fig.~\ref{Fig3.5}
displays the calculated LSP density in a sampling of phenomenological
models \cite{sample}
where we see that an interesting cosmological density is quite
plausible for LSP masses
\beq
20\ \hbox{GeV} \lsim m_\chi \lsim 300 \ \hbox{GeV}
\label{E79}
\eeq
The next problem is how to detect the LSP, and this is where polarized
structure function measurements may have a role to play.

\subsection{Searches for Neutralinos}

    Several strategies to search for cosmological relic neutralinos have
been proposed.  One is to look for the products of their annihilations
in our galactic halo \cite{haloann}.
Here the idea is that two self-conjugate $\chi$
particles may find each other while circulating in the halo, and have
a one-night stand and annihilate each other:
$\chi\chi\rightarrow\ell\ell,\bar q q$,
leading to a flux of stable particles such as $\bar p, e^+,\gamma,\nu$
in the cosmic rays.  Several experiments have searched for cosmic-ray
antiprotons \cite{previous},
with the results shown in Fig.~\ref{Fig3.5}.  At low energies there
are only upper limits, but there are several positive detections at
higher energies, which are comparable with the flux expected from
secondary production by primary matter cosmic rays
\cite{Gaisser}. As also seen
in Fig.~\ref{Fig3.5}, relic LSP annihilation in our galactic halo might produce
an observable flux of low-energy cosmic ray antiprotons somewhat below
the present experimental upper limits \cite{Freese}.
These may be interpreted as
suggesting that
\beq
\rho_\chi \lsim 10\,\rho_{halo}
\label{E80}
\eeq
NASA and the DOE have recently approved a satellite experiment called
AMS \cite{AMS},
which should be able to improve significantly the present upper
limits on low-energy antiprotons, and may be able to start constraining
significantly supersymmetric models.  It is also possible to derive
limits on such models from the present experimental measurements of
the cosmic-ray $e^+$ and $\gamma$ fluxes \cite{Bergstrom},
but these are not yet very constraining.

    A second LSP detection strategy is to look for $\chi\chi$
annihilation inside the Sun or Earth.  Here the idea is that a relic
LSP wandering through the halo may pass through the Sun or Earth
\cite{solar},
collide with some nucleus inside it, and thereby lose recoil energy.
This could convert it from a hyperbolic orbit into an elliptic one,
with a perihelion or perigee below the solar or terrestrial radius.
If so, the initial capture would be followed by repeated scattering
and energy loss, resulting in a quasi-isothermal distribution within the
Sun or Earth.  The resulting LSP population would grow indefinitely,
\`a la Malthus, unless it were controlled either by emigration,  namely
evacuation from the surface, or by civil war, namely annihilation within
the Sun or Earth.  Evaporation is negligible for $\chi$ particles
weighing more than a few GeV \cite{Gould},
so the only hope is annihilation.  The
neutrinos produced by any such annihilation events would escape from
the core, leading to a high-energy solar neutrino flux ($E_{\nu}
\gappeq 1$ GeV).  This could be detected either directly in an
underground experiment, or indirectly via a flux of upward-going muons
produced by neutrino collisions in the rock.

    The polarized structure function measurements enter in the estimate
of the $\chi$ capture rate, which enters in the following
general formula \cite{EFR} for the neutrino flux:
\beq
R_\nu{=}2.7\times 10^{-2} f\left(m_\chi/m_p\right)
\left({\sigma (\chi p \rightarrow \chi p)
\over 10^{-40}\, \hbox{cm}^2}\right)
\left({\rho_\chi\over 0.3\,\hbox{GeV}\,\hbox{cm}^{-3}}\right)
\kern-0.2em
\left({300 \ \hbox{km\,s}^{-1}\over \bar v_{\chi}}\right)
\times F_\nu
\label{E81}
\eeq
where we have simplified to the case of capture by the Sun, where
proton targets predominate.  Here $f$ is a kinematic function,
$\rho_{\chi}$ and $\bar v_{\chi}$ are the local density and mean velocity
of the halo LSPs, and $F_\nu$ represents factors associated with
the neutrino interaction rate in the apparatus.  The factor which
interests us here is the elastic LSP-proton scattering cross section
$\sigma (\chi\,p \rightarrow \chi\,p)$.

\begin{figure}[htb]
\begin{center}
\mbox{\epsfig{file=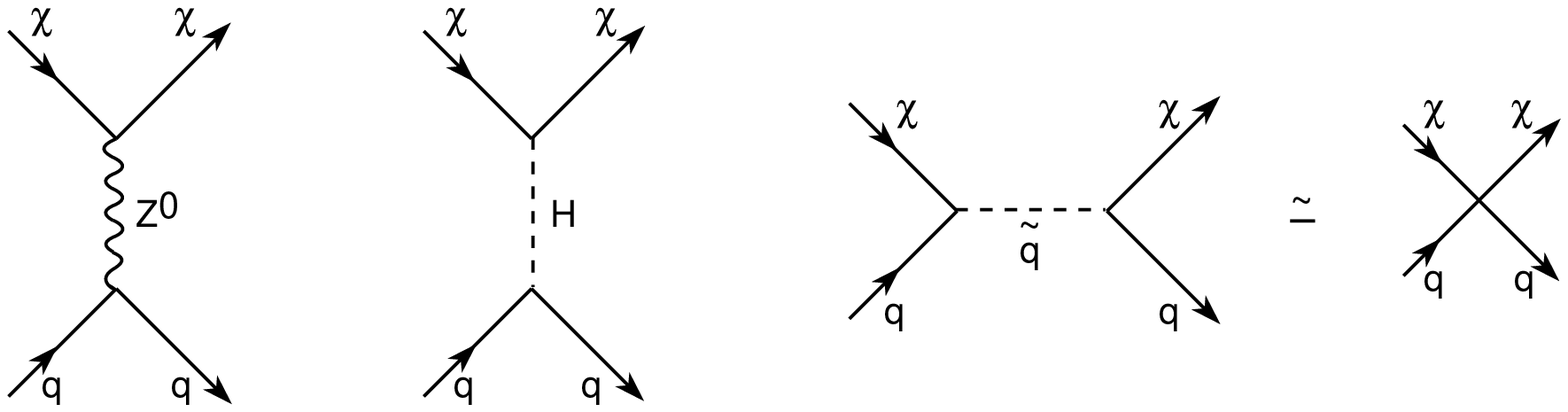,width=10.0truecm,angle=0}}
\end{center}
\mycapt{The exchanges of massive particle such as the $Z^0$, Higgs
boson and squarks give rise to an effective four-fermion
interaction between the neutralinos $\chi$ and quarks inside a
proton target.
\label{Fig3.6}
}
\end{figure}

    To see how the polarized structure functions enter into the
estimation of the elastic scattering cross section \cite{EFR},
first note
that the LSP interacts with hadrons via an effective four-fermion
interaction of the general form $\chi\chi\bar q q$, which is
mediated by the exchanges of massive particles such as the $Z^0$,
Higgs bosons and squarks $\tilde q$, as seen in Fig.~\ref{Fig3.6}.
This four-fermion interaction
is similar in many ways to the original Fermi four-fermion weak
interaction mediated by $W^{\pm}$ exchange. The matrix elements
of the latter interaction between nucleon states are governed by
the familiar $\beta$-decay constant $g_A$, which may be written
in the form
\beq
|g_A| = \Delta u - \Delta d
\label{E82}
\eeq
Analogous expressions in terms of the $\Delta q$ exist for the
spin-dependent part of the $\chi$-nucleon scattering matrix
element. For example, if the $\chi$ were to be a pure photino
state, we would have
\beq
a_p = {4\over9} \Delta u + {1\over9}\Delta d + {1\over9}\Delta s
\label{E83}
\eeq
It is amusing to note that this is exactly the same combination
of the $\Delta q$ that appears in charged-lepton scattering off
a proton target \cite{EFR}.
In a sense, the EMC and its successors have
been measuring $\tilde \gamma$-nucleon scattering!

The EMC and subsequent measurements indicate values of the
$\Delta q$ that are rather different from those predicted by
the \naive\ quark model, which means that the dark matter
scattering matrix elements are also rather different, and hence
also the upper limits on the halo dark matter density that can
be deduced from a given search for high-energy neutrinos from
the core of the Sun or the Earth.
By now, as discussed in the first
lecture, the determination of the $\Delta q$ is quite
precise, and, as discussed earlier in this lecture, we no
longer believe that the LSP $\chi$ can be a pure $\tilde \gamma$.
Consider, for example, the plausible case of an essentially pure
$U(1)$ gaugino LSP $\tilde B$: its scattering matrix elements
on protons and neutrons are given by \cite{BjSRalphas}
\bea
a_p &\simeq& {17\over36} \Delta u + {5\over 36}
(\Delta d + \Delta s)
\nonumber\\
\label{E84}\\
a_n &\simeq &\phantom{aaa}(\Delta u \leftrightarrow \Delta d)
\nonumber
\eea
We see that, in this case, the uncertainties
\cite{BjSRalphas} from polarized
structure measurements are likely to be much smaller than
those from other components in equation (\ref{E81}).

\begin{figure}[htb]
\begin{center}
\mbox{\epsfig{file=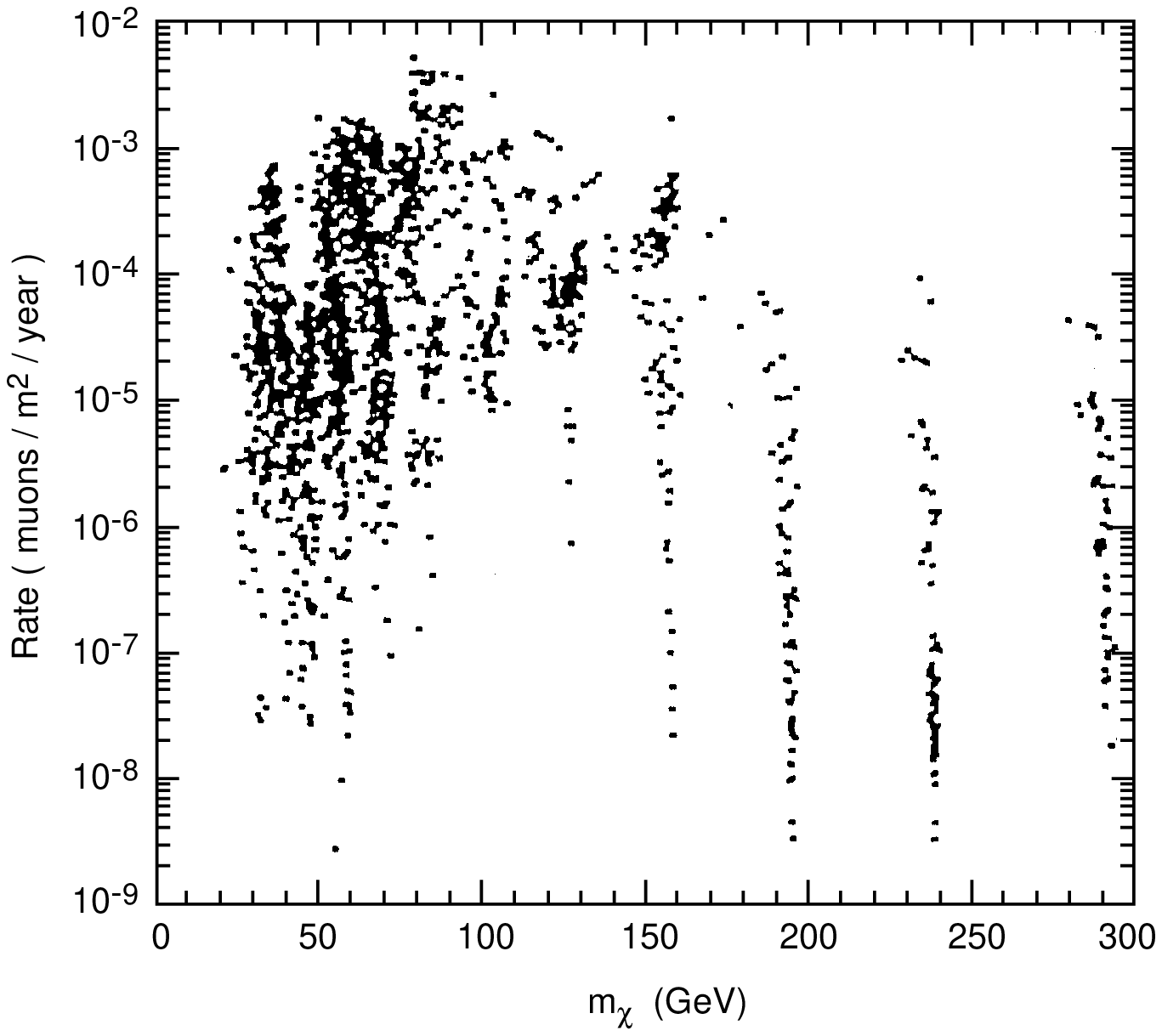,width=10.0truecm,angle=0}}
\end{center}
\mycapt{The flux of upward-going muons expected from $\chi \chi$
annihilation inside the Sun in a sampling of supersymmetric 
models \protect\cite{sample}.
\label{Fig3.7}
}
\end{figure}
Some typical rate estimates for upward-going muons originating
from high-energy solar neutrinos in
a sampling of supersymmetric
models are shown in Fig.~\ref{Fig3.7}:: while some models
are already excluded by unsuccessful searches, most are not
\cite{sample}.
\begin{figure}[htb]
\begin{center}
\mbox{\epsfig{file=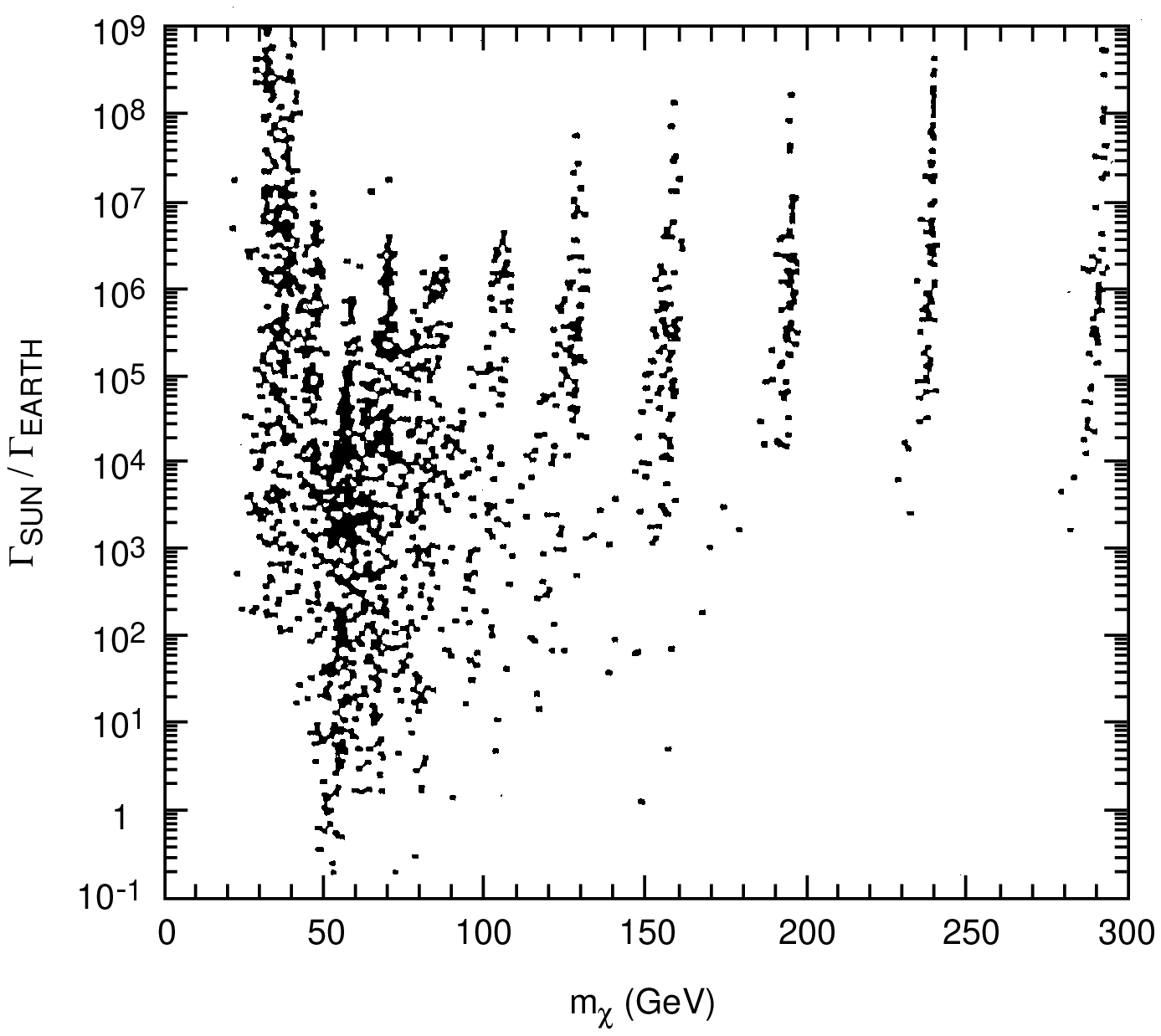,width=10.0truecm,angle=0}}
\end{center}
\mycapt{A comparison of the muon fluxes from the centre of the Sun
and Earth, as found in a sampling of supersymmetric models
\protect\cite{sample}.
\label{Fig3.8}
}
\end{figure}
We see in Fig.~\ref{Fig3.8} that searches for solar signals usually
constrain models more than searches for terrestrial signals,
though this is not a model-independent fact
\cite{sample}. In the long run,
it seems that a search for upward-going neutrino-induced muons
with a $1$ km$^2$ detector could almost certainly detect LSP
annihilation \cite{Halzen}, if most of the cold dark matter is indeed
composed of LSPs.

    The third LSP search strategy is to look directly for
LSP scattering off nuclei in the laboratory \cite{GW}. It is easy to
see that the typical recoil energy
\beq
\Delta E < m_\chi v^2 \simeq 10
\left({m_\chi\over 10\,\hbox{GeV}}\right)\,\hbox{keV}
\label{E85}
\eeq
deposited by elastic $\chi$-nucleus scattering would probably
lie in the range of $10$ to $100$ keV. The type of spin-dependent
interaction, mediated by $Z^0$ or $\tilde q$ exchange,
 that we discussed in previous paragraphs is likely
to dominate for light nuclei
\cite{Flores}, whereas coherent spin-dependent
interactions mediated by $H$ and $\tilde q$ exchange are likely
to dominate scattering off heavy nuclei
\cite{Griest}. As we have already
discussed, the spin-dependent interactions on individual nucleons
are controlled by the $\Delta q$: translating these into matrix
elements for interactions on nuclei depends on the decomposition
of the nuclear spin, which must be studied using the shell model
\cite{Flores} or some other theory of nuclear structure
\cite{nuclear}. The spin-independent
interactions on individual nucleons are related to the different
quark and gluon contributions to the nucleon mass, which is also
an interesting phenomenological issue related to the $\pi$-nucleon
$\sigma$-term discussed in Lecture II. Again, the issue of nuclear
structure arises when one goes from the nucleon level to coherent
scattering off a nuclear target.

\begin{figure}[htb]
\begin{center}
\mbox{\epsfig{file=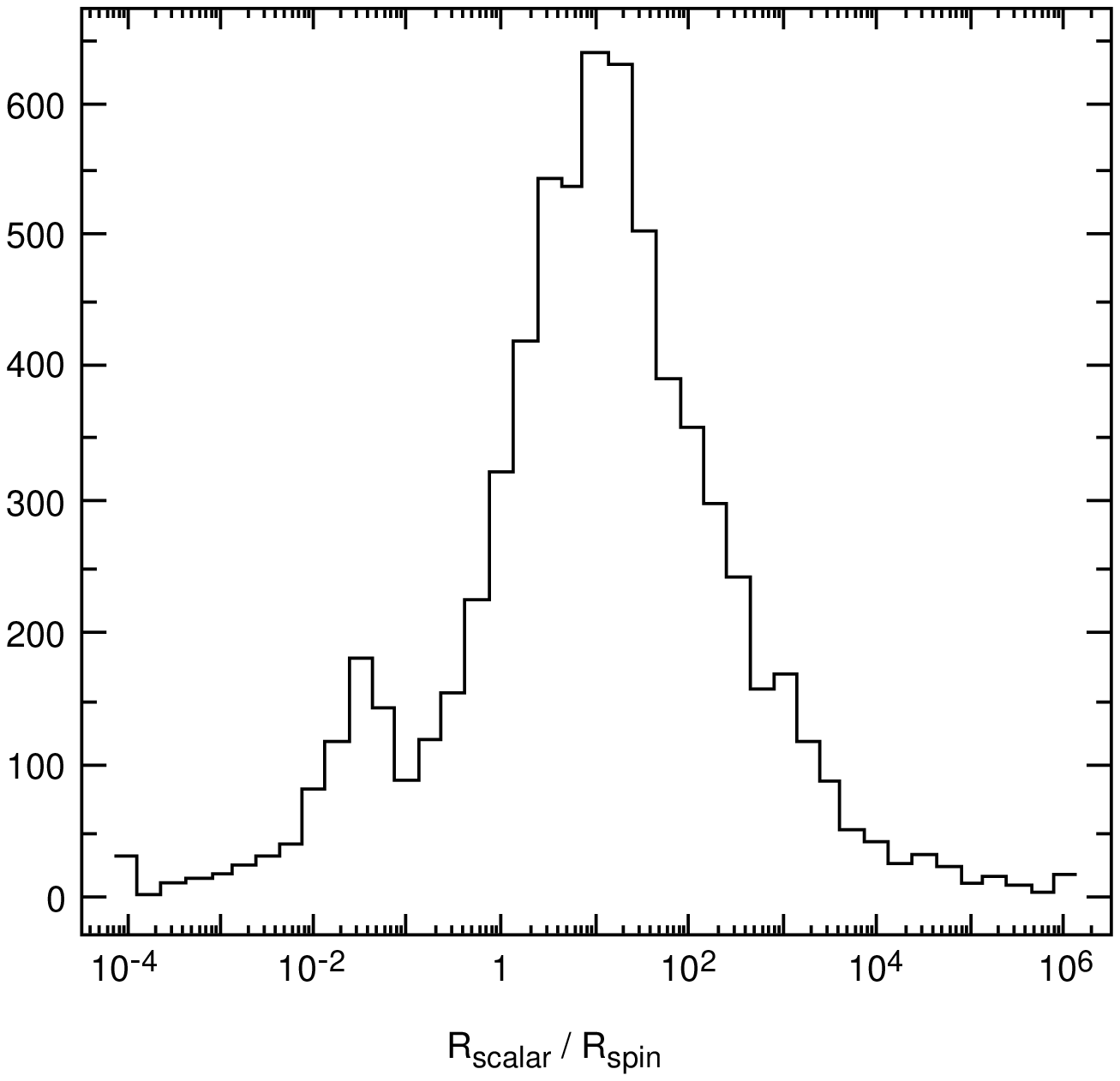,width=10.0truecm,angle=0}}
\end{center}
\mycapt{A comparison of the spin-dependent and spin-independent
interaction rates of relic neutralinos $\chi$ with Germanium in a
sampling of supersymmetric models \protect\cite{sample}.
\label{Fig3.9}
}
\end{figure}
\begin{figure}[htb]
\begin{center}
\mbox{\epsfig{file=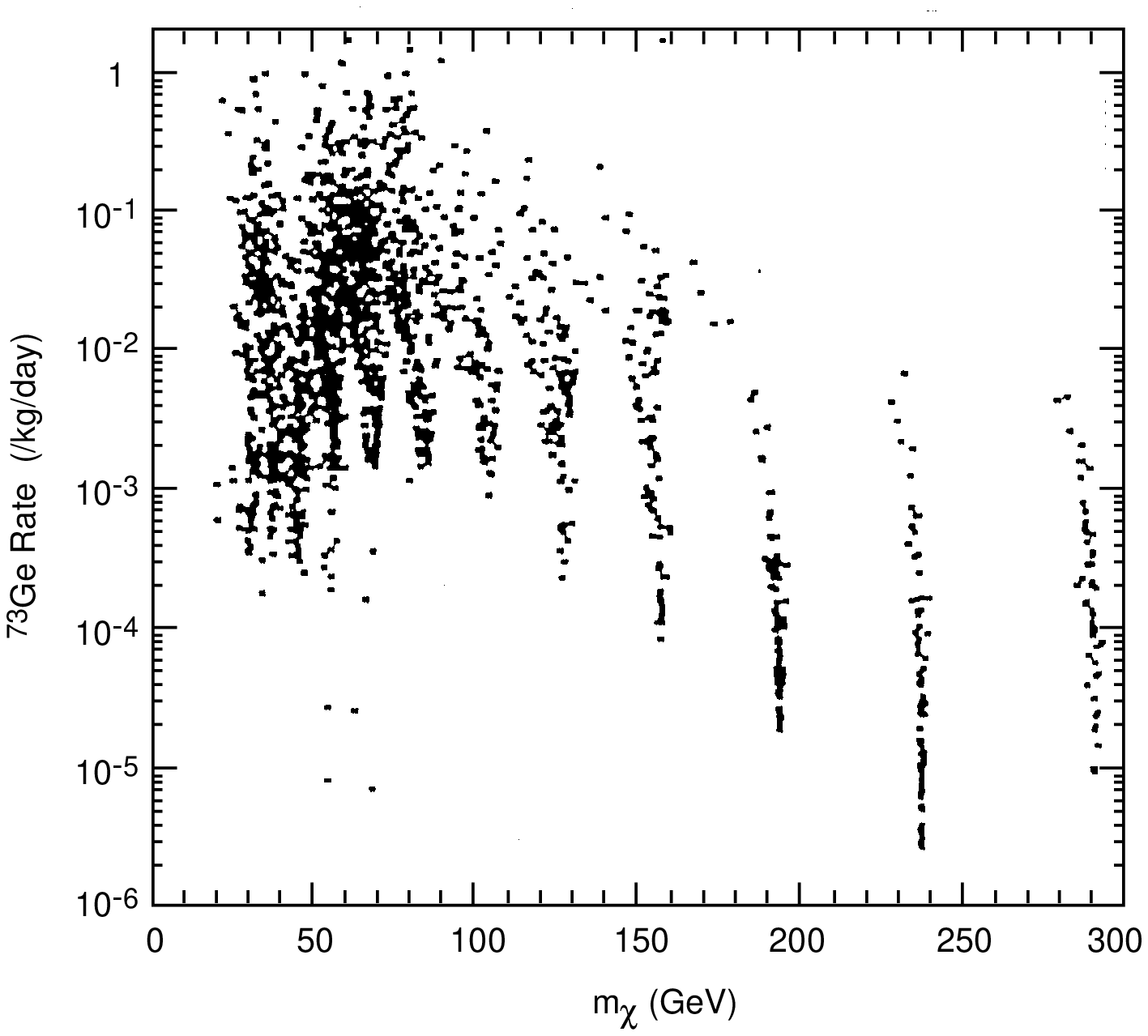,width=10.0truecm,angle=0}}
\end{center}
\mycapt{Total scattering rates on Germanium in a sampling of
supersymmetric \break models~\protect\cite{sample}.
\label{Fig3.10}
}
\end{figure}
We will not discuss here the details of such nuclear calculations,
but present in Figs~\ref{Fig3.9} and \ref{Fig3.10}
the results of a sampling of
different supersymmetric models
\cite{sample}. We see in Fig.~\ref{Fig3.9} that the
spin-independent contribution tends to dominate over the
spin-dependent one in the case of Germanium, those this is not
universally true, and would not be the case for scattering off
Fluorine \cite{Flores}.
In Fig.~\ref{Fig3.10} we plot the scattering rates off $^{73}$Ge,
where we see that there are many models in which more than
$0.01$ events/kg/day are expected, which may be observable.
Thus the direct search for cold dark matter scattering in the
laboratory may be a useful complement to the searches for
supersymmetry at accelerators.

\subsection{Axions}

The axion \cite{axion} is our second-favourite candidate for the cold
dark matter. It was invented to guarantee conservation of
$P$ and $CP$ in the strong interactions, which would otherwise
be violated by the $\theta$ parameter:
\beq
{\cal L}_{QCD} \,\ni\,
{\theta \,g^2 \over 32 \pi^2} \,\epsilon^{\mu\nu\rho\sigma}
\,G_{\mu\nu} G_{\rho\sigma}
\label{E86}
\eeq
which is known experimentally \cite{RPP} to be smaller than about $10^{-9}$.
The $\theta$ parameter relaxes to zero in any extension of
the Standard Model which contains the axion, a light pseudoscalar
boson with mass and
couplings to matter that are scaled inversely by the
axion decay constant $f_a$:
\beq
m_a \sim {\Lambda_{QCD} \,m_q \over f_a}\,,\quad
g_{a\kern-0.1em\bar{f} \kern-0.1em f} \sim {m_f\over f_a}\,,\qquad
g_{a\gamma\gamma} \sim {1\over f_a}
\label{E87}
\eeq
The fact that no axion has been seen in any accelerator experiment
tells us that
\beq
f_a \gsim 1\,\hbox{TeV}
\label{E88}
\eeq
and hence that any axion must be associated with physics beyond
the scale of the Standard Model.

Axions would have been produced in the early Universe in the
form of slow-moving coherent waves that could constitute
cold dark matter. The relic density of these waves has been
estimated as \cite{adensity}
\beq
\Omega_a \,\simeq\,
\left({0.6\times 10^{-5}\,\hbox{eV}\over m_a}\right)^{7/6}
\left({200\,\hbox{MeV}\over \Lambda_{QCD}}\right)^{3/4}
\left({75\over H_0}\right)^2
\label{E89}
\eeq
which is less than unity if
\beq
f_a \lsim 10^{12}\,\hbox{GeV}
\label{E90}
\eeq
In addition to these coherent waves, there may also be axions
radiated from cosmic strings
\cite{radiation}, which would also be non-relativistic
by now, and hence contribute to the relic axion density and
strengthen the limit in equation (\ref{E90}).

The fact that the Sun shines photons rather than axions, or, more
accurately but less picturesquely, that the standard solar model
describes most data, implies the lower limit \cite{Sun}
\beq
f_a \gsim 10^{7}\,\hbox{GeV}
\label{E91}
\eeq
This has been strengthened somewhat by unsuccessful searches for
the axio-electric effect, in which an axion ionizes an atom. More
stringent lower bounds on $f_a$ are provided by the agreements
between theories of Red Giant and White Dwarf stars with the
observations \cite{stars}\ :
\beq
f_a \gsim 10^{9}\,\hbox{GeV}
\label{E92}
\eeq
Between equations (\ref{E90}) and (\ref{E92})
there is an open window in which the
axion could provide a relic density of interest to astrophysicists
and cosmologists.

Part of this window may be closed by the observations of the
supernova SN1987a, which is where the polarized structure
function measurements come into play. According to the standard
theory of supernova collapse to form a neutron star,
$99 \%$ of the binding energy released in the collapse to the
neutron star escapes as neutrinos.
This theory agrees \cite{neutrinos}
with the observations of SN1987a made
by the Kamiokande \cite{Kam}
and IMB experiments \cite{IMB}, which means that most of
the energy could not have been carried off by other invisible
particles such as axions.

\begin{figure}[H]
\begin{center}
\mbox{\epsfig{file=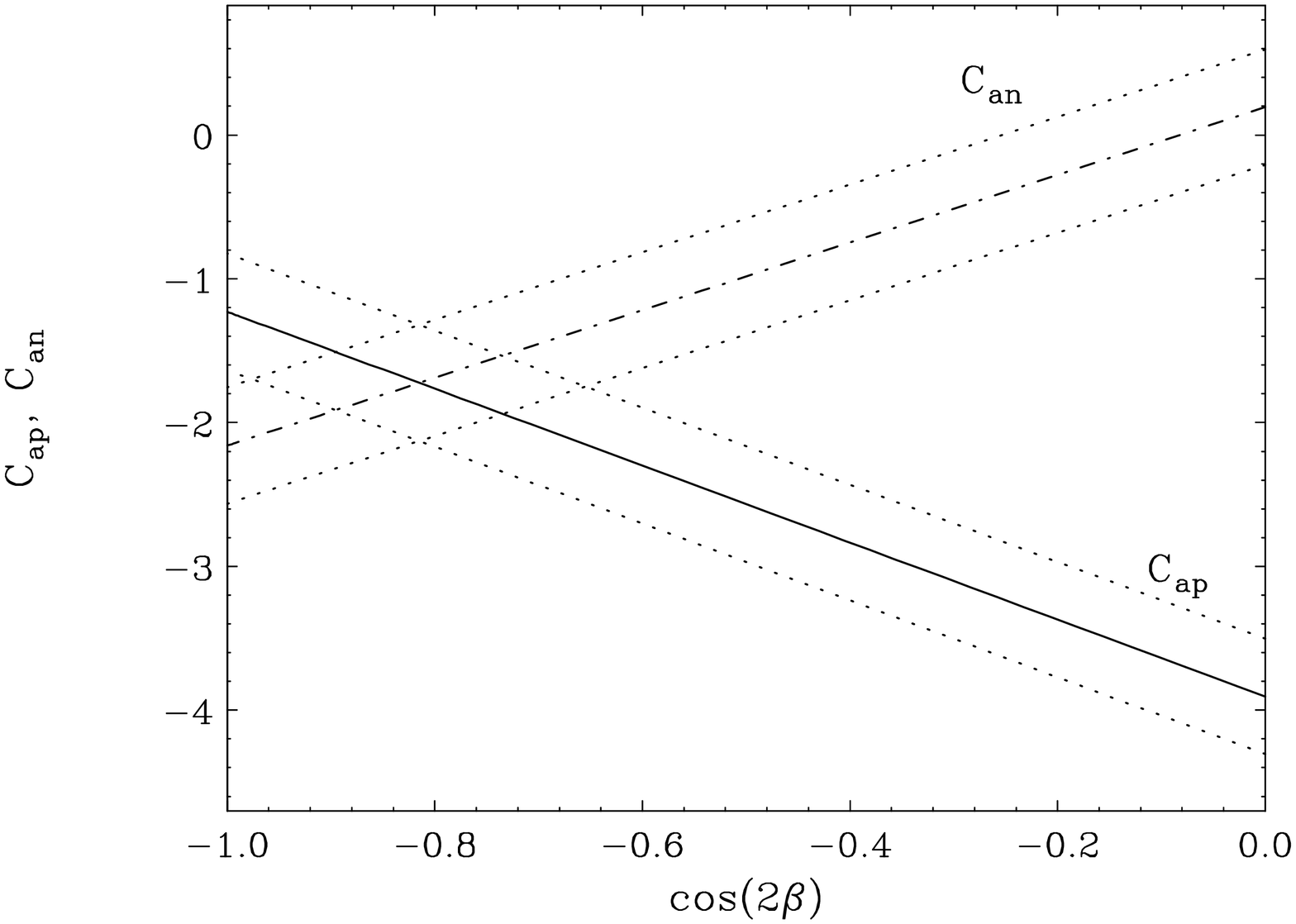,width=11.0truecm,angle=90}}
\end{center}
\mycapt{
Axion couplings to the nucleon, as determined in 
Ref.~\protect\bcite{BjSRalphas} 
using polarized structure function data.
\label{Fig3.11}
}
\end{figure}

Since the axion is a light pseudoscalar boson, its couplings to
nuclear matter are related by a generalized Goldberger-Treiman relation
to the corresponding axial-current matrix elements, and these are
in turn determined by the corresponding $\Delta q$
\cite{GTaxion}. Specifically,
we find for the axion couplings to individual nucleons that
\bea
C_{ap} =  \,2[{-} 2.76\, \Du - 1.13\, \Dd + 0.89\, \Ds
-\cos 2 \beta \, (\Du - \Dd - \Ds) ]\,,
\nonumber\\
\label{E93}\\
C_{an} = \,2[{-} 2.76\, \Dd - 1.13\, \Du + 0.89\, \Ds
-\cos 2 \beta \, (\Dd - \Du - \Ds) ]
\phantom{\,,}
\nonumber
\eea
Evaluating the $\Delta q$ at a momentum scale around $1$ GeV, as is
appropriate in the core of a neutron star, we estimate
\cite{BjSRalphas} that
\bea
C_{ap} &=& ({-}3.9 \pm 0.4) - (2.68 \pm 0.06)
\cos 2 \beta
\nonumber\\
\label{E94}\\
C_{an} &=& (0.19 \pm 0.4)\, + \,(2.35 \pm 0.06)
\cos 2 \beta
\phantom{,}
\nonumber
\eea
which are plotted in  Fig.~\ref{Fig3.11}.
As in the case of LSP scattering, the uncertainties associated with
polarized structure function measurements are by now considerably
smaller than other uncertainties, in this case particularly those
associated with the nuclear equation of state. The total axion
emission rate from the core of a neutron star is approximately
proportional to the combination
\beq
{C_{an}}^2 + 0.8 \,(C_{an} + C_{ap})^2 + 0.5 \,{C_{ap}}^2
\label{Csquares}
\eeq
\begin{figure}[htb]
\begin{center}
\mbox{\epsfig{file=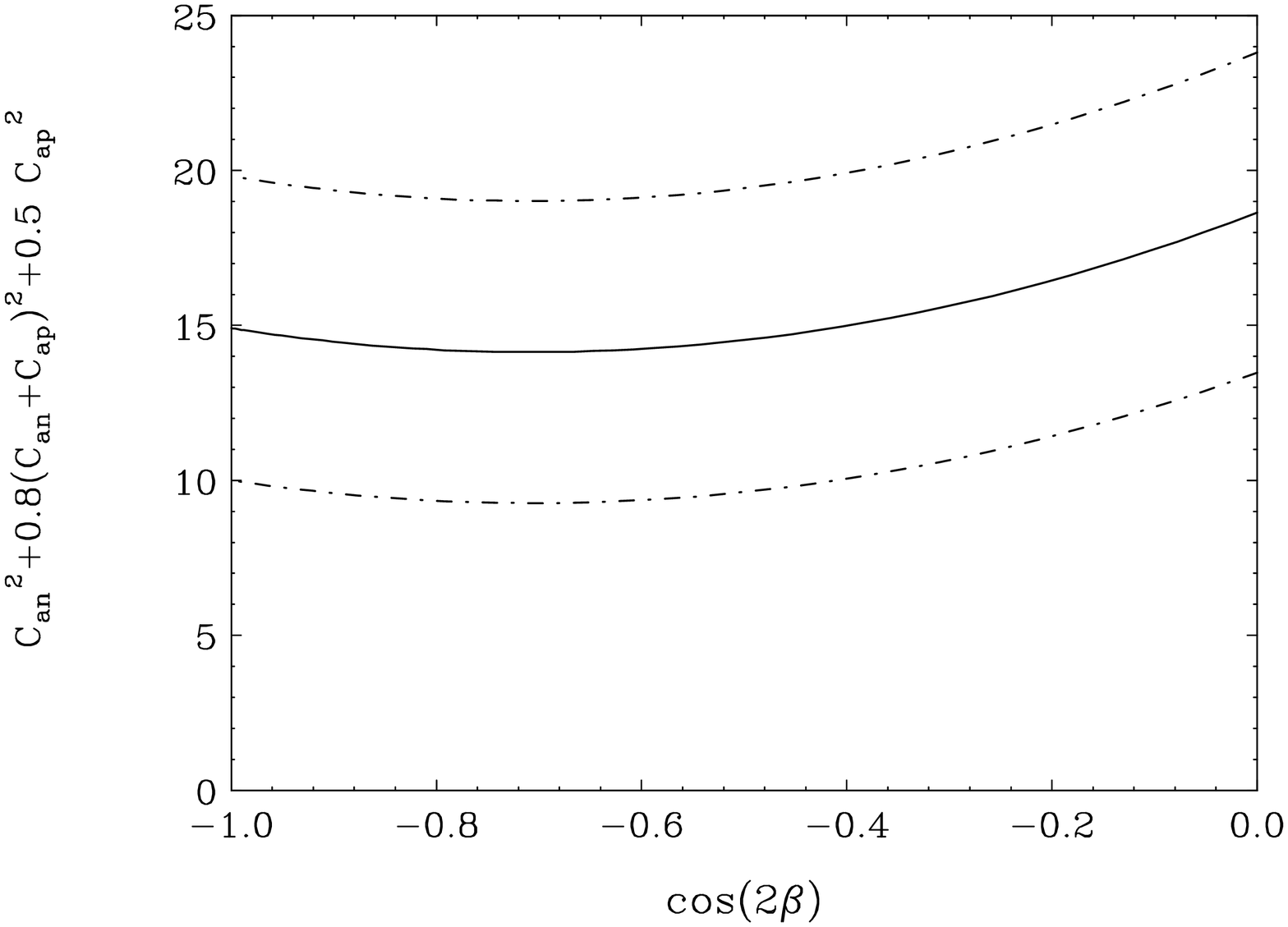,width=11.0truecm,angle=90}}
\end{center}
\mycapt{The total axion
emission rate from the core of a neutron star, proportional to
the combination
${C_{an}}^2 + 0.8 \,(C_{an} + C_{ap})^2 + 0.5 \,{C_{ap}}^2$,
as determined in Ref.~\protect\bcite{BjSRalphas}.
\label{Fig3.12}
}
\end{figure}
which is plotted in  Fig.~\ref{Fig3.12}, together with the associated
with the errors in the $\Delta q$. Estimating axion emission
rates in this way
\cite{newsn}, it seems that part of the previous axion
window is still open, and an experiment
\cite{axexp} is underway which should
be able to detect halo axions if they occupy this window.

\subsection{The Role of Spin in the Forthcoming Revolutions in
Particle Physics and Cosmology}

Measurements of the spin decomposition may not only overturn
our understanding of nucleon structure, but may also help usher
in the revolutions in particle physics and cosmology for which
we all yearn. A ``Standard Model" for
 the formation of structure
in the Universe is now emerging, in which Gaussian perturbations
laid down during an inflationary epoch are amplified by gravitational
instabilities fed by cold dark matter particles. The COBE
observations of primordial fluctuations in the microwave background
spectrum \cite{COBE}
were perhaps the first observational evidence for this
picture, just as the discovery of neutral currents heralded the
establishment of the Standard Model of elementary particles.
Confirmation of this emerging picture of structure formation
would be provided by the observation of one or more species of
dark matter particle, such as massive neutrinos and LSPs or
axions. The discovery of any of these would also cause a
revolution in particle physics by taking us finally
beyond the Standard Model, as well as telling
us how galaxies and clusters were formed. As we have seen in this
last lecture, polarized structure function measurements control
the couplings to matter of both LSPs and axions, and could well
play an important role in their detection.

\bigskip
\noindent
{\large \it Long live the Spin Revolution!}
\vfill\eject
%-------- end of lecture III ------------------------
\begin{flushleft}
{\large\bf Acknowledgements}
\end{flushleft}
We
thank Michelle Mazerand for her help in preparing the manuscript.
The research described in these lectures
was supported in part by the Israel Science Foundation
administered by the Israel Academy of Sciences and Humanities,
and by
a Grant from the G.I.F., the
German-Israeli Foundation for Scientific Research and
Development.

\end{document}